\documentclass[aps,prfluids,twocolumn,reprint,superscriptaddress,floatfix,longbibliography]{revtex4-2}
\usepackage[T1]{fontenc}
\usepackage{ae,aecompl}
\usepackage{graphicx,color}
\usepackage{amsmath,amsfonts,amssymb,cancel} 
\usepackage{times}
\usepackage{bm}
\usepackage{gensymb}
\allowdisplaybreaks
\usepackage{newtxtext,newtxmath}
\usepackage{hyperref}
\PassOptionsToPackage{hyphens}{url}
\usepackage{color}
\definecolor{Brown}{rgb}{0.647,0.165,0.165}
\definecolor{NavyBlue}{rgb}{0.0,0,0.5}
\definecolor{Burgundy}{rgb}{0.5,0.0,0.125}
\hypersetup{
    breaklinks=true,
    colorlinks=true,
    citecolor={Burgundy},
    linkcolor={NavyBlue},
    urlcolor={Brown}
}

\def\apj{Astrophys.\ J.}

\def\mnras{Mon.\ Not.\ R.\ Astron.\ Soc.}
\def\aap{Astron.\ Astrophys.}
\def\apjl{Astrophys.\ J.\ Lett.}

\def\physrep{Phys.\ Rep.}
\def\pre{Phys.\ Rev.\ E}
\def\prl{Phys.\ Rev.\ Lett.}

\def\apjs{ApJS}

\def\ssr{SSRv}

\def\araa{Ann.\ Rev.\ Astron.\ Astrophys.}

\newcommand{\lb}{\ell_b} 
\newcommand{\lu}{\ell_u} 
\newcommand{\ld}{\ell_0} 

\newcommand{\cs}{c_\mathrm{s}}           
\newcommand{\Mach}{\mathcal{M}}      
\newcommand{\Pm}{\text{Pm}}
\renewcommand{\Re}{\text{Re}} 
\newcommand{\Rm}{\text{Rm}} 

\renewcommand{\vec}[1]{\bm{#1}}	
\newcommand{\dd}{\mathrm{d}}        

\newcommand{\brms}{b_{\rm rms}}

\newcommand{\urms}{u_{\rm rms}}

\newcommand{\w}{\omega}
\newcommand{\jrms}{j_{\rm rms}}
\newcommand{\wrms}{\omega_{\rm rms}}

\renewcommand{\i}{{\mathrm{i}}}

\newcommand{\ku}{\mathcal{K}}
\newcommand{\cosuw}{\cos(\theta)_{\vec{u}, \vec{\omega}}}
\newcommand{\cosub}{\cos(\theta)_{\vec{u}, \vec{b}}}
\newcommand{\cosjb}{\cos(\theta)_{\vec{j}, \vec{b}}}
\newcommand{\kpar}{k_{\parallel}}
\newcommand{\kbdj}{k_{\vec{b} \cdot \vec{j}}}
\newcommand{\kbcj}{k_{\vec{b} \times \vec{j}}}
\newcommand{\krms}{k_{\mathrm{rms}}}
\newcommand{\lambdacomp}{{\lambda}_{\mathrm{comp}}}
\newcommand{\lambdanull}{{\lambda}_{\mathrm{null}}}
\newcommand{\lambdastre}{{\lambda}_{\mathrm{stre}}}
\newcommand{\ecomp}{{\vec{e}}_{\mathrm{comp}}}
\newcommand{\enull}{{\vec{e}}_{\mathrm{null}}}
\newcommand{\estre}{{\vec{e}}_{\mathrm{stre}}}
\newcommand{\cosecompb}{\cos(\theta)_{{\vec{e}}_{\mathrm{comp}}, \vec{b}}}

\newcommand{\cosestreb}{\cos(\theta)_{{\vec{e}}_{\mathrm{stre}}, \vec{b}}}

\newcommand\Eq[1]{Eq.~\ref{#1}}
\newcommand\Fig[1]{Fig.~\ref{#1}}
\newcommand\Sec[1]{Sec.~\ref{#1}}
\newcommand\Tab[1]{Table~\ref{#1}}

\newcommand\rev[1]{#1}

\graphicspath{{./}{figs/}}

\begin{document} 
\title{Saturation mechanism of the fluctuation dynamo in supersonic turbulent plasmas}
\author{Amit Seta}\thanks{amit.seta@anu.edu.au}
\affiliation{Research School of Astronomy and Astrophysics, Australian National University, Canberra, ACT 2611, Australia}
\author{Christoph Federrath}
\affiliation{Research School of Astronomy and Astrophysics, Australian National University, Canberra, ACT 2611, Australia}
\date{\today}

\begin{abstract} 
Magnetic fields in several astrophysical objects are amplified and maintained by a dynamo mechanism, which is the conversion of the turbulent kinetic energy to magnetic energy. A dynamo that amplifies magnetic fields at scales less than the driving scale of turbulence is known as the fluctuation dynamo. We aim to study the properties of the fluctuation dynamo in supersonic turbulent plasmas, which is of relevance to the interstellar medium of star-forming galaxies, structure formation in the Universe, and laboratory experiments of laser-plasma turbulence. Using numerical simulations of driven turbulence, we explore the global and local properties of the exponentially growing and saturated (statistically steady) state of the fluctuation dynamo for subsonic and supersonic turbulent flows. First, we confirm that the fluctuation dynamo efficiency decreases with compressibility. Then, we show that the fluctuation dynamo generated magnetic fields are spatially intermittent and the intermittency is higher for supersonic turbulence, but in both cases, the level of intermittency decreases as the field saturates. We also find a stronger back reaction of the magnetic field on the velocity for the subsonic case as compared to the supersonic case. Locally, we find that the level of alignment between vorticity and velocity, velocity and magnetic field, and current density and magnetic field in the saturated stage is enhanced in comparison to the exponentially growing phase for the subsonic case, but only the current density and magnetic field alignment is enhanced for the supersonic case. Finally, we show that both the magnetic field amplification (mainly due to weaker stretching of magnetic field lines) and diffusion decreases when the field saturates, but the diffusion is enhanced relative to amplification. This occurs throughout the volume in the subsonic turbulence, but primarily in the strong-field regions for the supersonic case. This leads to the saturation of the fluctuation dynamo. Overall, both the amplification and diffusion of magnetic fields are affected by the exponentially growing magnetic fields and thus a drastic change in either of them is not required for the saturation of the fluctuation dynamo.
\end{abstract}

\maketitle

\section{Introduction} \label{sec:intro}
It is important to study the properties of magnetic fields in supersonic turbulent plasmas because of their applications to astrophysics \citep{BrandenburgS2005,Federrath2016} and recently possible laboratory experiments of laser-plasma turbulence \citep{BottEA2020II}. In the Sun, turbulent plasma in the convection zone is slightly supersonic \citep{CatteneoHT1990, ProctorW2014} and affects surface dynamics \citep{Martinez2013}. On galactic scales, turbulence in the interstellar medium (ISM) of star-forming galaxies is driven supersonically at a range of scales by a variety of mechanisms such as supernova explosions, gravitational collapse, accretion, and jets from young stellar objects and active galactic nuclei \citep{MacLowK2004, Elmegreen2009, FederrathEA2017a}. Magnetic fields in the supersonic turbulent plasma of the ISM play a crucial role in the present-day \citep{KrumholzF2019} and primordial \citep{SurEA2010, FederrathEA2011b, Klessen2019, ShardaEA2021} star formation. The turbulence driven by structure formation in galaxy clusters can also be supersonic \citep{VazzaEA2017} and this affects cluster magnetic field structure, which in turn controls the acceleration and propagation of relativistic particles \citep{BrunettiJ2014}.

Physically, the strength and structure of observed magnetic fields in astrophysical objects \citep{GovoniF2004, Martinez2013, Beck2016} and derived magnetic fields in plasma turbulence experiments with a dynamically insignificant initial magnetic field \citep{TzeferacosEA18,BottEA2020} can be largely explained by a turbulent dynamo, the mechanism by which the turbulent kinetic energy is converted to magnetic energy \citep{BrandenburgS2005, Rincon2019}. \rev{Turbulence is prevalent in most astrophysical systems as the Reynolds number, $\Re = \urms \ld /\nu$ (where $\ld$ is the driving scale of turbulence, $\urms$ is the root mean square (rms) velocity, and $\nu$ is the viscosity), is usually very high.} A dynamo that amplifies magnetic fields at scales less than the driving scale of turbulence is known as the {\it fluctuation or small-scale dynamo}. In a turbulent (or even random) flow, the fluctuation dynamo exponentially amplifies (kinematic stage) a weak seed field of any form \citep{Kazantsev1968, SetaF2020} to dynamically significant strengths when the magnetic Reynolds number, $\Rm = \urms \ld/\eta$ ($\eta$ is the resistivity, which controls magnetic diffusion), is greater than a critical value \citep[$\gtrsim 100$ as shown in][]{RuzmaikinS1981, MeneguzziFP1981, HaugenBD2004, SchekochihinEA2007, FederrathEA2014}. \rev{The critical value of the magnetic Reynolds number also depends on the $\Pm~(= \Rm / \Re)$ \citep{SchekochihinEA2007, BrandenburgEA2018} and in this paper we explore the $\Pm \gtrsim 1$ regime, which is applicable to both the subsonic, hot, and the supersonic, cold phases of the ISM \citep[see Table~2 in][]{Ferriere2020}.} This magnetic field amplification is primarily due to the stretching of magnetic field lines by turbulent motions \citep{VainshteinZ1972,Eyink10,SetaBS2015}. Once the field becomes strong enough, it back reacts on the turbulent flow via the Lorentz force and then the dynamo saturates (saturated stage). The saturation mechanism, primarily studied for subsonic turbulence, is due to a combination of reduced amplification and diffusion \citep{SchekochihinEA02, SetaEA2020}. In this paper, using driven turbulence periodic box magnetohydrodynamic (MHD) simulations, we aim to study the saturation mechanism of the fluctuation dynamo in supersonic turbulent plasmas.

The fluctuation dynamo in subsonic turbulent plasma has been studied analytically \citep{Kazantsev1968, RuzmaikinS1981, ZeldovichEA1984, ZeldovichRS1990, KulsrudA1992, Subramanian1999,SchekochihinBK02, SchekochihinEA02, BoldyrevC2004}, numerically \citep{MeneguzziFP1981,SchekochihinEA2004, HaugenBD2004, CattaneoT2009, ChoEA2009, Beresnyak2012, BhatS2013, SetaEA2020, AchikanathEA2021}, and recently via experiments \citep{TzeferacosEA18,BottEA2020}. These studies confirm exponential growth of magnetic fields and show the following magnetic field properties: the saturated magnetic energy is a fraction of the turbulent kinetic energy, the magnetic power spectra in the kinematic stage seems to follow a power law with an exponent $3/2$, and the magnetic field in the saturated stage has a higher correlation length than in the kinematic stage. Although not as extensively as for subsonic turbulence, the fluctuation dynamo in supersonic turbulent plasma is also studied analytically \citep{KazantsevRS1985, SchoberEA2015, AfonsoMV2019} and numerically \citep{HaugenBM2004, FederrathEA2011,FederrathEA2014, SetaF2020}. These studies show that with increasing compressibility, the critical magnetic Reynolds number increases and the fraction of turbulent kinetic energy getting converted to magnetic energy, \rev{per unit time}, decreases. Thus, the overall efficiency of the dynamo decreases in supersonic turbulent plasmas as compared to their subsonic counterparts. However, the effect of compressibility on the local interaction of the magnetic and velocity fields and the saturation mechanism is not known yet. We aim to explore such questions with this study. Furthermore, some of the properties of the fluctuation dynamo are also seen in recent large-scale cosmological simulations of galaxies \cite{PakmorEA2017,RiederT2016,RiederT2017} and galaxy clusters \citep{VazzaEA2018,MarinacciEA2018,DominguezEA2019}. The turbulence in these cosmological simulations would also be supersonic in regions with shocks and understanding the physics of the fluctuation dynamo in supersonic turbulent plasmas would further help understand magnetic fields during cosmological evolution.

The remainder of this paper is organised as follows. In \Sec{sec:met}, we describe our numerical methods and parameters of the study. The results are presented and discussed in \Sec{sec:glo} and \Sec{sec:loc}. In \Sec{sec:glo}, we describe the difference in the global (spectral and structural) properties of magnetic fields in the kinematic and saturated stages as a function of the compressibility of the turbulent flow. In \Sec{sec:loc}, we study the local interaction of the velocity and magnetic fields in the kinematic and saturated stages for subsonic and supersonic flows. Finally, we summarise and conclude our results in \Sec{sec:con}.

\section{Methods} \label{sec:met}
To study the physics of the fluctuation dynamo in supersonic turbulent plasmas, we use a modified version of the FLASH code (version~4) \citep{FryxellEA2000,DubeyEA2008} to numerically solve the equations of compressible MHD (\Eq{eq:ce} - \Eq{eq:div}) for an isothermal gas \rev{(an isothermal equation of state is adopted for simplicity)} in a triply periodic cartesian ($xyz$) domain of size $L$ with a uniform grid and $512^3$ grid points. We use the HLL3R (3-wave approximate) Riemann solver \citep{WaaganFK2011} to solve the following equations:
\begin{align}
	&\frac{\partial \rho}{\partial t} + \nabla \cdot (\rho \vec{u}) = 0,  \label{eq:ce} \\
	&\frac{\partial (\rho \vec{u})}{\partial t} + \nabla \cdot \left(\rho~\vec{u} \otimes \vec{u} - \frac{1}{4 \pi} \vec{b} \otimes \vec{b}\right) + 
	\nabla \left(\cs^2 \rho + \frac{\vec{b}^2}{8 \pi}\right) =  \nonumber \\ 
	& \hspace{0.675\columnwidth} \nabla \cdot (2 \nu \rho \vec{\tau}) + \rho \vec{F}, \label{eq:ns} \\
	&\frac{\partial \vec{b}}{\partial t} = \nabla \times (\vec{u} \times \vec{b}) + \eta \nabla^2 \vec{b},  \label{eq:ie} \\
	&\nabla \cdot \vec{b} = 0 \label{eq:div},
\end{align}
where $\rho$ is the density, $\vec{u}$ is the velocity field, $\vec{b}$ is the magnetic field, $\cs$ is the constant sound speed, $\tau_{ij} = (1/2) \, (u_{i,j} + u_{j,i} - (2/3) \, \delta_{ij} \, \nabla \cdot \vec{u})$ is the traceless rate of strain tensor, $\vec{F}$ is the prescribed acceleration field constructed using the Ornstein-Uhlenbeck process for the turbulent driving, and $\nu$ and $\eta$ are constant viscosity and resistivity, respectively. 

The turbulent flow is driven solenoidally \rev{($k \cdot F_{k} = 0$, where $k$ is the wavenumber and $F_{k}$ is the forcing vector in $k$ space)} on large scales ($1 \le kL/2 \pi \le 3$) using a parabolic function of power with a peak at $kL/2 \pi = 2$ and zero power at $kL/2 \pi = 1, 3$ \citep{FederrathEA2010, SetaF2020}.  Thus, the effective driving scale of turbulence $\ld$ is approximately equal to $L/2$. \rev{The correlation time of the forcing is set to the eddy turnover time of the turbulent flow, $t_0 = L/(2 \urms)$. We use purely solenoidal driving instead of compressive driving because solenoidal driving gives a higher dynamo efficiency \citep{FederrathEA2011, AchikanathEA2021}.} The diffusion of velocity and magnetic fields is characterised by the hydrodynamic ($\Re = \urms \ld/\nu$, where $\urms$ is the rms of the turbulent velocity) and magnetic ($\Rm = \urms \ld/\eta$) Reynolds numbers. The compressibility of the medium is quantified using the turbulent Mach number, $\Mach=\urms/\cs$.

We initialize our simulations with zero velocity, a uniform density ($\rho_0$), and a very weak random seed field (plasma beta $= \cs^2 \rho_0/(\brms / 8 \pi) = 2.5 \times 10^{13}$, where $\brms$ is the rms magnetic field strength). We select $\Re=2000$ for all our runs and vary $\Rm$ in the range $2000~\text{--}~6000$. Thus, the magnetic Prandtl number, $\Pm$, is always greater than or equal to one and varies in the range $1~\text{--}~3$. The main parameter of the study is the Mach number, which is varied from $0.1$ (subsonic) to $10$ (supersonic). We run all simulations till the dynamo saturates and the magnetic fields achieve a statistically steady state. For our set of selected parameters, the magnetic field saturates in less than $100~t_0$ ($t_0=\ld/\urms$ is the eddy turnover time). We then study the properties of the velocity and magnetic fields in the kinematic and saturated stages for the subsonic and supersonic turbulent flows.

\section{Global fluctuation dynamo properties} \label{sec:glo}
Here, we first compare the properties of the time evolution of magnetic fields, i.e., the growth rate in the kinematic phase and the ratio of magnetic to turbulent kinetic energy in the saturated stage with existing studies. Then, we discuss the spectral and structural properties of velocity and magnetic fields in the kinematic and saturated stages for turbulent flows with low ($\Mach=0.1$, subsonic) and high ($\Mach=10$, supersonic) Mach numbers.

\begin{figure}
    \includegraphics[width=\columnwidth]{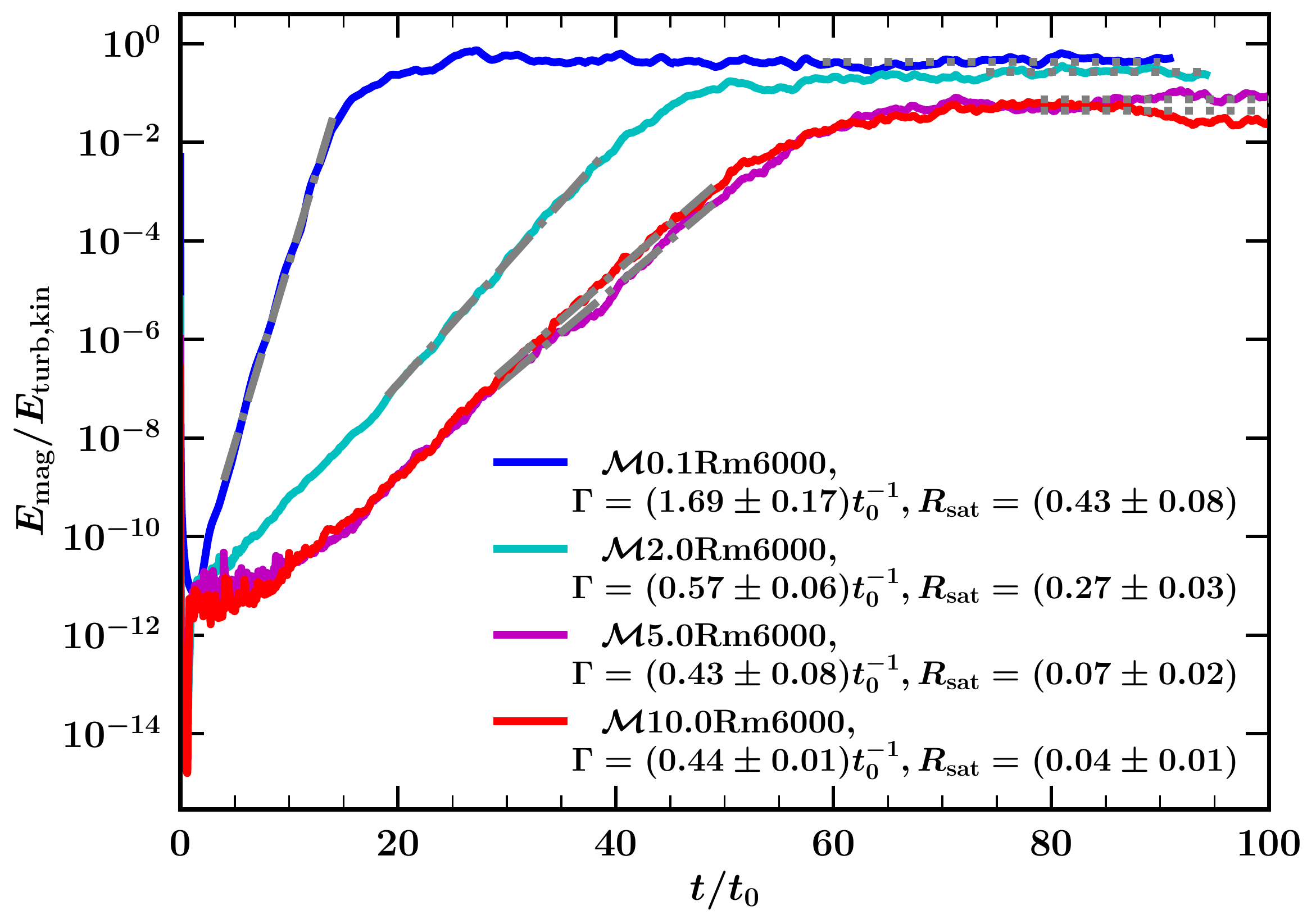}
    \caption{Time evolution (in terms of the eddy turnover time, $t_0$, which also varies with the Mach number) of the ratio of magnetic to turbulent kinetic energies, $E_{\rm mag}/E_{\rm turb,kin}$, for $\Re=2000$, $\Rm=6000$, and $\Mach=0.1$ (blue), $2$ (cyan), $5$ (magenta), and $10$ (red). For all the cases, after the initial transient decay phase, the magnetic field grows exponentially (kinematic stage, dashed-dotted grey line) and then saturates (saturated stage, dotted grey line). The growth rate, $\Gamma [t_0^{-1}]$ in the exponentially growing or kinematic phase decreases with compressibility till $\Mach=5$ and then increases for $\Mach=10$. The saturated value of the ratio $E_{\rm mag}/E_{\rm turb,kin}$, $R_{\rm sat}$, decreases with increasing Mach numbers. Thus, per unit time, a smaller fraction of the turbulent kinetic energy is converted to magnetic energy for supersonic flows as compared to the subsonic flows. See \Tab{tab:pro} for the values of $\Gamma$ and $R_{\rm sat}$ in other runs and \Fig{fig:gammasat} for the dependence of $\Gamma$ and $R_{\rm sat}$ on $\Mach$. }
    \label{fig:ts} 
\end{figure}
\subsection{Growth rate, saturated level, structure, and spectra} \label{sec:glo1}
\Fig{fig:ts} shows the evolution of the ratio of magnetic to turbulent kinetic energies, $E_{\rm mag}/E_{\rm turb,kin}$, as a function of time normalised by the eddy turnover time, $t_0$, for various Mach numbers ($0.1, 2, 5,$ and $10$) with $\Re=2000$ and $\Rm=6000$.  After the initial transient phase, the magnetic energy amplifies exponentially (kinematic stage) for all the cases. Then when the magnetic field becomes strong enough to react back on the flow, the exponential increase slows down, and finally the magnetic energy saturates to a statistically steady value (saturated stage).  This happens for all the runs and the corresponding growth rate, $\Gamma [t_0^{-1}]$, in the exponentially growing or kinematic stage and the saturated level of $E_{\rm mag}/E_{\rm turb,kin}$, $R_{\rm sat}$, in the saturated stage are given in \Tab{tab:pro}.

\begin{table*}
    \centering
	\caption{A summary of parameters and derived properties for all simulations. Note that all the simulations are performed on a uniform grid with $512^3$ points in a numerical domain of size $L^3$. For all the runs, the flow is driven solenoidally on larger scales ($1 \le kL/2\pi \le 3$) and the hydrodynamic Reynolds number, $\Re$, is fixed to be $2000$. The columns are as follows: 1. simulation name, 2. Mach number of the turbulent flow, $\Mach$, 3. magnetic Reynolds number, $\Rm$, 4. growth rate in the exponentially growing or kinematic phase in units of inverse eddy turnover time, $t_0^{-1}$, $\Gamma$, 5. ratio of the magnetic to kinetic energy in the saturated stage, $R_{\rm sat}$, 6. kurtosis of the $x$-component of the magnetic field in the kinematic stage, $\ku_{b_x}$ (kin), and 7. kurtosis of the $x$-component of the magnetic field in the saturated stage, $\ku_{b_x}$ (sat). The errors reported in columns 4. and 5. are from fitting an exponential and a constant function, respectively, to $E_{\rm mag}/E_{\rm turb,kin}$ in the kinematic and saturated stages (see \Fig{fig:ts} for an example). The errors reported in columns 6. and 7. are standard deviation obtained after averaging the kurtosis of the distribution over $10~t_0$ in the kinematic and saturated stages, respectively.}
	\label{tab:pro}
	\begin{tabular}{lcccccc} 
		\hline
		\hline
        Simulation Name & $\Mach$ &  $\Rm$ & $\Gamma \, [t_0^{-1}]$ & $R_{\rm sat}$ & $\ku_{b_x}$ (kin) & $\ku_{b_x}$ (sat)
        \\
        \hline
$\Mach 0.1\Rm 2000$ & $0.1$ & $2000$& $0.88 \pm 0.11$& $0.33 \pm 0.06$& $9.68 \pm 0.50$& $5.16 \pm 0.40$ \\
$\Mach 0.1\Rm 3000$ & $0.1$ & $3000$& $1.23 \pm 0.12$& $0.38 \pm 0.06$& $10.38 \pm 0.60$& $5.04 \pm 0.32$ \\
$\Mach 0.1\Rm 4000$ & $0.1$ & $4000$& $1.42 \pm 0.13$& $0.38 \pm 0.07$& $10.32 \pm 0.45$& $4.95 \pm 0.31$ \\
$\Mach 0.1\Rm 5000$ & $0.1$ & $5000$& $1.59 \pm 0.14$& $0.44 \pm 0.06$& $10.44 \pm 0.57$& $4.88 \pm 0.42$ \\
$\Mach 0.1\Rm 6000$ & $0.1$ & $6000$& $1.69 \pm 0.17$& $0.43 \pm 0.08$& $10.15 \pm 0.66$& $4.95 \pm 0.28$ \\
\\
$\Mach 2.0\Rm 2000$ & $2.0$ & $2000$& $0.24 \pm 0.02$& $0.09 \pm 0.01$& $19.57 \pm 2.46$& $7.49 \pm 0.47$ \\
$\Mach 2.0\Rm 3000$ & $2.0$ & $3000$& $0.43 \pm 0.04$& $0.15 \pm 0.02$& $19.15 \pm 1.82$& $6.32 \pm 0.42$ \\
$\Mach 2.0\Rm 4000$ & $2.0$ & $4000$& $0.43 \pm 0.03$& $0.14 \pm 0.02$& $19.92 \pm 1.53$& $5.91 \pm 0.60$ \\
$\Mach 2.0\Rm 5000$ & $2.0$ & $5000$& $0.51 \pm 0.06$& $0.16 \pm 0.02$& $20.48 \pm 1.51$& $5.51 \pm 0.24$ \\
$\Mach 2.0\Rm 6000$ & $2.0$ & $6000$& $0.57 \pm 0.06$& $0.27 \pm 0.03$& $19.73 \pm 1.79$& $5.07 \pm 0.30$ \\
\\
$\Mach 5.0\Rm 2000$ & $5.0$ & $2000$& $0.26 \pm 0.11$& $0.02 \pm 0.01$& $49.01 \pm 2.53$& $23.96 \pm 4.73$ \\
$\Mach 5.0\Rm 3000$ & $5.0$ & $3000$& $0.35 \pm 0.02$& $0.04 \pm 0.01$& $48.71 \pm 4.95$& $16.08 \pm 1.92$ \\
$\Mach 5.0\Rm 4000$ & $5.0$ & $4000$& $0.36 \pm 0.08$& $0.05 \pm 0.01$& $48.91 \pm 2.42$& $13.85 \pm 1.43$ \\
$\Mach 5.0\Rm 5000$ & $5.0$ & $5000$& $0.41 \pm 0.02$& $0.06 \pm 0.02$& $50.79 \pm 2.15$& $13.28 \pm 1.89$ \\
$\Mach 5.0\Rm 6000$ & $5.0$ & $6000$& $0.43 \pm 0.08$& $0.07 \pm 0.02$& $47.19 \pm 3.35$& $11.08 \pm 1.34$ \\
\\
$\Mach 10.0\Rm 2000$ & $10.0$ & $2000$& $0.36 \pm 0.05$& $0.02 \pm 0.01$& $74.95 \pm 9.07$& $38.20 \pm 5.47$ \\
$\Mach 10.0\Rm 3000$ & $10.0$ & $3000$& $0.37 \pm 0.05$& $0.03 \pm 0.01$& $73.31 \pm 2.63$& $24.74 \pm 3.92$ \\
$\Mach 10.0\Rm 4000$ & $10.0$ & $4000$& $0.46 \pm 0.08$& $0.03 \pm 0.01$& $76.63 \pm 4.39$& $30.29 \pm 4.97$ \\
$\Mach 10.0\Rm 5000$ & $10.0$ & $5000$& $0.41 \pm 0.09$& $0.04 \pm 0.01$& $83.90 \pm 4.55$& $18.73 \pm 3.22$ \\
$\Mach 10.0\Rm 6000$ & $10.0$ & $6000$& $0.44 \pm 0.01$& $0.04 \pm 0.01$& $83.03 \pm 2.17$& $24.14 \pm 6.15$ \\
        
		\hline
		\hline
   \end{tabular}
\end{table*}

\begin{figure*}
    \includegraphics[width=\columnwidth]{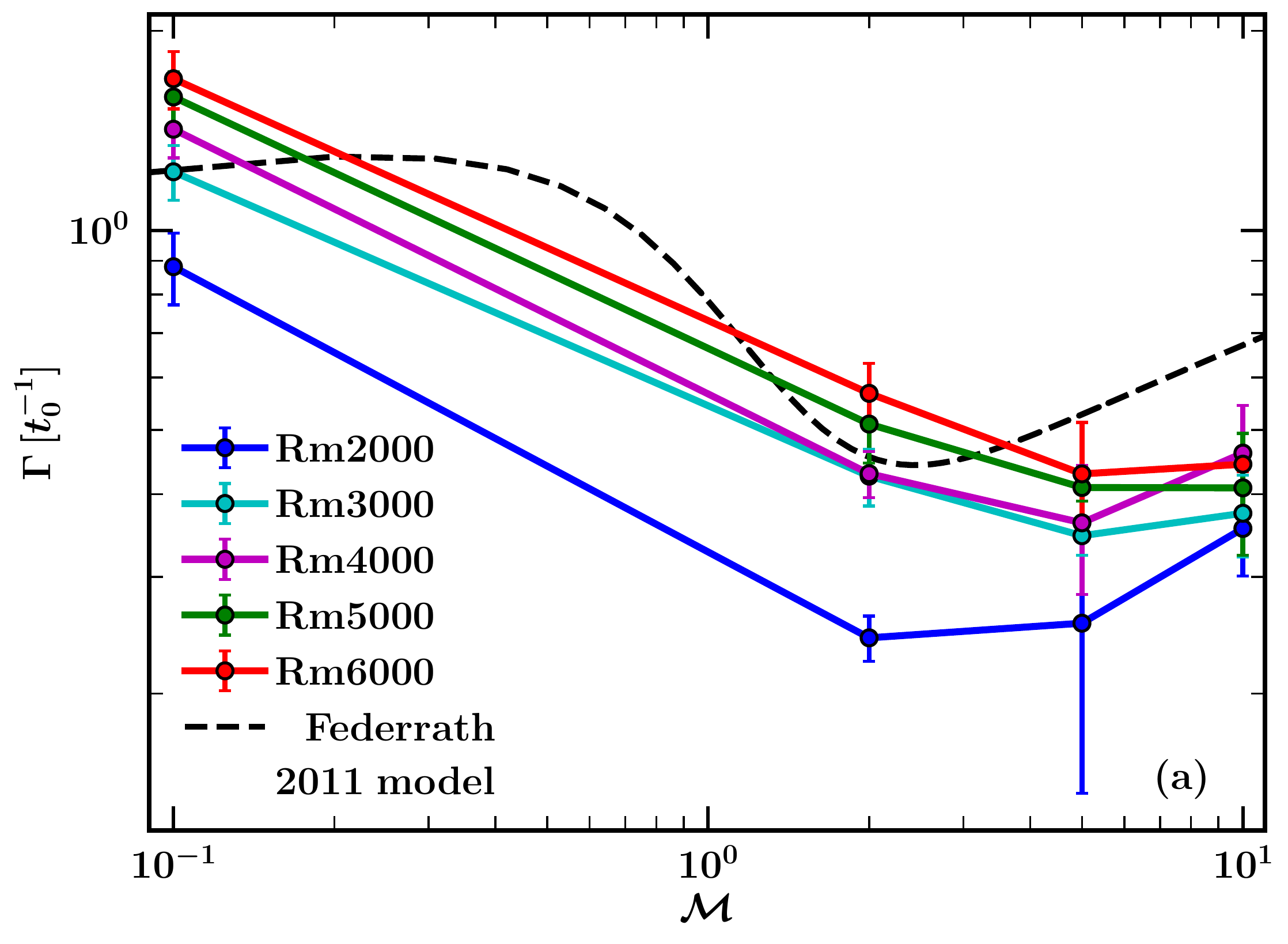}
    \includegraphics[width=\columnwidth]{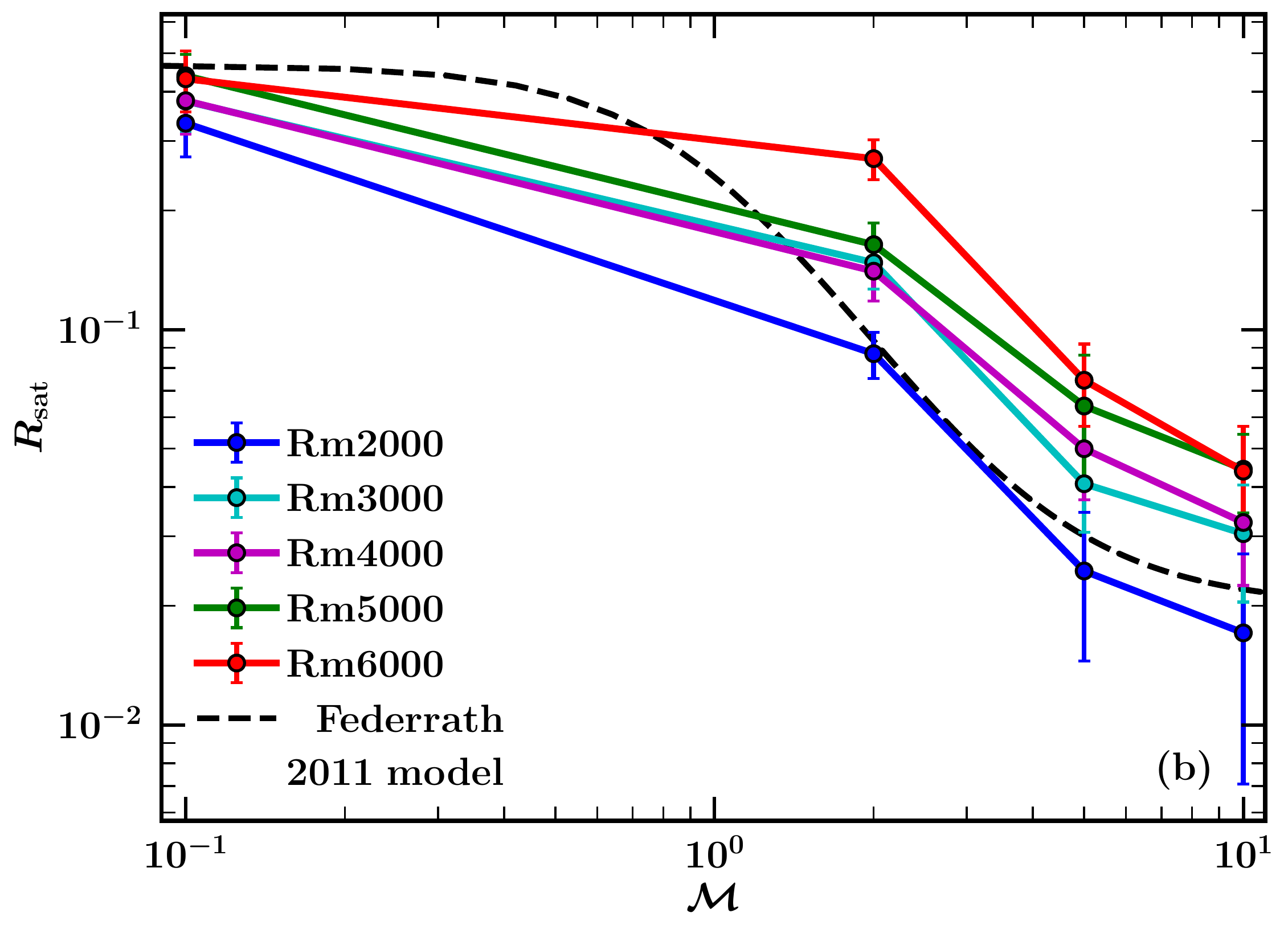}
       \caption{The growth rate in the kinematic phase, $\Gamma \, [t_0^{-1}]$, (a) and the ratio of magnetic to turbulent kinetic energies in the saturated stage, $R_{\rm sat}$, (b) as a function of the Mach number of the turbulent flow, $\Mach$, for all $\Rm$s. The growth rate as a function of the Mach number first decreases till $\Mach=5$ and then increases for $\Mach=10$. The overall trend with $\Mach$ does not change much with $\Rm$. For $R_{\rm sat}$, as the Mach number of the turbulent flow increases, the ratio decreases. This confirms that a smaller fraction of the turbulent kinetic energy is converted to magnetic energy, per unit time, for the supersonic turbulence as compared to the subsonic case. The variation of $\Gamma$ and $R_{\rm sat}$ with $\Mach$ is roughly consistent with the empirical model (shown in dashed, black line) of Federrath et al.~2011 \citep[see Eq.~3 and Tab.~1 in][]{FederrathEA2011}.}
       \label{fig:gammasat} 
 \end{figure*}
\Fig{fig:gammasat} shows the growth rate, $\Gamma \, [t_0^{-1}]$, (\Fig{fig:gammasat}~(a)) and the saturated level of $E_{\rm mag}/E_{\rm turb,kin}$, $R_{\rm sat}$, (\Fig{fig:gammasat}~(b)) as a function of $\Mach$. The growth rate decreases till $\Mach=5$ but then increases for $\Mach=10$ and the overall trend is consistent with the empirical model in the literature \citep{FederrathEA2011}. \Fig{fig:gammasat}~(b) shows the dependence of the saturated level, $R_{\rm sat}$, \rev{on $\Mach$}. \rev{For all values of $\Rm$, $R_{\rm sat}$ decreases with $\Mach$}. This shows that as the compressibility increases, a smaller fraction of turbulent kinetic energy is converted to magnetic energy, \rev{per unit time}. \rev{Thus, the efficiency of the fluctuation dynamo decreases with increasing compressibility.} Here too, the trend of $R_{\rm sat}$ with $\Mach$ roughly agrees with the known empirical model \citep{FederrathEA2011}. \rev{These empirical models for the dependence of $\Gamma$ and $R_{\rm sat}$ on $\Mach$ are constructed from numerical simulations in \citep{FederrathEA2011}, where most of the simulations used numerical viscosity and resistivity, whereas here, all our simulations include explicit (viscous and resistive) diffusion terms (\Eq{eq:ns} and \Eq{eq:ie}). This might be the reason for differences in the results (coloured lines vs.~dashed black line in \Fig{fig:gammasat}), but the overall trend is roughly consistent with the empirical models.} Thus, \rev{generally}, the growth rates and the saturated levels agree with previous results. We now show and discuss velocity and magnetic structures in the kinematic and saturated stages for $\Mach=0.1$ and $\Mach=10$ with a fixed $\Rm$ of $6000$. 

\begin{figure*}
    \centering
    \includegraphics[width=2\columnwidth]{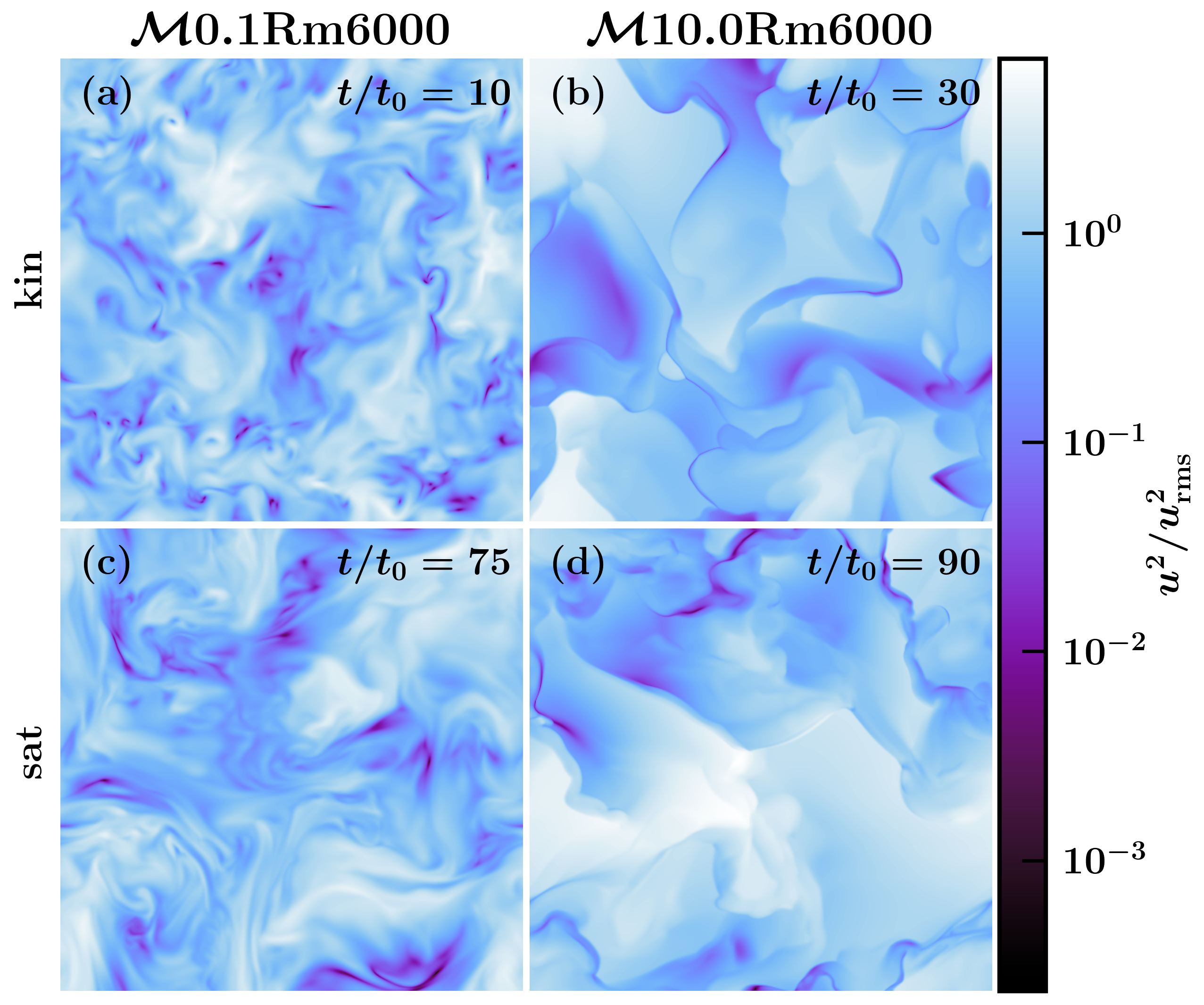}
       \caption{Two-dimensional slices of the velocity field at $z=L/2$ with colour showing the square of velocity normalised to its root mean square (rms) value, $u^2/\urms^2$, for subsonic ($\Mach0.1\Rm6000$) and supersonic ($\Mach10.0\Rm6000$) cases in their kinematic ($t/t_0=10$, a and $t/t_0 = 30$, b) and saturated ($t/t_0=75$, c, and $t/t_0=90$, d) stages. Visually, the velocity structures are larger in the saturated stage for $\Mach0.1\Rm6000$ than in the kinematic stage (effect of the back reaction of the strong magnetic field on the turbulent flow). The corresponding difference for $\Mach10.0\Rm6000$ is smaller. As compared to the subsonic case, the velocity field in the supersonic case can be locally high in strength. This is due to random strong shocks for the $\Mach=10$ case.}
       \label{fig:vstruc2d} 
 \end{figure*}
 \begin{figure*}
    \centering
    \includegraphics[width=2\columnwidth]{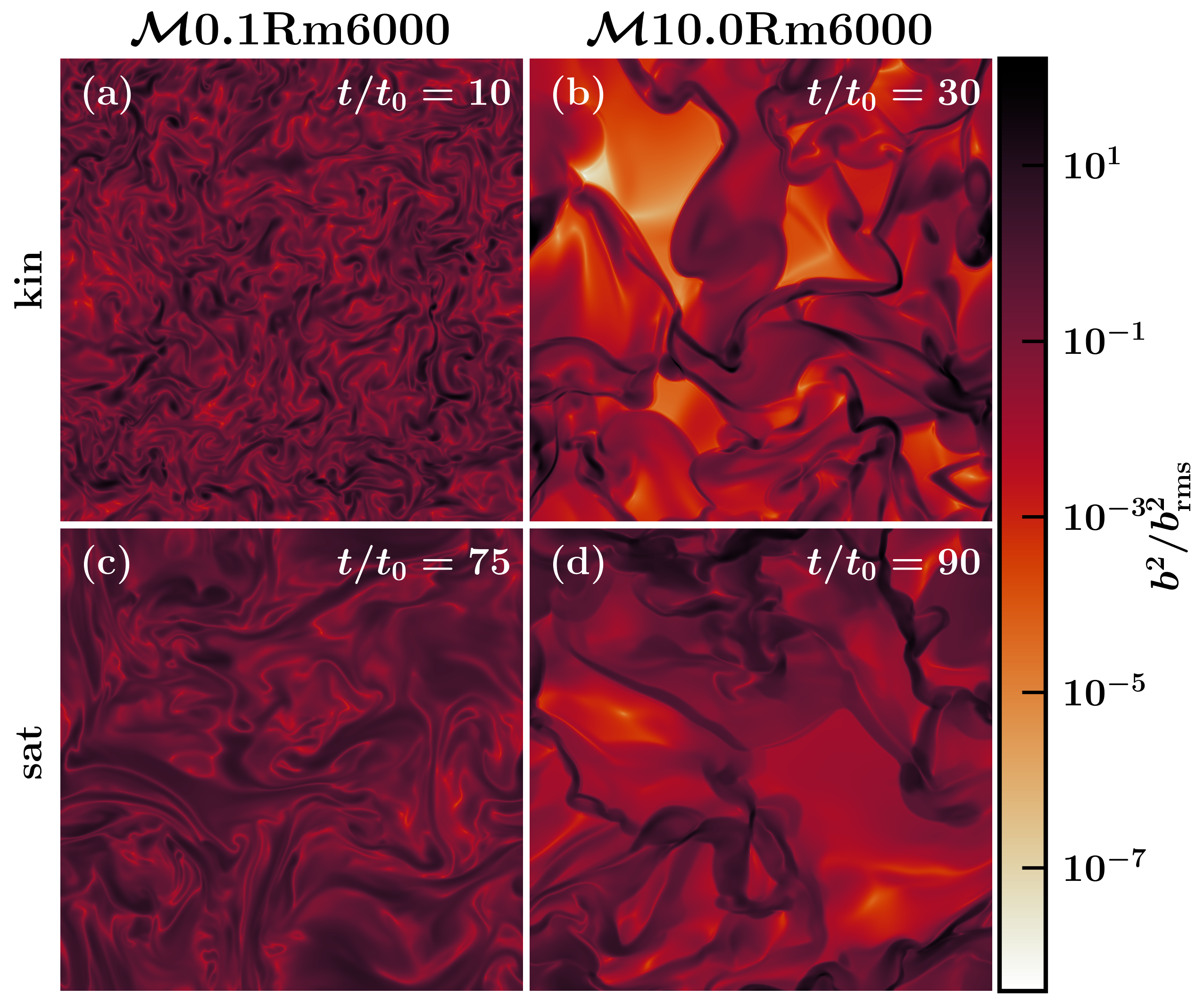}
       \caption{Same as \Fig{fig:vstruc2d} but for $b^2/\brms^2$. In comparison to the kinematic stage (a, b), the magnetic structures in the saturated stages (c, d) are seen to be larger in size for both the subsonic and supersonic turbulence. Visually, in the kinematic stage and on an average, the structures are of a larger size for the supersonic turbulence in comparison to the subsonic case. This difference is less for the saturated stage.}
       \label{fig:bstruc2d} 
 \end{figure*}
 \begin{figure*}
    \centering
    \includegraphics[width=2\columnwidth]{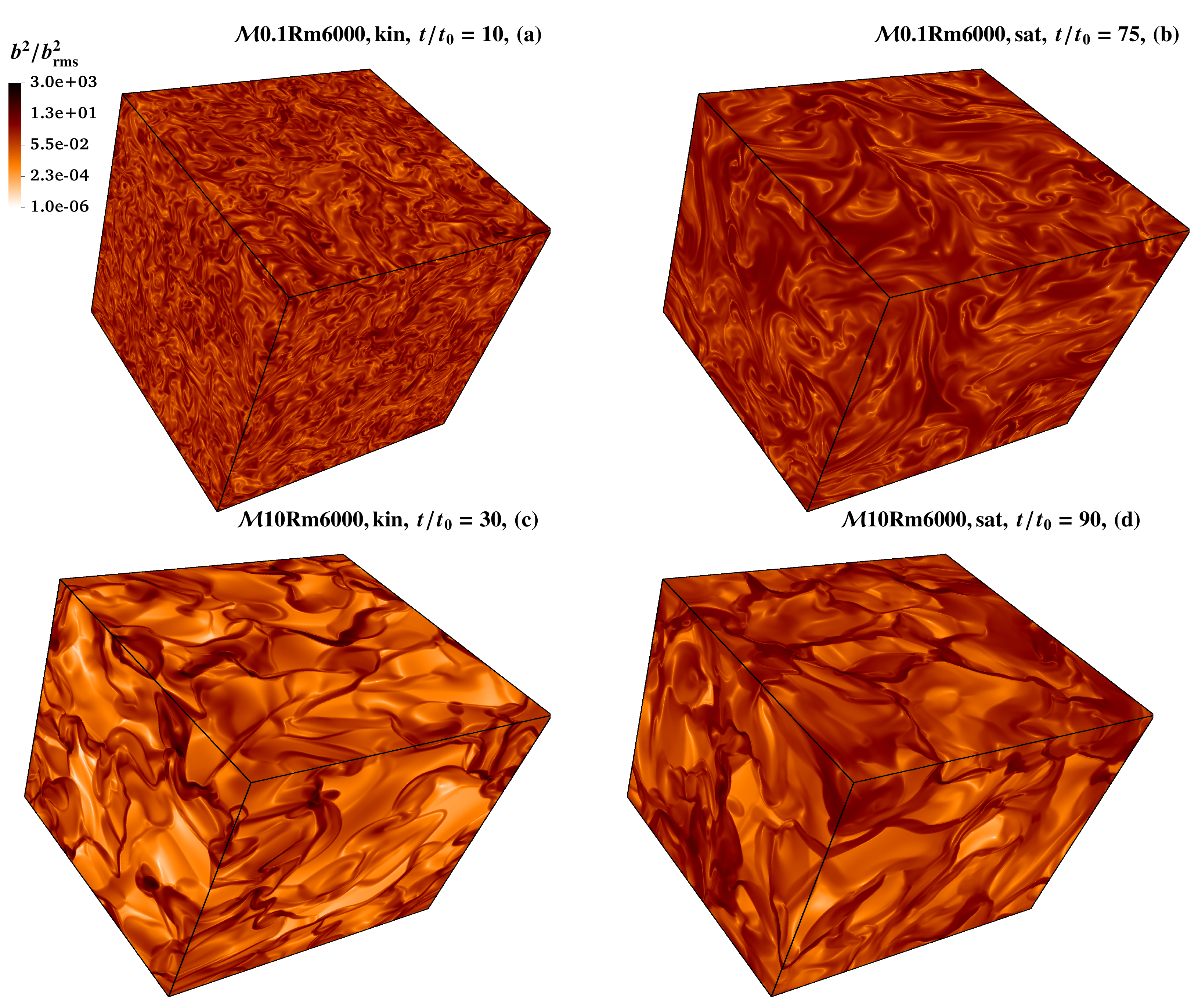}
       \caption{Same as \Fig{fig:bstruc2d} but now showing three-dimensional magnetic structures. Visually, the magnetic structures for the supersonic turbulence are of a larger size and this is more evident in the kinematic stage. The magnetic fields, for both subsonic and supersonic turbulent flows, seen to have larger structures in their respective saturated stage as compared to the kinematic stage.}
       \label{fig:bstruc3d} 
 \end{figure*}

\rev{\Fig{fig:vstruc2d} and \Fig{fig:bstruc2d} shows two-dimensional velocity and magnetic field structures in the kinematic and saturated stages for subsonic and supersonic turbulence. Both the velocity and magnetic fields show random distributions with complex structures and without any significant mean trend. This is expected for the turbulent flows and the magnetic fields they amplify. The velocity structures for subsonic flow look larger in size in the saturated stage (\Fig{fig:vstruc2d}~(b)) as compared to the kinematic stage (\Fig{fig:vstruc2d}~(a)). This is due to the back reaction of the strong magnetic field on the turbulent flow (the back reaction is negligible in the kinematic stage). Such a difference is smaller for the supersonic case. The structures in supersonic turbulence (\Fig{fig:vstruc2d}~(b, d)), in both the kinematic and saturated stages, are of a larger size than in the subsonic case (\Fig{fig:vstruc2d}~(a, c)) but the supersonic case can have locally strong velocity because of random shocks. The difference in structure between the kinematic and saturated stages is more pronounced for the magnetic field. For both the Mach numbers, magnetic fields in the saturated stages (\Fig{fig:bstruc2d}~(c, d)) have larger structures than in their corresponding kinematic stages (\Fig{fig:bstruc2d}~(a, b)). \Fig{fig:bstruc3d} shows the three-dimensional magnetic structures in both the kinematic and saturated stages for both Mach numbers. Overall, for both Mach numbers, the magnetic fields have visually larger structures in the saturated stage than in their respective kinematic stage.} We now discuss the spectral properties of velocity and magnetic fields in the kinematic and saturated stages for subsonic and supersonic turbulent flows.

\begin{figure*}
    \includegraphics[width=\columnwidth]{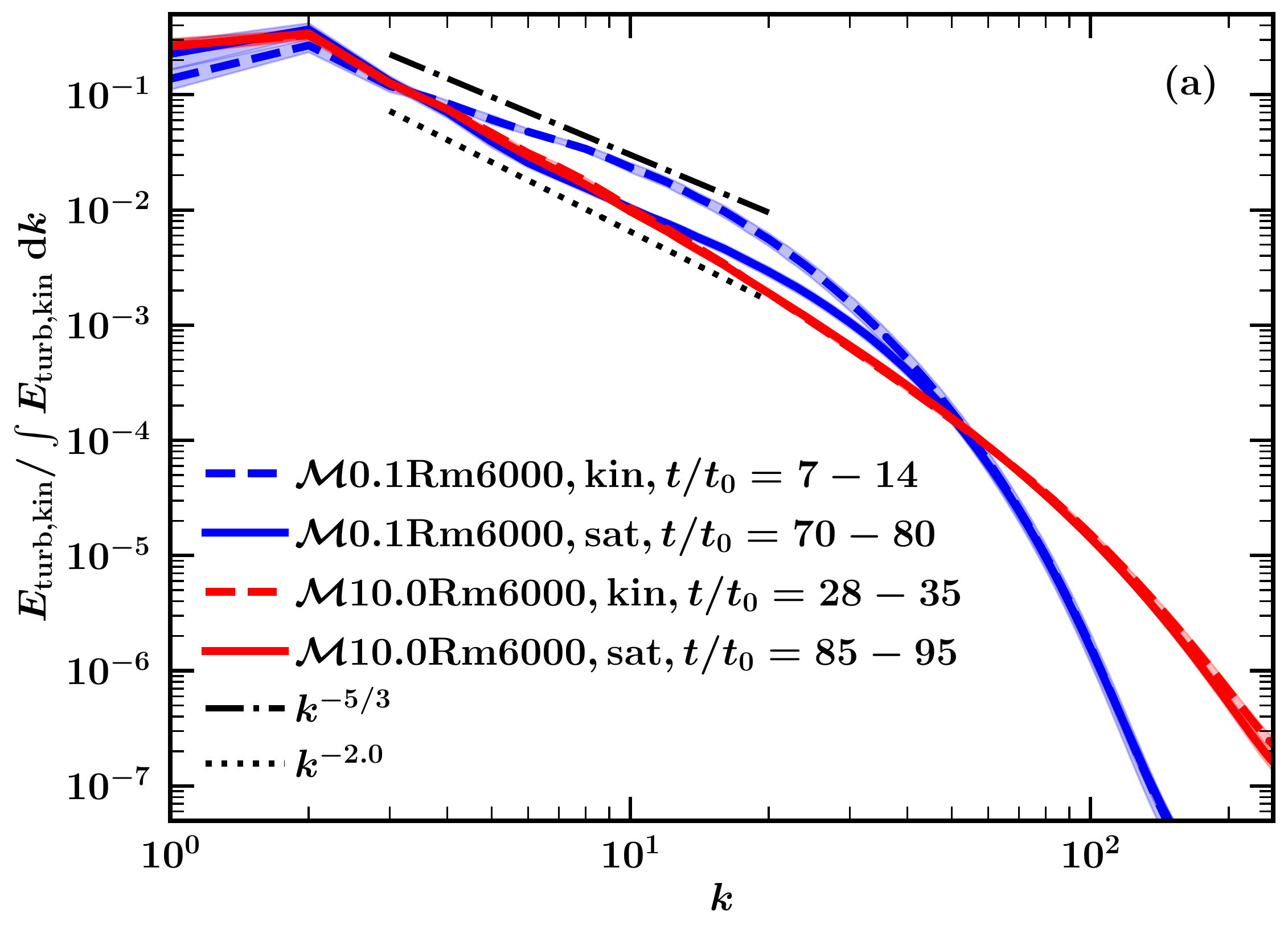}
    \includegraphics[width=\columnwidth]{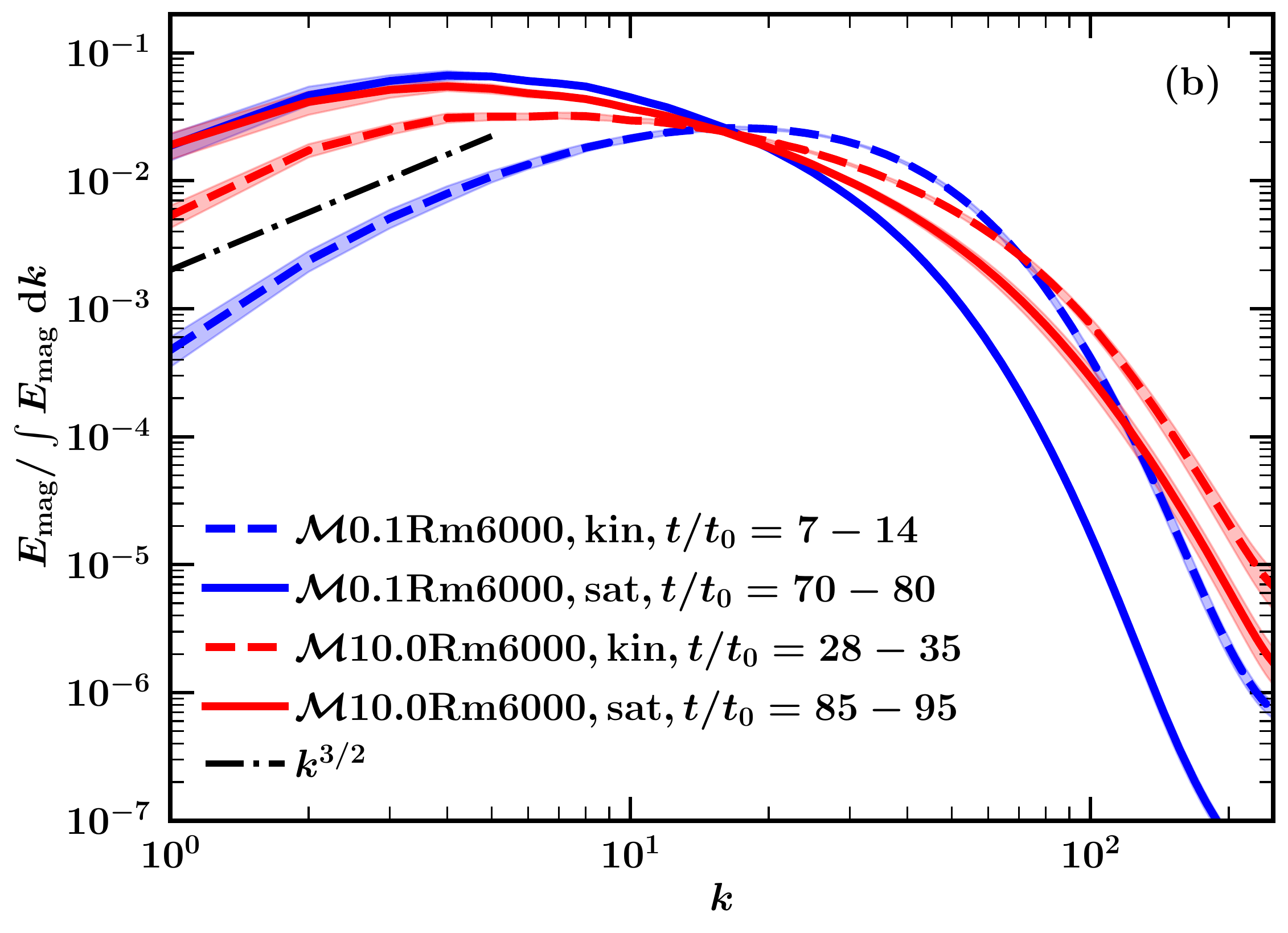}
    \includegraphics[width=\columnwidth]{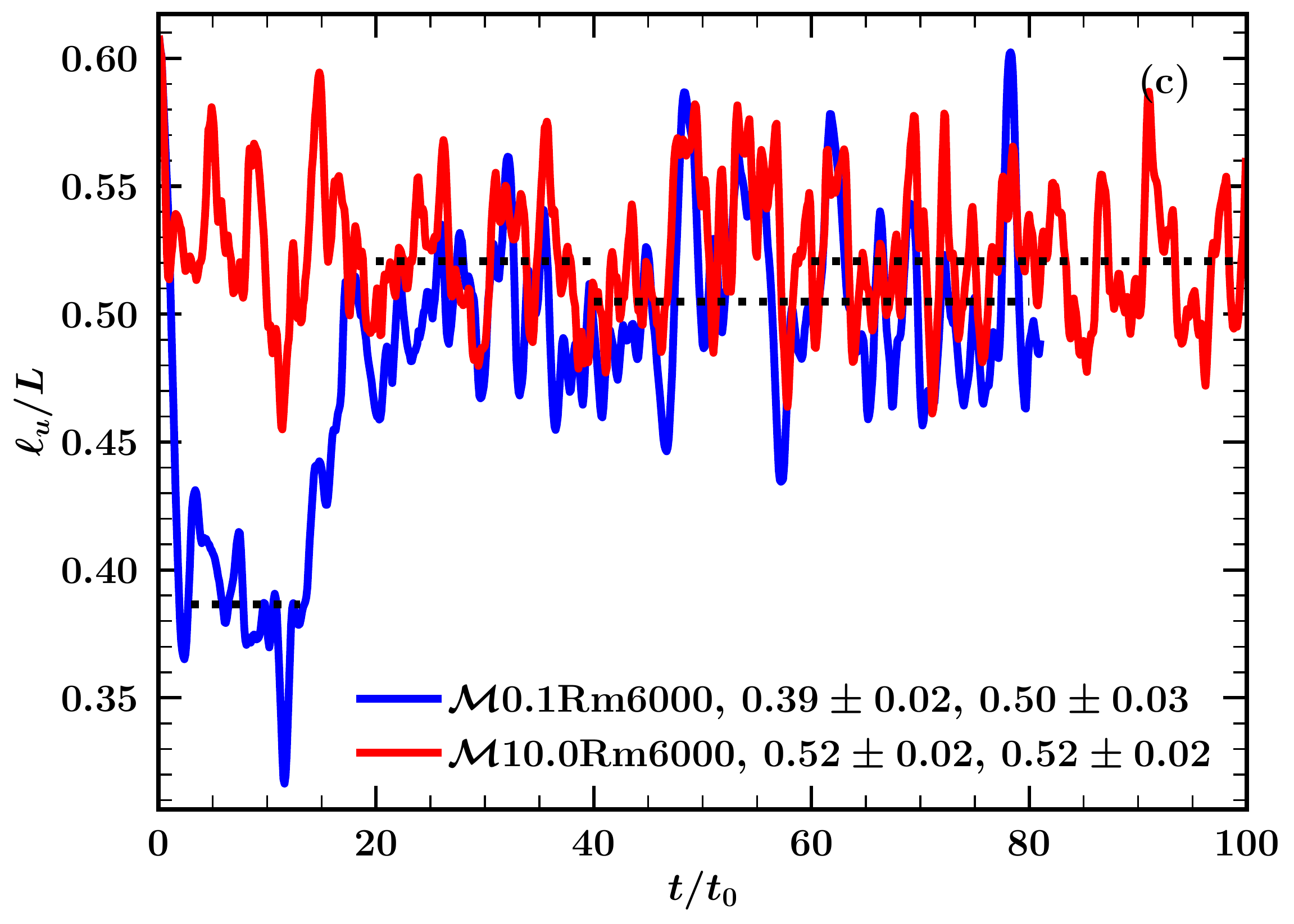}
    \includegraphics[width=\columnwidth]{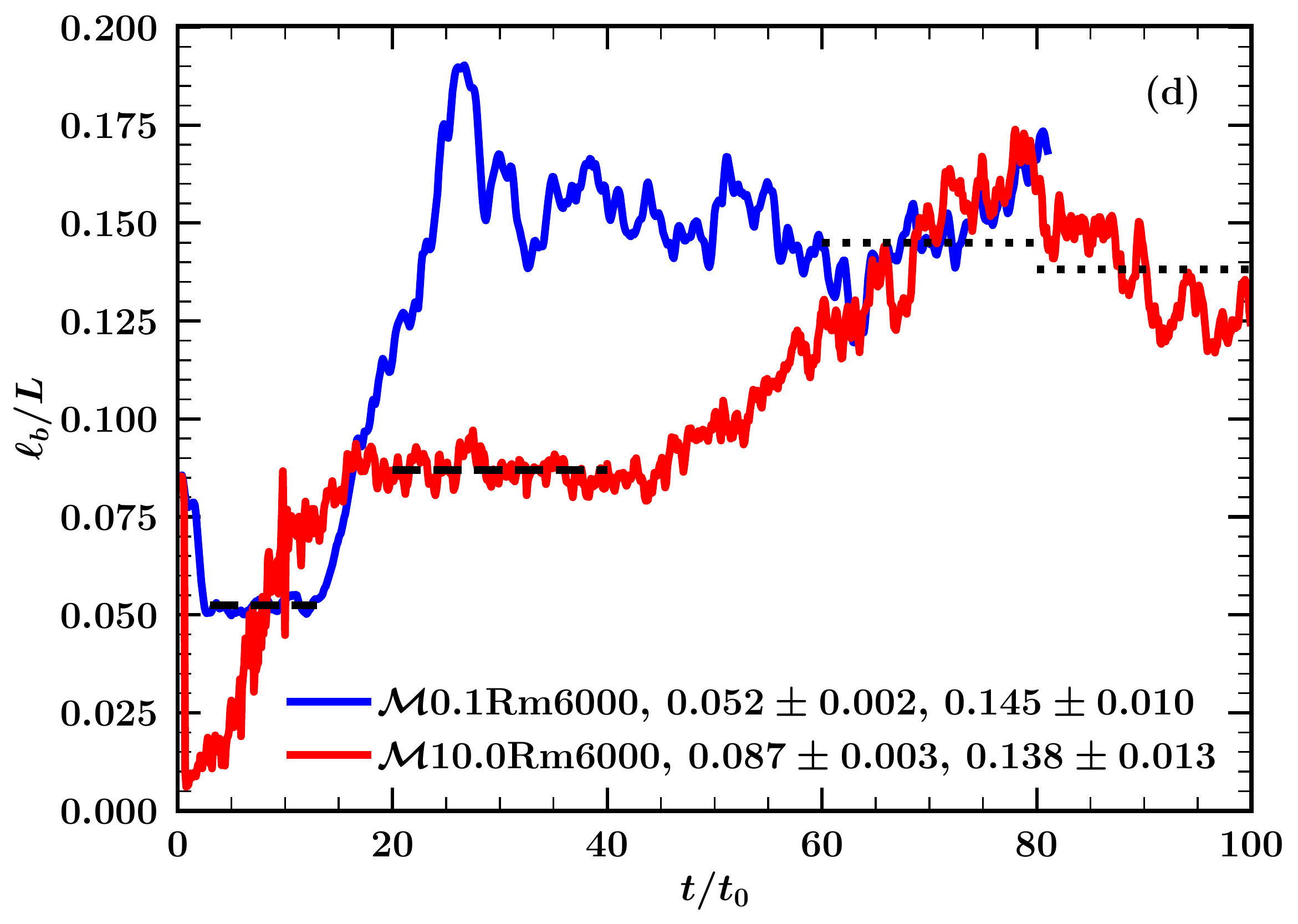}
    \caption{\rev{{\it Top panels:} the shell-integrated power spectrum of the turbulent kinetic (a) and magnetic (b) energies normalised by their total energies for $\Mach0.1\Rm6000$ (subsonic, blue) and $\Mach10\Rm6000$ (supersonic, red) cases. In each panel, the dashed and solid coloured line shows the spectra averaged over the kinematic ($7 \le t/t_0 \le 14$ for the subsonic case and $28 \le t/t_0 \le 35$ for the supersonic case) and saturated ($70 \le t/t_0 \le 80$ for the subsonic case and $85 \le t/t_0 \le 95$ for the supersonic case) stages, respectively, with the shaded region indicating their one-sigma variations. The turbulent kinetic energy spectra, in the range $3 \le k \le 20$, for the subsonic turbulence in the kinematic stage seem to be consistent with the Kolmogorov $k^{-5/3}$ (shown in a black, dashed-dotted line) spectrum \citep{Kolmogorov1941} and for the supersonic turbulence with the Burgers $k^{-2}$ (shown in a black, dotted line) spectrum \citep{Burgers1948}. The supersonic case has larger power on smaller scales compared to the subsonic one and this is because of strong shocks for $\Mach=10$ run (see \Fig{fig:vstruc2d} for locally strong structures in velocity fields). As the magnetic field saturates, the turbulent kinetic energy spectrum for the subsonic case steepens (due to the effect of back reaction of the strong magnetic fields on the flow) and such an effect is not that significant for the supersonic case. The magnetic energy spectra in the kinematic stage, at lower wavenumbers, for the subsonic case agree better with the Kazantsev $k^{3/2}$ (shown in black, dashed-dotted line) spectrum \citep{Kazantsev1968} than the supersonic case. As the dynamo saturates, spectra for both the cases become flat on larger scales. However, the magnetic energy for the supersonic case is higher at smaller scales in comparison to the subsonic case. This is also because of strong shocks and higher turbulent kinetic energy at smaller scales for supersonic flows. {\it Bottom panels:} the time evolution of the velocity ($\lu$, (c)) and magnetic ($\lb$, (d)) field correlation scales in units of the numerical domain length, $L$, for subsonic ($\Mach0.1\Rm6000$, blue) and supersonic ($\Mach10\Rm6000$, red) cases computed using \Eq{eq:corrlen}. The dashed and dotted black lines shows the average value of each length scale in the kinematic and saturated stages, respectively, and the corresponding values with one-sigma fluctuations are given in the legend. For the subsonic case, after the initial transient decay,  $\lu/L$ remains statistically steady at a value of $\approx 0.4$ in the kinematic stage ($7 \le t/t_0 \le 14$), but then increases as the field saturates and fluctuates around $\approx 0.5$. For the supersonic case, $\lu$ always fluctuates around $L/2$ after the initial transient phase. Thus, the back-reaction of the growing magnetic fields on the velocity is evident here for the low Mach number run. The magnetic field correlation scale, $\lb$, decreases in the initial transient phase and then remains roughly constant in the kinematic stage for both subsonic ($\lb/L \approx 0.05$) and supersonic ($\lb/L \approx 0.09$) cases. In fact, in the kinematic stage, $\lb$ is higher for the supersonic case (as can also be seen from \Fig{fig:bstruc2d} and \Fig{fig:bstruc3d}). Then, $\lb$ for both Mach numbers increases as the magnetic fields become strong enough to react back on the flow (see \Fig{fig:ts} for corresponding times). Finally, in the saturated stage, $\lb$ saturates to a statistically steady value, which is roughly the same for both Mach numbers ($\lb/L \approx 0.14$).}}
\label{fig:spec} 
\end{figure*}
The top panels of \Fig{fig:spec} show the shell-integrated turbulent kinetic ($E_{\rm turb,kin}$, \Fig{fig:spec}~(a)) and magnetic energy ($E_{\rm mag}$, \Fig{fig:spec}~(b)) spectra, which were averaged over a few eddy turnover times ($7\,t_0$ for the kinematic stage and $10\,t_0$ for the saturated stage) for the subsonic ($\Mach0.1\Rm6000$) and supersonic ($\Mach10\Rm6000$) cases. The turbulent kinetic energy spectra, over a range of wavenumbers ($3 \le k \le 20$), are seen to be consistent with the Kolmogorov $k^{-5/3}$ spectrum \citep{Kolmogorov1941} for the subsonic turbulence and Burgers $k^{-2}$ spectrum \citep{Burgers1948} for the supersonic turbulence (agreement is slightly better for the supersonic case). At smaller scales, the turbulent kinetic energy is higher for the supersonic case as compared to the subsonic case due to strong shocks. \rev{For the subsonic case, as the magnetic field saturates, the turbulent kinetic energy spectrum steepens and it shows the effect of the back-reaction due to the Lorentz force on the turbulent flow. However, this effect on the turbulent kinetic energy spectra for the supersonic case is not that significant.} The magnetic energy spectra vary between the kinematic and saturated stages for both subsonic and supersonic flows (\Fig{fig:spec}~(b)). \rev{At lower wavenumbers, the magnetic spectra in the kinematic stage seem to follow $k^{3/2}$ spectra as expected from Kazantsev's analytical work \citep{Kazantsev1968}. The agreement seems to be better for the subsonic flow than the supersonic case and this is because Kazantsev's theory is derived assuming incompressible flows. However, from extensions to Kazantsev's theory \citep{SchekochihinBK02}, the slope of the magnetic power spectrum for the supersonic case in the kinematic stage might still be $3/2$ \citep{FederrathEA2014}.} For both cases, the spectra flatten as the magnetic field saturates. On smaller scales, the magnetic energy is higher for the supersonic case in comparison to the subsonic one. This is primarily due to strong shocks and higher turbulent kinetic energy at smaller scales for supersonic turbulence.

We use the turbulent kinetic ($E_{\rm turb,kin} (k)$) and magnetic ($E_{\rm mag} (k)$) energy spectra to compute the correlation (approximately average) length of the velocity ($\lu$) and magnetic ($\lb$) fields as 
\begin{equation}
    \frac{\lu (\text{or}~ \lb)}{L} = \frac{\int_0^{\infty} k^{-1} E_{\rm turb,kin} (\text{or}~ E_{\rm mag}) ~\dd k}{\int_0^{\infty}  E_{\rm turb,kin} (\text{or}~ E_{\rm mag}) ~\dd k}.
    \label{eq:corrlen}
\end{equation}
\rev{The bottom panels of \Fig{fig:spec} show the velocity ($\lu/L$, \Fig{fig:spec}~(c)) and magnetic field ($\lb/L$, \Fig{fig:spec}~(d)) correlation scale as a function of the normalised time, $t/t_0$. After the initial transient phase, $\lu$ fluctuates around $L/2$ immediately for the supersonic case, but is initially lower ($\lu/L \approx 0.4$) for the subsonic case in its kinematic stage. For subsonic flows, as the magnetic field grows, $\lu$ increases and reaches a value of $L/2$ in the saturated stage. Thus, the back reaction of the growing magnetic field on the velocity field enhances its correlation length. This is seen only for the subsonic case \rev{(can also be concluded from the steepening of the turbulent kinetic energy spectra in \Fig{fig:spec}~(a)}). For the magnetic field, after the initial transient phase, the correlation length remains roughly constant in the kinematic phase ($\lb/L \approx 0.05$ and $0.09$ for the subsonic and supersonic case, respectively) and then increases as the magnetic fields become strong enough to react back on the turbulent flow. In the saturated stage, $\lb$ reaches a statistically steady value ($\lb/L \approx 0.14$, similar for both Mach numbers). Overall, the magnetic field correlation length increases due to the growing magnetic field for both the subsonic and supersonic cases (this also agrees with the larger size of magnetic structures seen in \Fig{fig:bstruc2d} and \Fig{fig:bstruc3d} for the saturated stage in comparison to the kinematic stage for both cases). We now quantify the non-Gaussianity of these magnetic structures.} 

\subsection{Magnetic intermittency} \label{sec:glo2}
\begin{figure*}
    \includegraphics[width=\columnwidth]{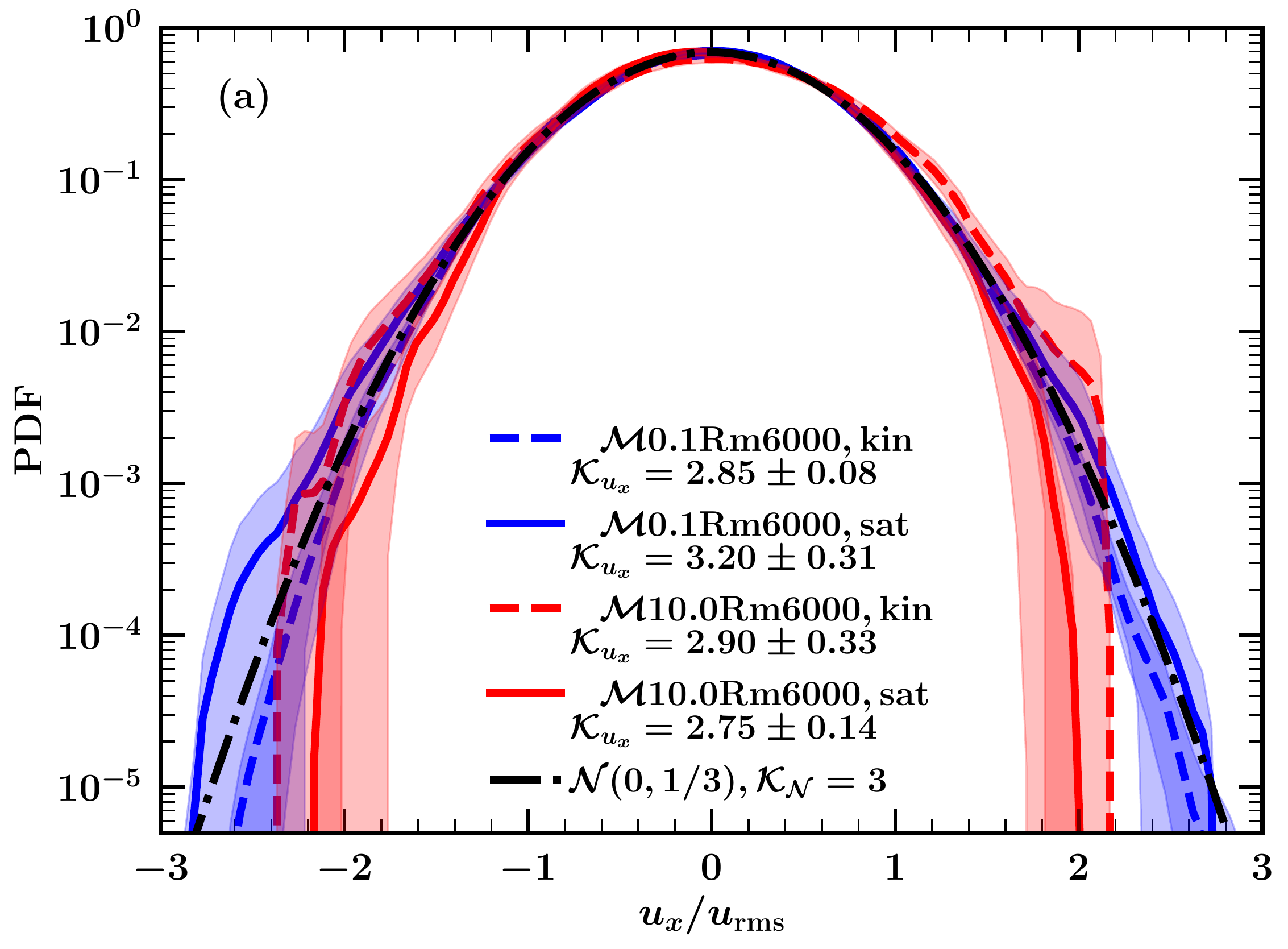}
    \includegraphics[width=\columnwidth]{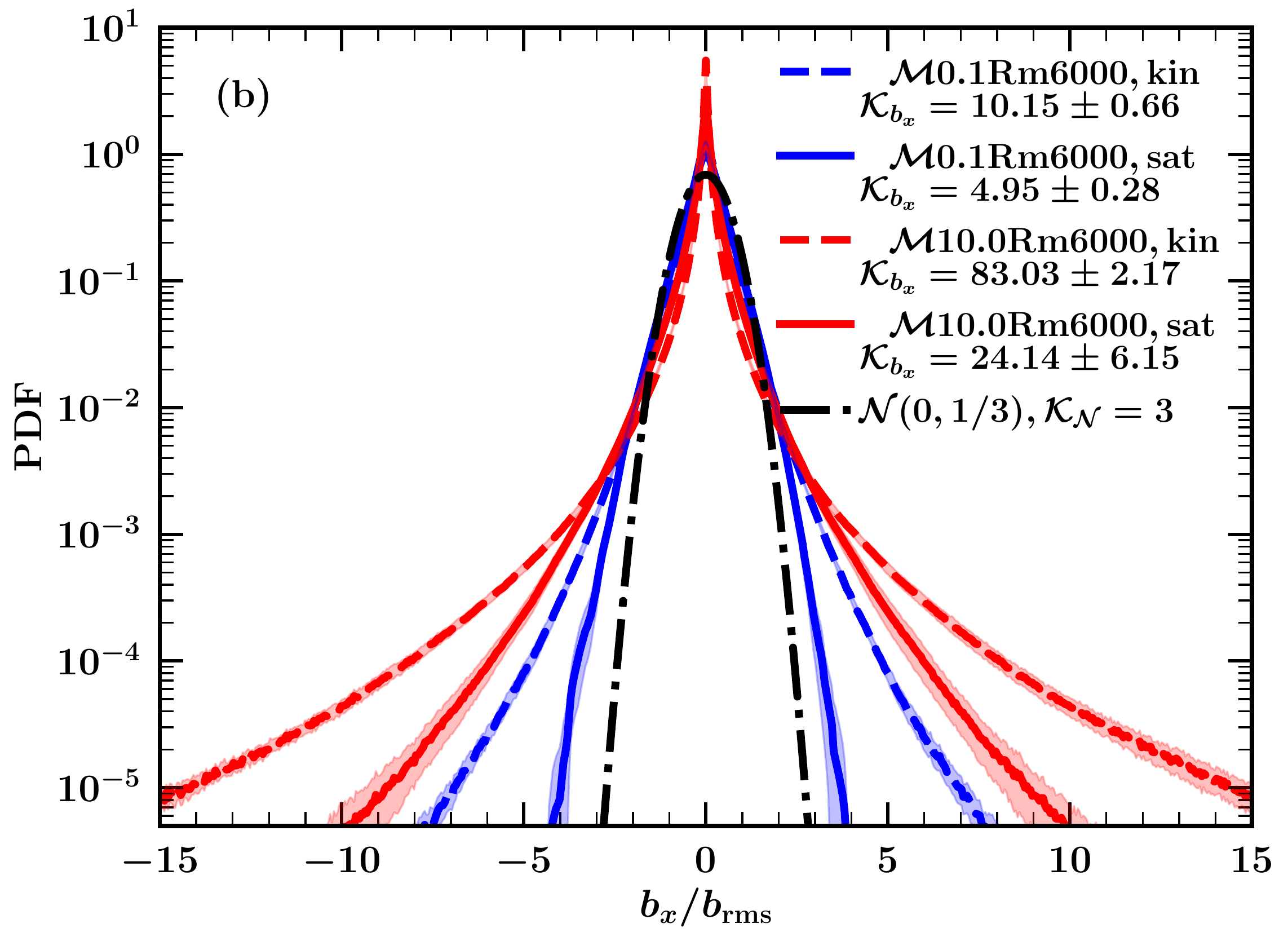}
    \caption{Probability distribution function (PDF) of the $x$ component of the normalised velocity, $u_x/\urms$ (a), and $x$ component of the normalised magnetic field, $b_x/\brms$ (b), for the subsonic (blue) and supersonic (red) turbulence in their respective kinematic (kin, dashed) and saturated (sat, solid) stages of the fluctuation dynamo. The lines show the mean obtained after averaging over ten eddy turnover times (in the range $7 \le t/t_0 \le 17$ and $70 \le t/t_0 \le 80$ for the subsonic case and $25 \le t/t_0 \le 35$ and $85 \le t/t_0 \le 95$ for the supersonic case) and the shaded regions of the same colour show one-sigma deviations. The computed kurtosis ($\ku_{u_x}$ and $\ku_{b_x}$) with its corresponding error (estimated from the time averaging) for each case is shown in the legend. For both the subsonic  and supersonic flows, in the kinematic and saturated stages, the velocity PDF is close to a Gaussian distribution ($\mathcal{N}$, the dashed-dotted, black line). Also, the computed kurtosis is always very close to that of a Gaussian distribution ($\ku_{\mathcal{N}} = 3$). Thus, the random velocity field in solenoidally driven turbulence follows a Gaussian distribution, immaterial of the Mach number and the dynamo stage. The magnetic field is always non-Gaussian (the dashed-dotted, black line shows the Gaussian distribution, $\mathcal{N}$). This is also confirmed from their computed kurtosis values, they are significantly higher than three. The magnetic field is always more non-Gaussian (or spatially intermittent) in the supersonic turbulence than in the subsonic case. Furthermore, for both Mach numbers, the magnetic field in the kinematic stage is more intermittent than their respective saturated stages.}
    \label{fig:vxbxpdf} 
\end{figure*}
The fluctuation dynamo generated magnetic fields are non-Gaussian or spatially intermittent and we aim to quantify the intermittency using the statistical measure kurtosis, $\ku$, which for any random function $f$ with mean zero is defined as 
\begin{equation}
    \ku_{f} = \frac{\langle f \rangle^4}{\langle f^2 \rangle^2},
    \label{eq:kurt}
\end{equation} 
where $\langle \rangle$ denote the average over the entire domain. The kurtosis of a Gaussian distribution is three. 

First, in \Fig{fig:vxbxpdf}~(a), we show the probability distribution function (PDF) of a single component (here, $x$) of the random velocity field, $u_x/\urms$, obtained in subsonic and supersonic turbulence in the kinematic and saturated stages of the fluctuation dynamo. The distribution, for all four cases, is very close to a Gaussian (or normal) distribution (the computed kurtosis is very close to that of a Gaussian, three). Thus, random velocity in these driven turbulence simulations always follows an approximately Gaussian distribution \citep[also see][]{Federrath2013}.

\begin{figure*}
    \includegraphics[width=\columnwidth]{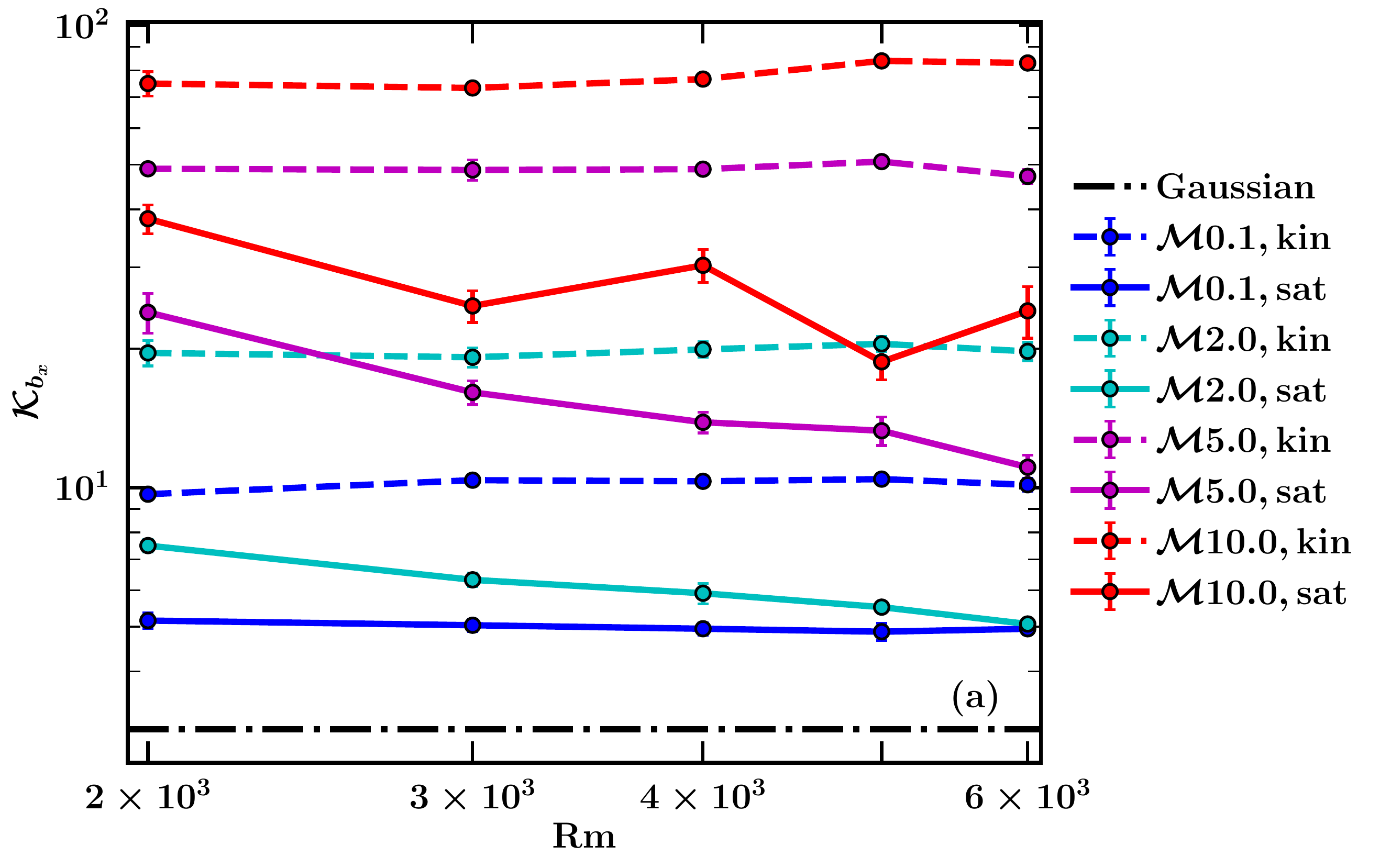}
    \includegraphics[width=0.9\columnwidth]{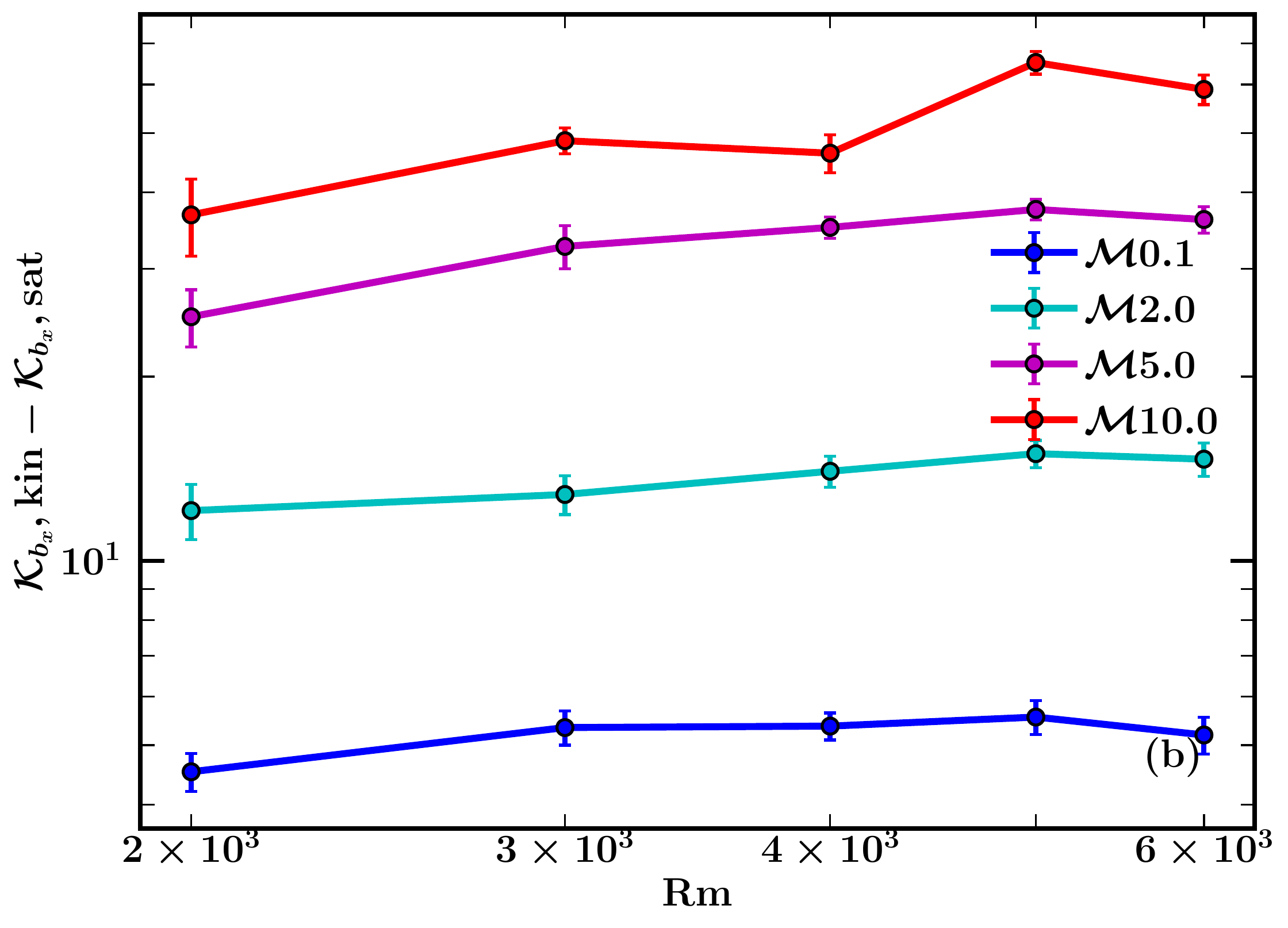}
       \caption{(a) The kurtosis of the $b_x/\brms$ distribution (see \Fig{fig:vxbxpdf}~(b)) as a function of $\Rm$ in the kinematic (kin, dashed) and saturated (sat, solid) stages of the fluctuation dynamo for $\Mach = 0.1$ (blue), $2$ (cyan), $5$ (magenta), and $10$ (red). The points show the mean obtained after averaging over ten eddy turnover times and the error bars of the same colour show one-sigma deviations (values are also provided in \Tab{tab:pro}). The kurtosis for all runs in both the kinematic and saturated stages are higher than the kurtosis of a Gaussian distribution (the dashed-dotted, black line). This confirms that all the distributions are non-Gaussian. The estimated kurtosis increases with $\Mach$ and this shows that the magnetic intermittency increases with the compressibility of the turbulent flow. Within this range of $\Rm$, the kurtosis for each stage does not seem to vary much with the $\Rm$. The kurtosis is always lower for the saturated stage in comparison to the kinematic stage and this difference (shown in (b)) increases with the Mach number of the turbulent flow.}
       \label{fig:bkurt} 
 \end{figure*}

\begin{figure*}
    
    \includegraphics[width=\columnwidth]{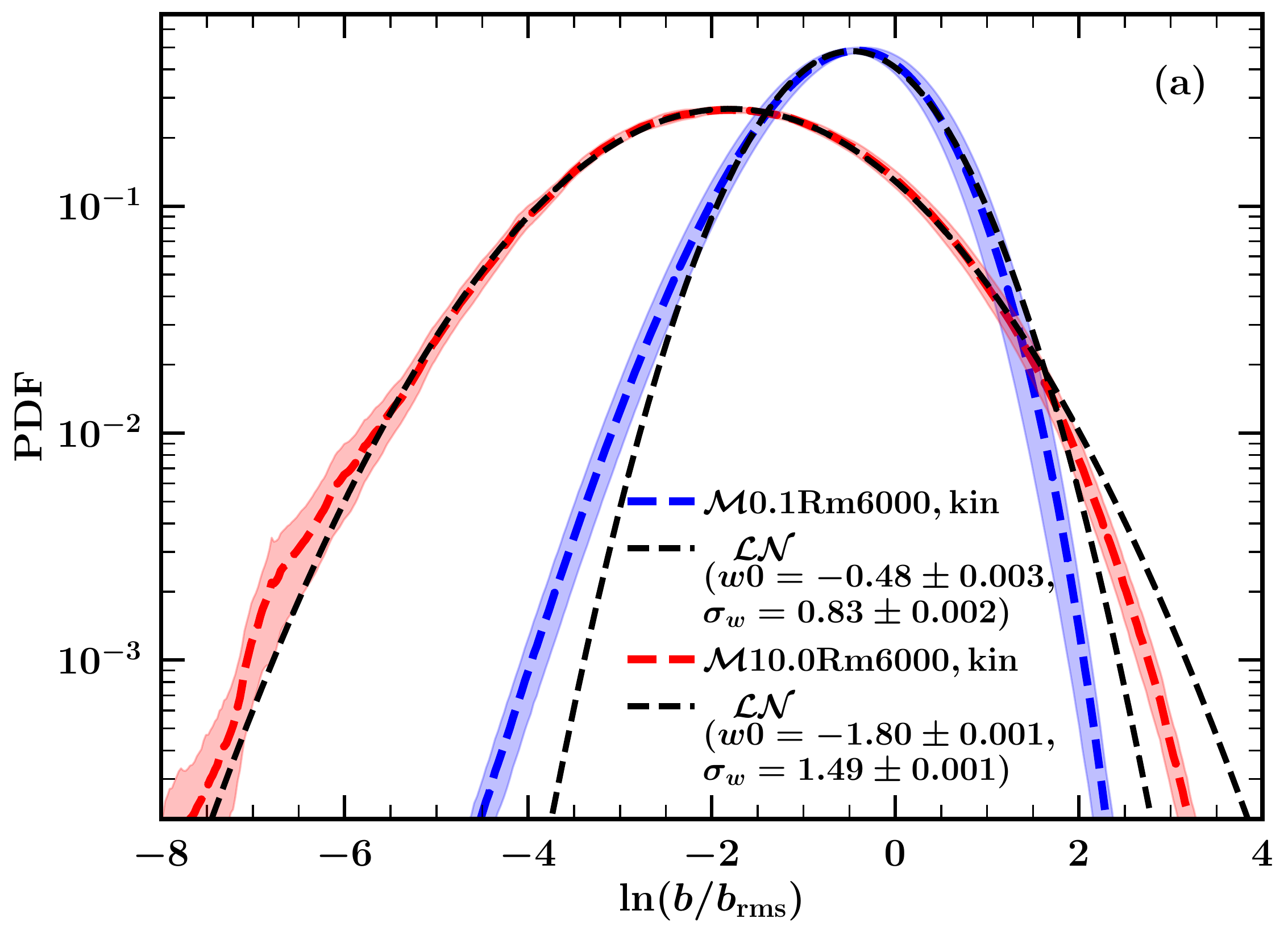}
    \includegraphics[width=\columnwidth]{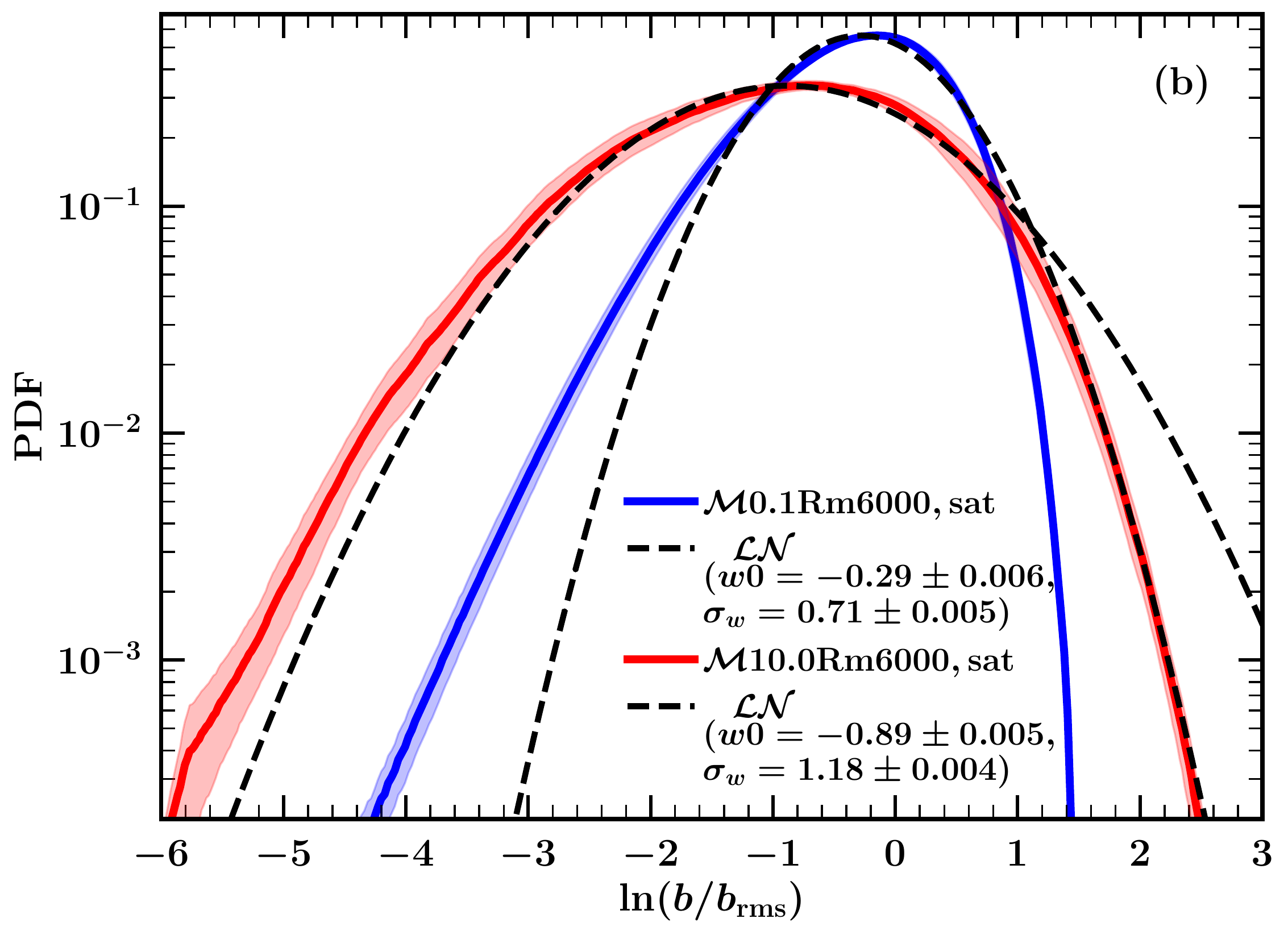}
       \caption{\rev{PDF of $\ln(b/\brms)$ for subsonic (blue) and supersonic (red) cases in their respective kinematic (kin, a) and saturated (sat, b) stages with black, dashed line showing the fit to a normal distribution (\Eq{eq:lnfit}). The normal distribution captures high probability regions better than the tails. In the kinematic stage, the lognormal fit to $b/\brms$ performs better for the supersonic case.  The fit worsens, for both Mach numbers, when the magnetic fields saturate (b).  Thus, PDF of $b/\brms$ in the kinematic stage can be represented by a near lognormal distribution. Furthermore, higher values of the fit parameters ($|w0|$ and $\sigma_{w}$) for the kinematic stage confirm that the magnetic intermittency decreases as the field saturates. Also, these parameters are always higher for the supersonic case as compared to the subsonic one. Thus, confirming that the degree of magnetic intermittency increases with the Mach number of the turbulent flow.}}
       \label{fig:blogpdfs} 
 \end{figure*}

Even though the turbulent velocity is normally distributed, the magnetic field it generates is non-Gaussian or spatially intermittent, as shown by \Fig{fig:vxbxpdf}~(b), which shows the PDF of a single component of the random magnetic field, $b_x/\brms$, in their kinematic and saturated stages for subsonic and supersonic turbulent flows. The magnetic field is always more intermittent for the supersonic turbulence than the subsonic one. Moreover, for both cases, the magnetic intermittency decreases as the dynamo saturates. These conclusions can also be confirmed via the computed kurtosis values (\Tab{tab:pro} and  \Fig{fig:bkurt}). \Fig{fig:bkurt}~(a) shows that the kurtosis, in the $\Rm$ range $2000~\text{--}~6000$, does not depend much on $\Rm$ but increases with $\Mach$ for both the kinematic and saturated stages. The difference in kurtosis of the magnetic fields between the kinematic and saturated stages increases with the Mach number of the turbulent flow, $\Mach$ (\Fig{fig:bkurt}~(b)). Thus, the non-Gaussian or intermittent nature of the dynamo generated magnetic fields is enhanced with compressibility.

\rev{\Fig{fig:blogpdfs} shows the PDF of $\ln(b/\brms)$ for the subsonic and supersonic cases in their respective kinematic (\Fig{fig:blogpdfs}~(a)) and saturated (\Fig{fig:blogpdfs}~(b)) stages. We fit the PDF of $\ln(b/\brms)$ with a normal distribution of form,
\begin{align}
\mathcal{LN} = (2 \pi \sigma_{w})^{-1/2} \exp\left(-\frac{(\ln(b/\brms) - w0)^{2}}{2 \sigma^{2}_{w}}\right),
\label{eq:lnfit}
\end{align} 
where $w0$ (mean) and $\sigma_{w}$ (standard deviation) are parameters of the fit. The black, dashed lines in \Fig{fig:blogpdfs} show the fitted distributions, and parameters of the fit are given in the legend. For the kinematic phase (\Fig{fig:blogpdfs}~(a)), the lognormal distribution for $b/\brms$ is a better fit to the supersonic case as compared to the subsonic one. Also, the fit worsens for the saturated stage. Overall, the lognormal distribution captures high probability regions quite well in the kinematic phase and performs better for supersonic turbulence. The parameters $|w0|$ and $\sigma_{w}$ are also higher in the kinematic stage as compared to the saturated stage and are always higher for the supersonic flows. This further confirms that the degree of magnetic intermittency decreases as the field saturates and increases with the compressibility of the turbulent flow.}

After characterising the global properties of velocity and magnetic fields in the fluctuation dynamo, in the next section, we study the local interaction of velocity and magnetic fields to explore the saturation mechanism of the fluctuation dynamo.

\section{Local dynamics and the saturation mechanism} \label{sec:loc}
To understand the saturation mechanism of the fluctuation dynamo in subsonic and supersonic turbulent plasmas, here, we investigate the local properties and interactions of velocity and magnetic fields. This is also motivated by the fact that the fluctuation dynamo generated magnetic field is spatially intermittent (\Sec{sec:glo2}) and thus we would expect that the back reaction of the magnetic field would be enhanced (locally and statistically) in strong-field regions (regions with magnetic field energy higher than the rms value, $b^2/\brms^2 > 1$).

\begin{figure}
    \includegraphics[width=\columnwidth]{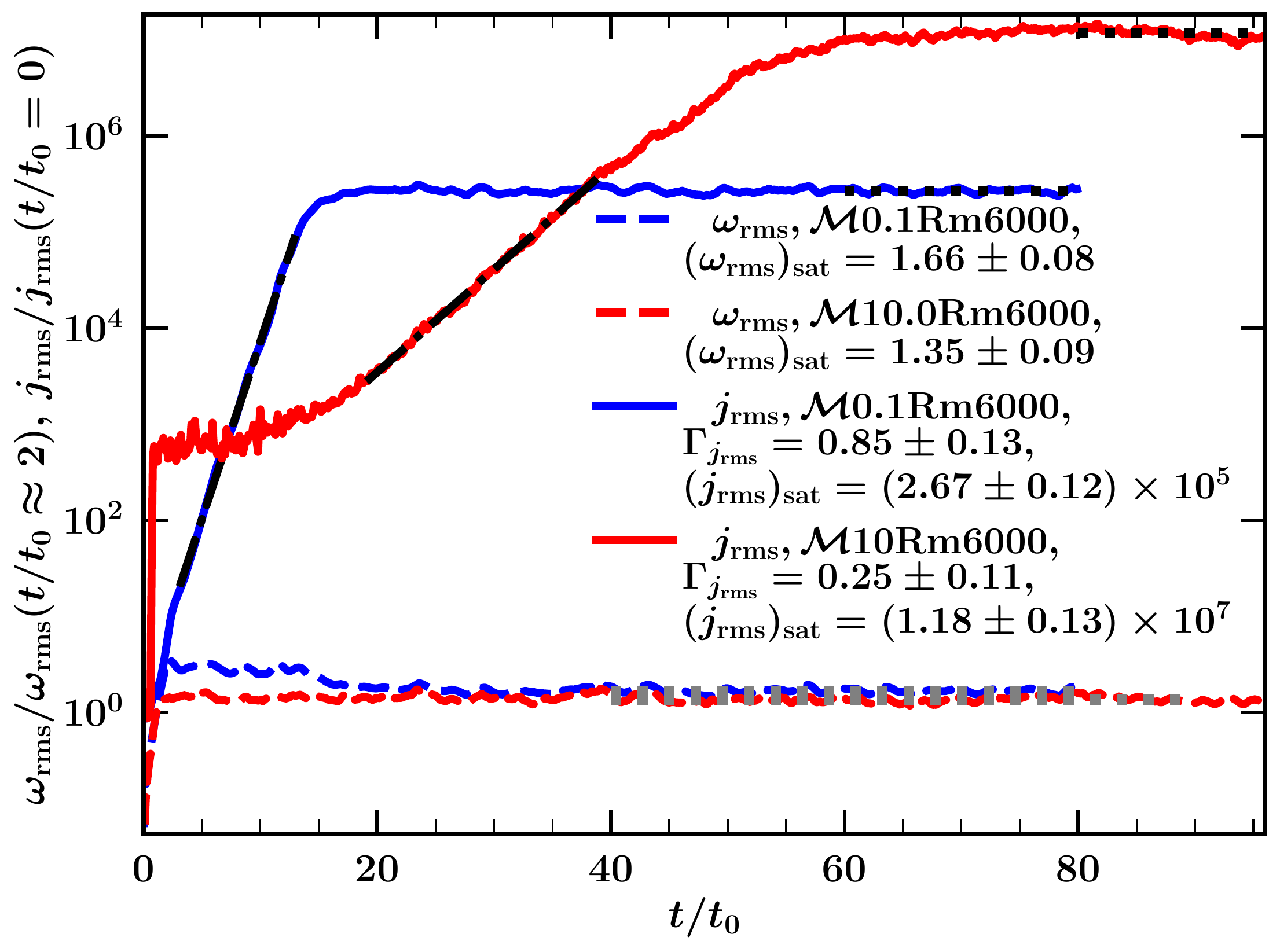}
    \caption{Temporal evolution of the (normalised) rms vorticity ($\wrms/\wrms(t/t_0\approx2)$, dashed) and current density ($\jrms/\jrms(t/t_0=0)$, solid) for subsonic ($\Mach0.1\Rm6000$, blue) and supersonic ($\Mach10\Rm6000$, red) turbulence. The vorticity remains roughly the same once the turbulence is established and is slightly higher for the subsonic case as compared to the supersonic (the mean value, $(\wrms)_{\rm sat}$, shown in the legend). The overall trend of the current density evolution is the same as the magnetic field, exponential amplification after the initial transient phase and then saturation. The growth rate, $\Gamma_{\jrms}$, is higher for the subsonic case but the saturated value, $(\jrms)_{\rm sat}$, is higher for the supersonic case. The growth rate of the current density is roughly half of the magnetic field growth rate (see \Fig{fig:ts}) for both the subsonic and supersonic flows.}
    \label{fig:wjts} 
\end{figure}
\begin{figure*}
    \includegraphics[width=\columnwidth]{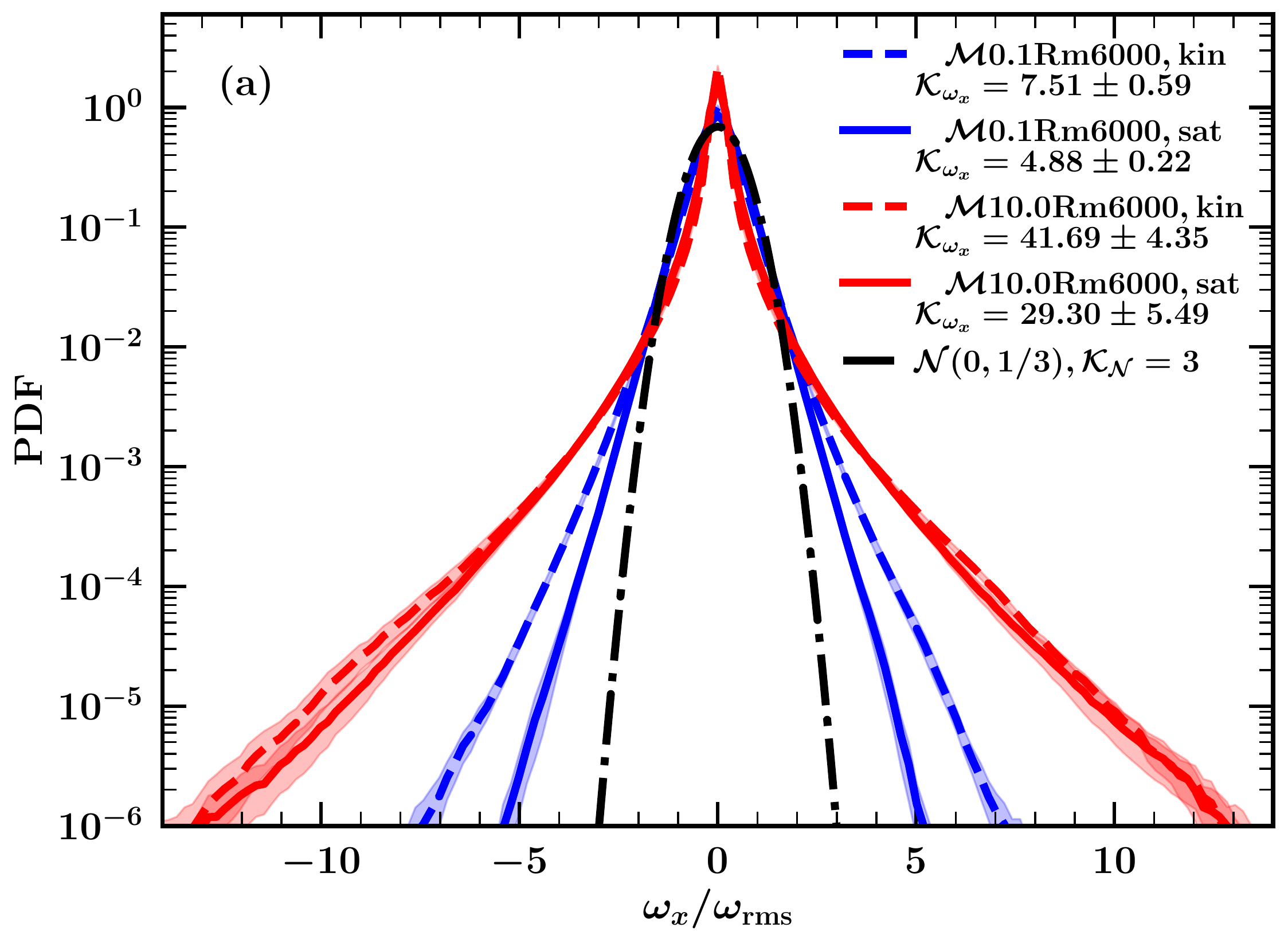}
    \includegraphics[width=\columnwidth]{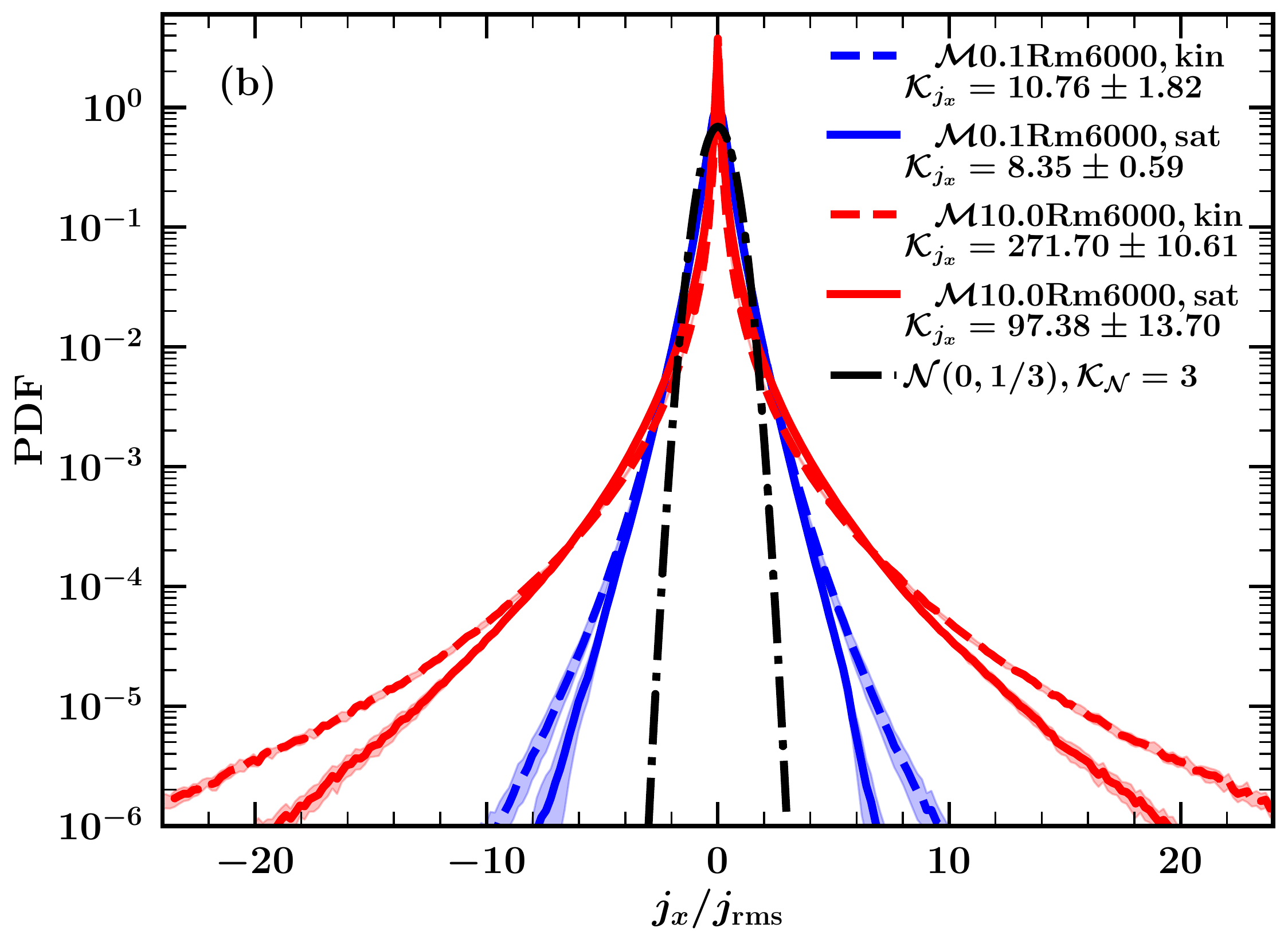}
       \caption{PDF of the normalised vorticity, $\w_x/\wrms$ (a), and current density, $j_x/\jrms$ (b), for subsonic (blue) and supersonic (red) flows in the kinematic (kin, dashed) and saturated (sat, solid) stages of the fluctuation dynamo (the lines show the mean over ten eddy turnover times and the shaded region show the one-sigma fluctuations). Both the vorticity and current density distributions are non-Gaussian or intermittent (the dashed-dotted, black line shows the Gaussian distribution) and the estimated kurtosis (shown in the legend for each case) is also higher than three. The vorticity distribution for the subsonic case is less intermittent in the saturated stage as compared to the kinematic stage and this is due to the back reaction of the amplified magnetic field on the velocity flow. Such a statistical difference in the kurtosis of the vorticity between the kinematic and saturated stages is not seen for the supersonic turbulence. The current density distribution, for both the subsonic and supersonic turbulence, is less intermittent in the saturated stage as compared to their respective kinematic stage \rev{but is always higher for the supersonic case.}}
       \label{fig:wjpdfs} 
 \end{figure*}
First, we look at the properties of the following two dynamically important quantities, vorticity ($\vec{\w}$) and current density ($\vec{j}$), which are defined as the curl of the velocity ($\vec{u}$) and magnetic field ($\vec{b}$),
\begin{equation}
    \vec{\w} = \nabla \times \vec{u}, \quad \vec{j} = \nabla \times \vec{b}.
    \label{eq:wj}
\end{equation}
Vorticity and current density characterise the local structure of the velocity and magnetic fields, respectively. The evolution of the rms (since the mean is zero) vorticity and current density is shown in \Fig{fig:wjts}. Once the turbulence is established, $\wrms$ is statistically steady and is slightly higher in the subsonic case. The current density evolution is very similar to that of the magnetic fields, i.e., exponential growth and then saturation. The growth rate of $\jrms$ is roughly half of the magnetic field growth rate for each case (\Fig{fig:ts} and \Tab{tab:pro}) and is higher for the subsonic flows. \rev{The saturation level of the rms current density, $\jrms$, is higher for the supersonic case compared to the subsonic case (note that the initial value, $\jrms(t/t_0=0)$, is the same for both subsonic and supersonic cases)}. \Fig{fig:wjpdfs} shows the pdfs of $\w_x/\wrms$ (\Fig{fig:wjpdfs}~(a)) and $j_x/\jrms$ (\Fig{fig:wjpdfs}~(b)) for the kinematic and saturated stages of the dynamo with subsonic and supersonic turbulent flows. Both vorticity and current density distributions are non-Gaussian and the non-Gaussianity (quantified by kurtosis) for both cases is higher for the supersonic turbulence than the subsonic one. For the subsonic case, the kurtosis of $\w_x/\wrms$ decreases as the field saturates and thus vorticities significantly stronger than the rms value are suppressed in the saturated stage in comparison to the kinematic stage. This is a direct consequence of the back reaction of magnetic fields on the velocity field, i.e., the structure of the flow is locally altered. Such a statistically significant difference in the vorticity distribution is not seen for the supersonic turbulent flow. \rev{This also agrees with the conclusion from studying the evolution of velocity correlation length in \Fig{fig:spec}~(c).} On the other hand, the current density distribution for both the subsonic and supersonic flows is less intermittent in the saturated stage compared to their respective kinematic stages \rev{(compare the computed kurtosis given in the legend of \Fig{fig:wjpdfs}(b))}. Thus, the magnetic field in the saturated stage is structurally different than in the kinematic stage  \rev{and for the supersonic turbulence in comparison to the subsonic case} (global differences shown in \Fig{fig:bstruc2d}, \Fig{fig:bstruc3d}, \Fig{fig:spec}, and \Fig{fig:vxbxpdf}). In the next subsection, we look at local interactions in terms of angles between the following vector quantities: vorticity, velocity, current density, and magnetic fields.

\subsection{Local interactions in terms of relevant angles}
\begin{figure*}
    \includegraphics[width=\columnwidth]{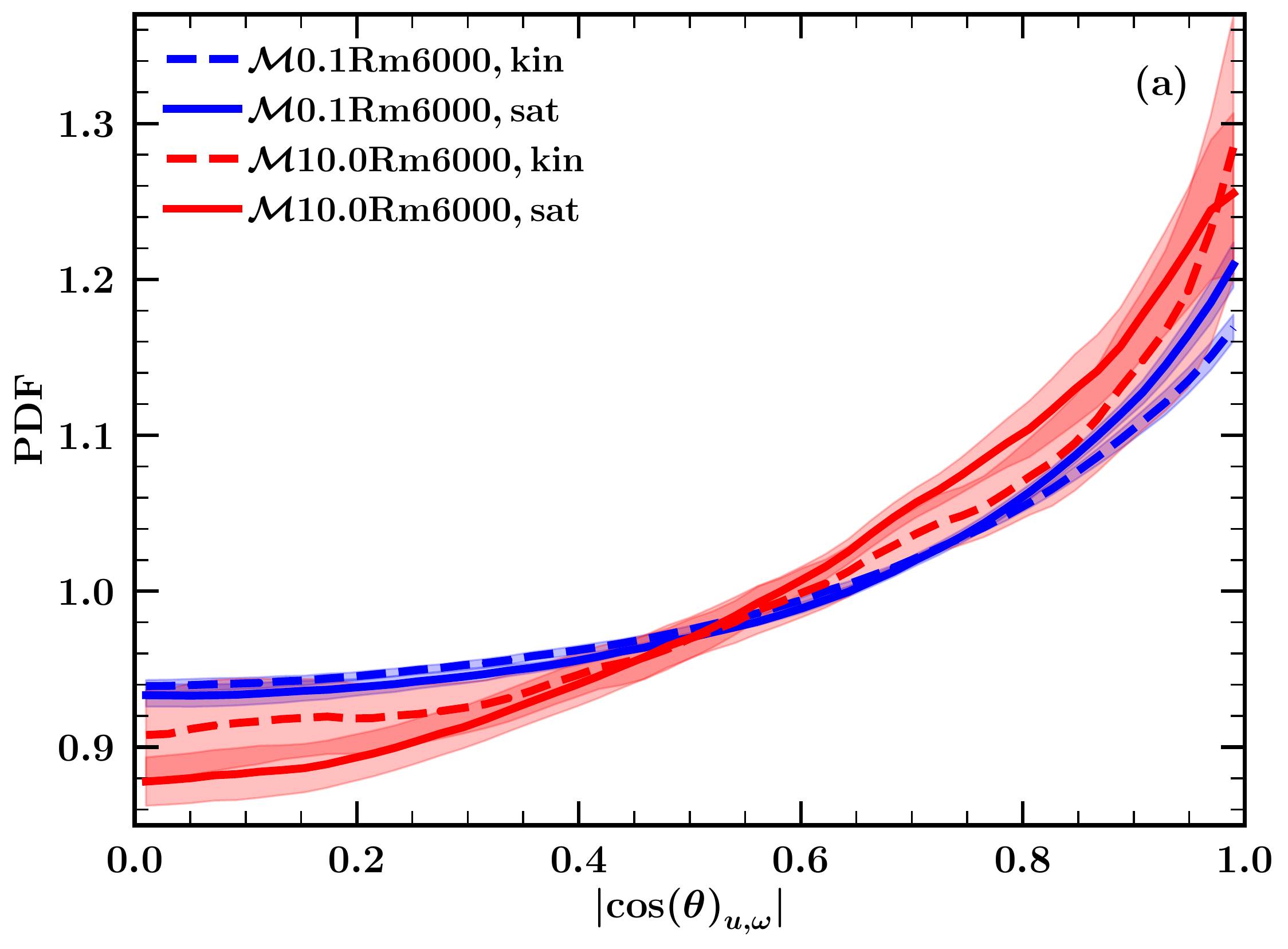}
    \includegraphics[width=\columnwidth]{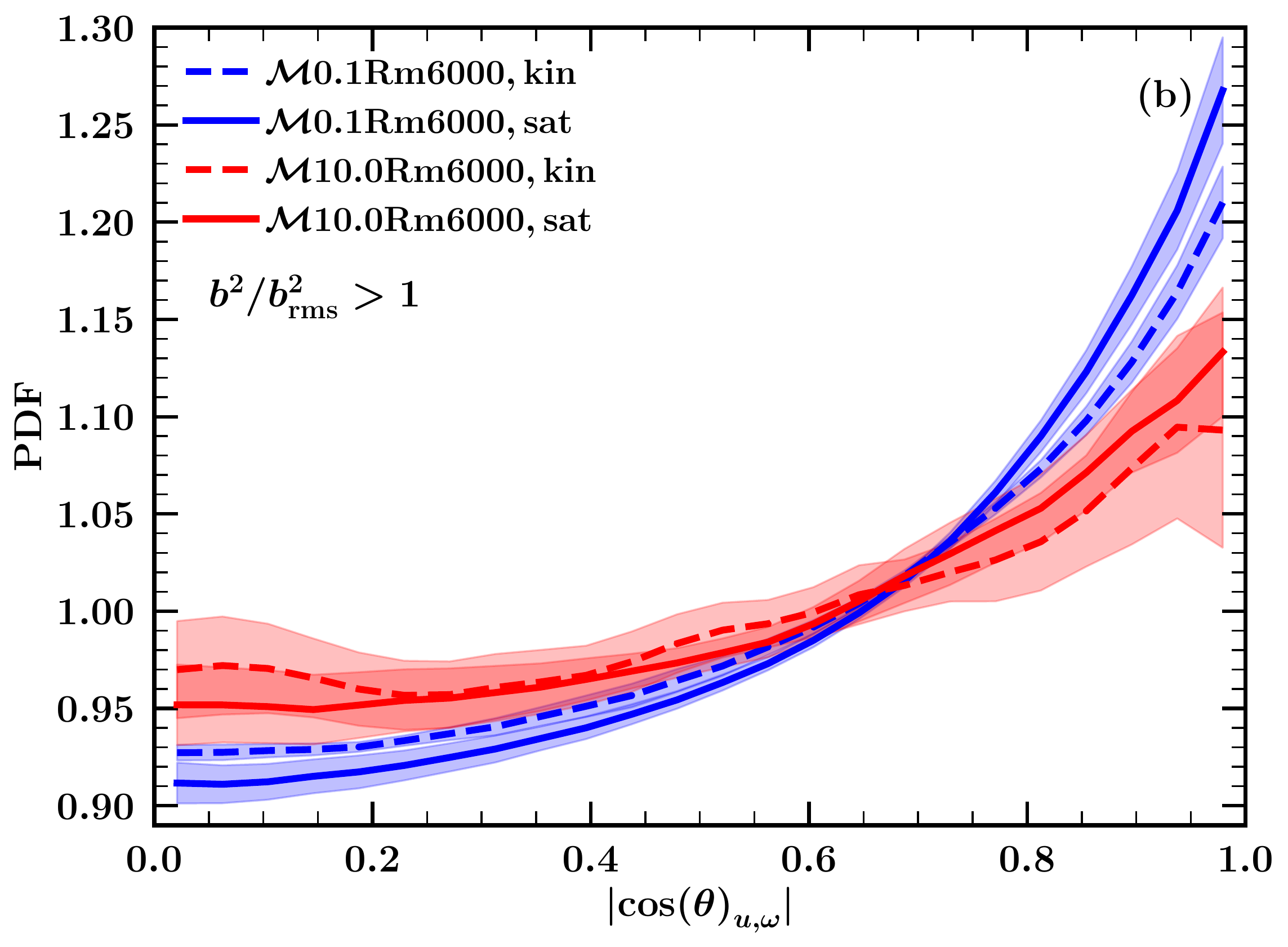} 
    \includegraphics[width=\columnwidth]{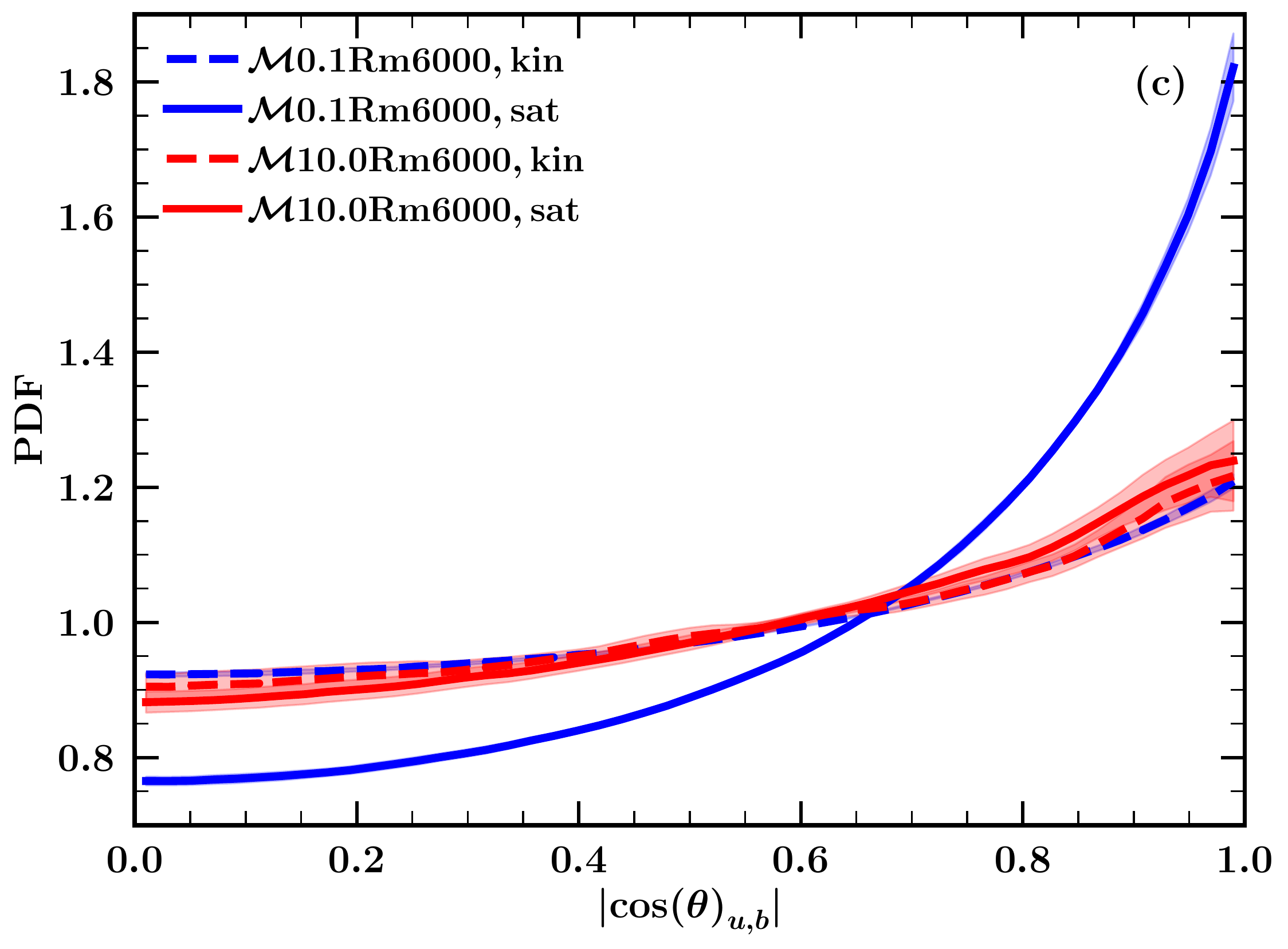}
    \includegraphics[width=\columnwidth]{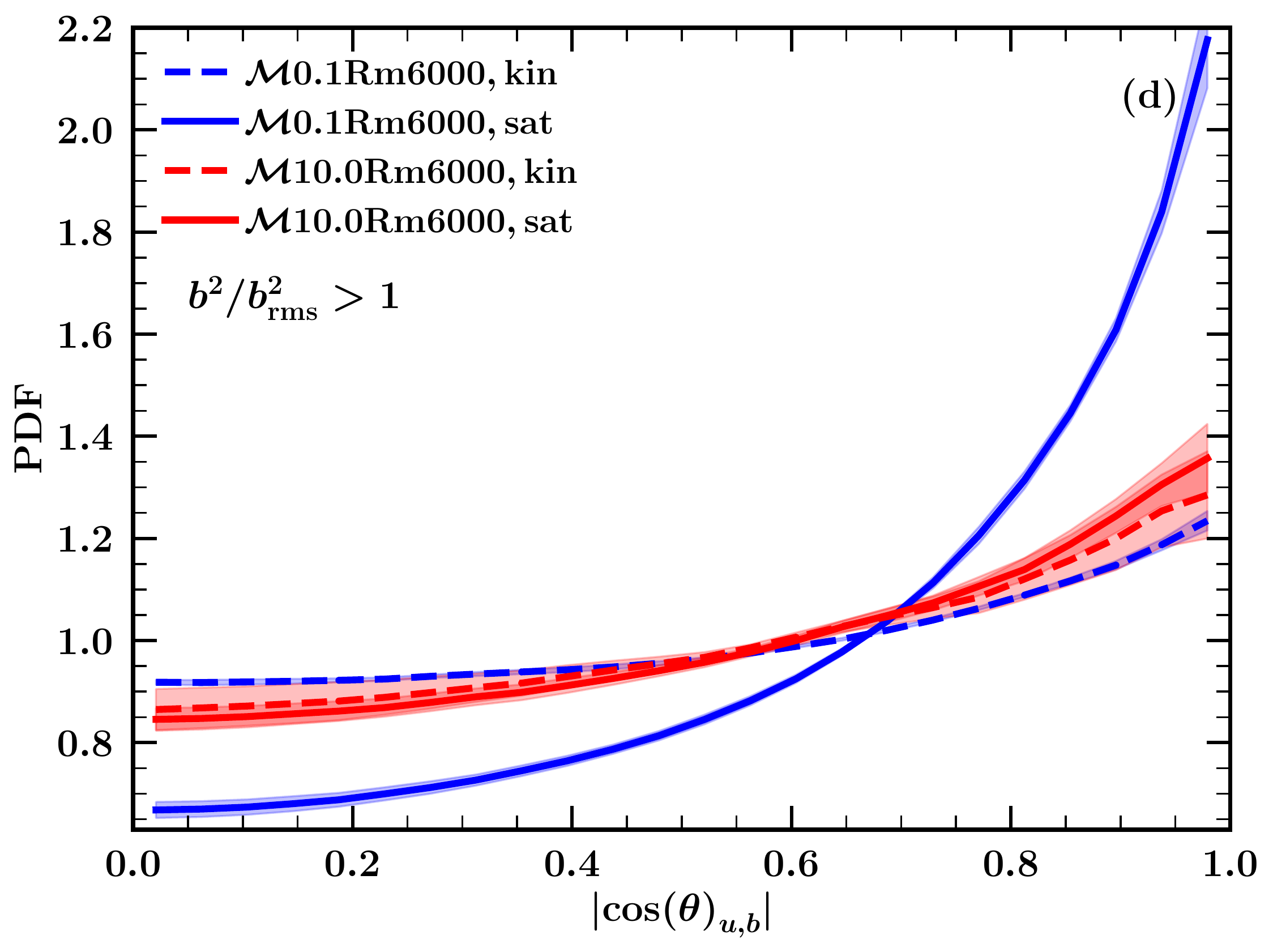}
    \includegraphics[width=\columnwidth]{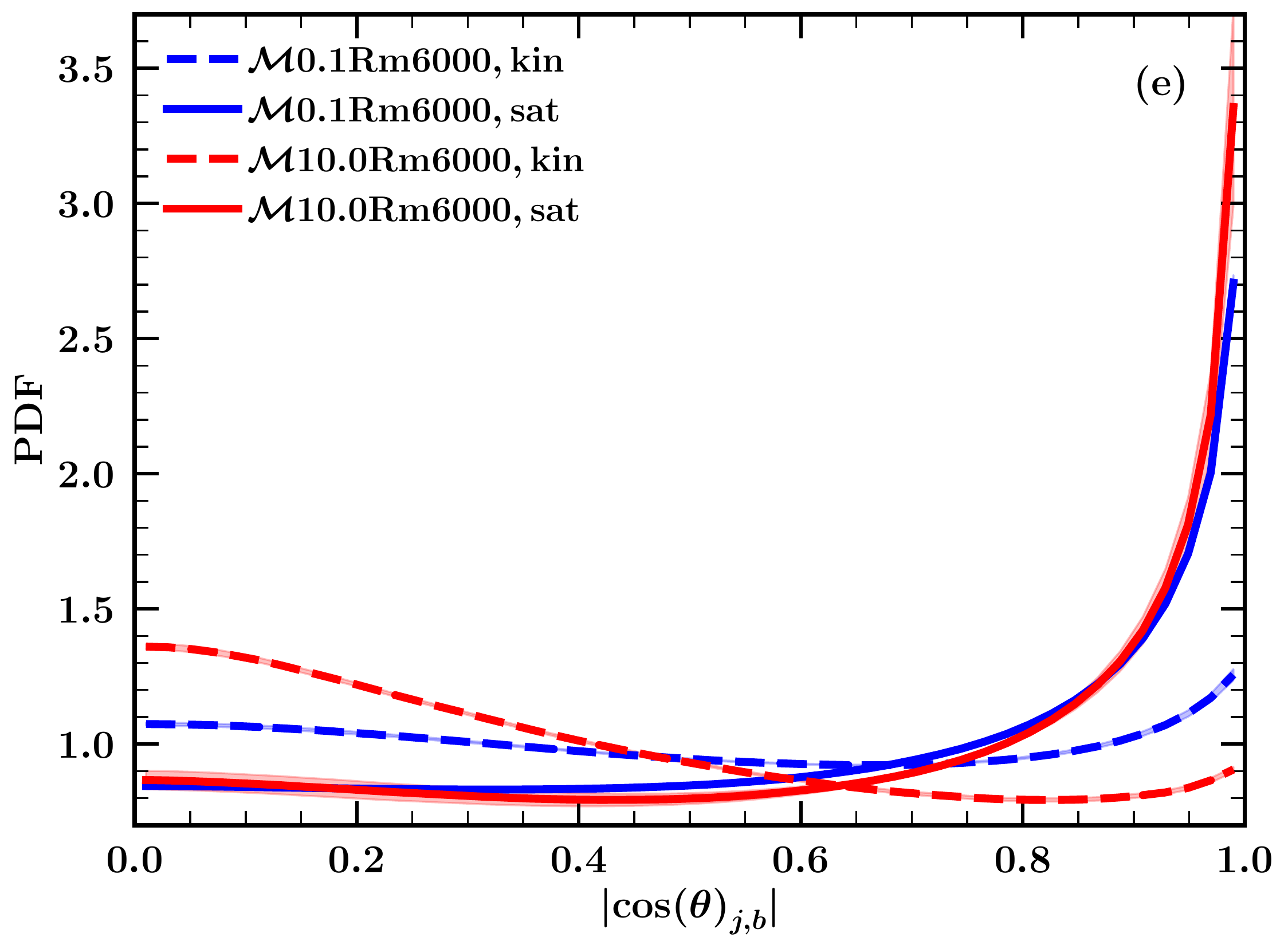}
    \includegraphics[width=\columnwidth]{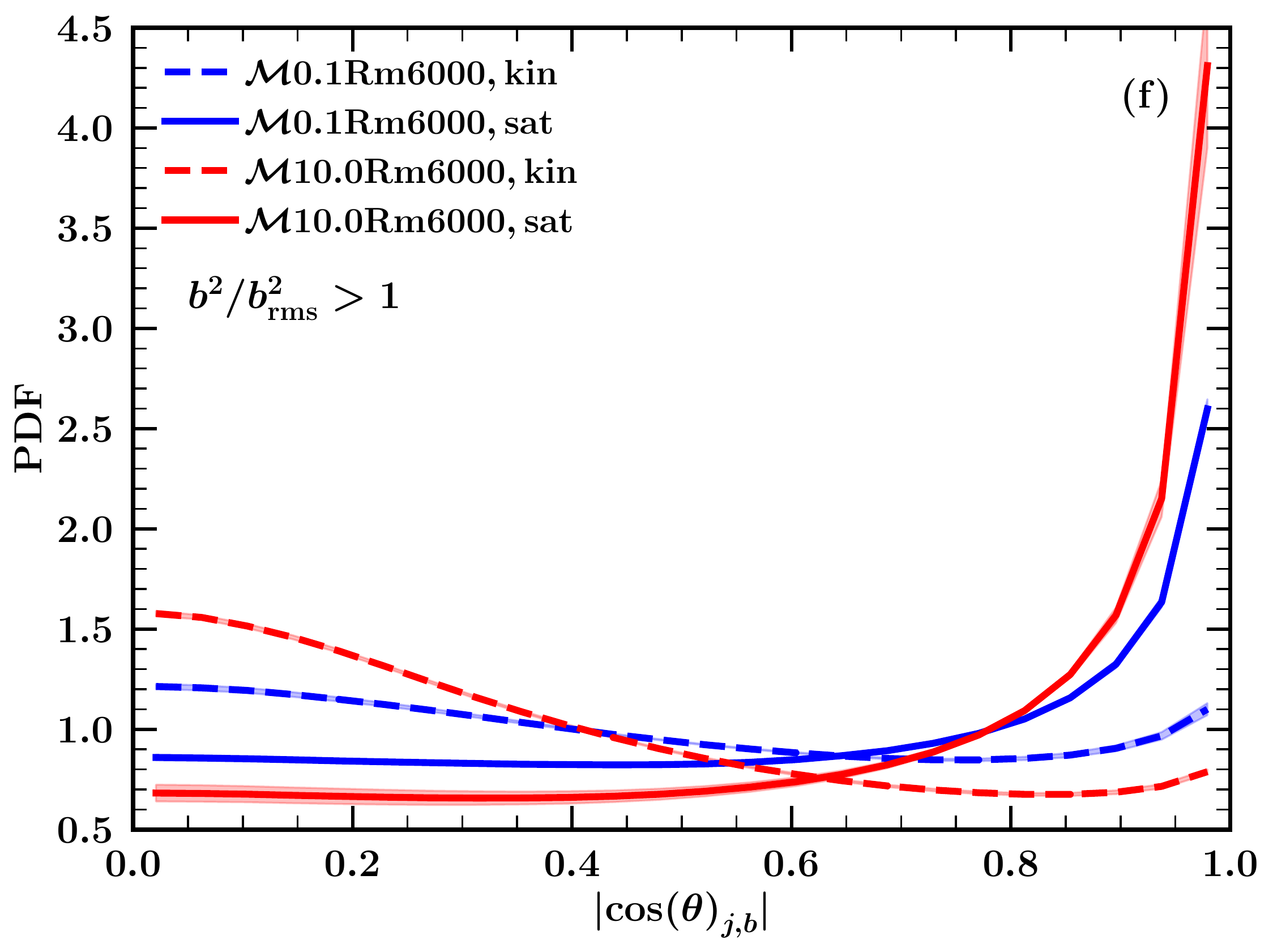}
      \caption{PDFs of the absolute value of the cosine of angle between the velocity and vorticity ($|\cosuw|$, a, b), velocity and magnetic field ($|\cosub|$, c, d),  and current density and magnetic field ($|\cosjb|$, e, f) for subsonic (blue) and supersonic (red) turbulence in the kinematic (kin, dashed) and saturated (sat, solid) stages of the fluctuation dynamo (the line shows the mean over ten eddy turnover times and the shaded regions shows the corresponding one-sigma variations). The left panels (a, c, e) show the PDFs over the entire domain and the right panels (b, d, f) show the conditional PDFs based only on the strong-field regions ($b^2/\brms^2 > 1$). The absolute value of the cosine of the angle is studied, because the distributions are symmetric about the $\cos(\theta) = 0$ for isotropic velocity and magnetic fields. $|\cos(\theta) | = 0$ implies an angle of $90 \degree$ (or $270 \degree$) between the two vectors and $|\cos(\theta) | = 1$ implies alignment (or anti-alignment) between them. For the subsonic turbulence, as the magnetic field saturates, velocity and vorticity, velocity and magnetic fields, and current density and magnetic fields are more aligned. Each trend gets more enhanced in the strong-field regions (b, d, f). \rev{The corresponding differences between the kinematic and saturated stages for the supersonic case are only significant for the angle between the current density and magnetic field. For the supersonic case, $|\cosuw|$ and $|\cosub|$ are statistically the same in the kinematic and saturated stage. In the saturated stage, the current density and magnetic field are more aligned (the level of alignment is enhanced in the strong-field regions) and the difference between the two stages is higher for the supersonic turbulence in comparison to the subsonic case.}}
       \label{fig:apdfs} 
 \end{figure*}
After confirming the change in the local structure of the velocity and magnetic field on saturation, here, we study the local interaction between those two vectors. First, the evolution of vorticity, obtained by taking the curl of the Navier-Stokes equation (the baroclinic term is zero for an isothermal equation of state), is \citep{Davidson2001, MeeB2006} 
\begin{equation}
    \frac{\partial \vec{\w}}{\partial t} = \nabla \times (\vec{u} \times \vec{\w}) + \nu \nabla^2 \vec{\w} + 2 \nu \nabla \times \left(\tau \ln \rho\right) + \nabla \times \left(\frac{\vec{j} \times \vec{b}}{c\rho} \right),
    \label{eq:vorevo}
\end{equation}
where the first term on the right-hand side represents growth of vorticity, the second term represents vorticity diffusion, the third term is due to density gradients, and the last term is due to the Lorentz force, $\vec{j} \times \vec{b}$ ($c$ is the speed of light). Since, we start with a zero velocity (and vorticity) and very weak magnetic fields, the vorticity is most likely generated by the third term when the density gradients develop and is then amplified by the first term \citep[see \Fig{fig:wjts} and in][]{MeeB2006, FederrathEA2011}. The amplification of the vorticity field, via the first term, depends on the angle between $\vec{u}$ and $\vec{\w}$.

Another two angles of interest are the angle between the velocity and magnetic field, which controls the magnetic field induction term (first term in the right-hand side of \Eq{eq:ie}) and the angle between the current density and magnetic field, which controls the Lorentz force (or the back reaction of the velocity field on the flow). Thus, we compute the cosine of the following three relevant angles,
\begin{align}
    \cosuw &= \frac{\vec{u} \cdot \vec{\w}}{|\vec{u}| |\vec{\w}|}, \\ \nonumber \, &\text{for the vorticity amplification term, $\nabla \times (\vec{u} \times \vec{\w})$}, \\ 
    \cosub &= \frac{\vec{u} \cdot \vec{b}}{|\vec{u}| |\vec{b}|}, \\ \nonumber \, &\text{for the magnetic induction term, $\nabla \times (\vec{u} \times \vec{b})$}, \\
    \cosjb &= \frac{\vec{j} \cdot \vec{b}}{|\vec{j}| |\vec{b}|}, \\ \nonumber \, &\text{for the Lorentz force, $\vec{j} \times \vec{b}$}.
\end{align}
$\cos(\theta) = 0$ implies an angle of $90 \degree$ (or $270 \degree$) between the two vectors and thus maximum effect of the physically relevant term (for example, angle of $90 \degree$ between $\vec{u}$ and $\vec{b}$ implies maximum induction). On the other hand, $\cos(\theta) = 1 (\text{or}~-1)$ implies alignment (or anti-alignment) between them and no effect of the term.  \Fig{fig:apdfs} shows the total and conditional (for strong magnetic field regions, $b^2/\brms^2 > 1$) PDFs of these three angles in the kinematic and saturated stages for the subsonic and supersonic turbulence. We only show the absolute value of these angles as these distributions are symmetric around $\cos(\theta)=0$ for isotropic random velocity and magnetic fields. For all three angles, as expected from isotropic random fields, all possible values of $|\cos(\theta)|$ have a non-zero probability. However, all angles are not equiprobable and there are clear differences in the distributions with the Mach number of the turbulent flow and the dynamo stage. We discuss these differences and their implications below.

For the subsonic turbulence, the vorticity is slightly more aligned with the velocity in the saturated stage as compared to the kinematic stage (the difference is not much in \Fig{fig:apdfs}~(a) but see \Fig{fig:apdfs}~(b)). This shows that as the magnetic field grows, the vorticity amplification term is probably suppressed, especially in the strong-field regions. This is a direct consequence of the back reaction of the magnetic field on the flow. Such a difference is not statistically significant for the supersonic turbulence. \rev{This is consistent with our previous results (\Fig{fig:spec}~(c) and \Fig{fig:wjpdfs}~(a)) that the effect of the back reaction on the velocity in the supersonic case is negligible.}

The angle between $\vec{u}$ and $\vec{b}$ controls the magnetic induction or amplification term and the alignment between $\vec{u}$ and $\vec{b}$ is statistically higher in the saturated stage as compared to the kinematic stage \rev{for subsonic turbulence (\Fig{fig:apdfs}~(c)) and the difference is not that significant for the supersonic turbulence (even for the strong-field regions, \Fig{fig:apdfs}~(d)). This mostly implies a reduction in induction due to enhanced alignment between $\vec{u}$ and $\vec{b}$ for the subsonic case but not so much for the supersonic case (because of locally strong shocks). Thus, for the subsonic turbulence,} the magnetic field evolves in such a way as to enhance the level of alignment between the velocity and magnetic field.

Finally, the back reaction of the magnetic field on the velocity field via the Lorentz force is controlled by the angle between $\vec{j}$ and $\vec{b}$, for which the alignment is enhanced in the saturated stage as compared to the kinematic stage (\Fig{fig:apdfs}~(e)). However, here, the difference is higher for the supersonic case as compared to the subsonic one. The effect is enhanced in the strong-field regions, i.e., the field is more aligned with the current density where the magnetic energy is higher than its rms value (\Fig{fig:apdfs}~(f)). \rev{In the saturated stage of the subsonic turbulence, the enhanced level of alignment between $\vec{j}$ and $\vec{b}$ would diminish the Lorentz force and the field advection by velocity would become dominant. This would give rise to a higher level of alignment between the velocity and magnetic field, which we see in \Fig{fig:apdfs}~(c). Such a difference is not seen in the case of supersonic turbulence because of the presence of strong shocks.}

\subsection{Characteristic magnetic scales}
\begin{figure*}
    \includegraphics[width=\columnwidth]{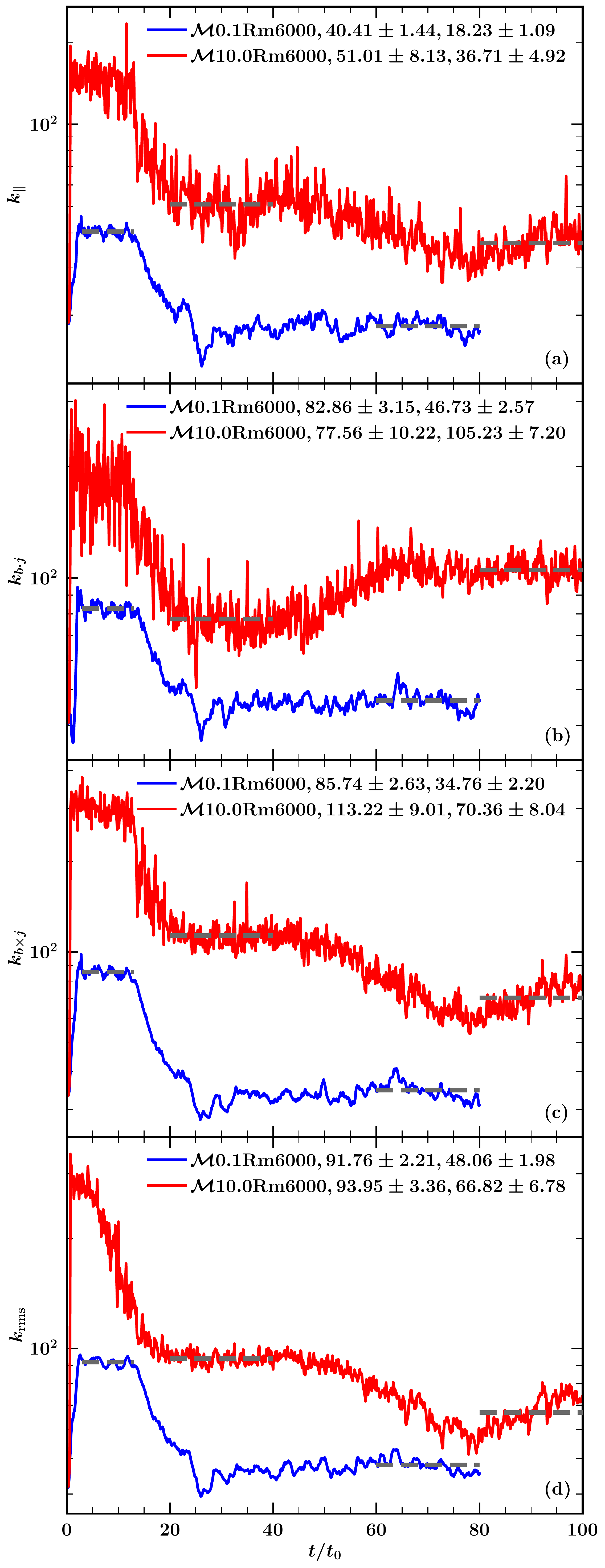}
    \caption{\rev{The evolution of characteristic magnetic length scales, $\kpar$ (a, \Eq{eq:kpar}, associated with the length scale for greatest  field line stretching), $\kbdj$ (b, \Eq{eq:kbdj}, associated with resistive dissipation length scale), $\kbcj$ (c, \Eq{eq:kbcj}, associated with the length scale for greatest field line compression), and $\krms$ (d, \Eq{eq:krms}, associated with overall magnetic field variation), as a function of the normalised time, $t/t_{0}$, for subsonic (blue, $\Mach0.1\Rm6000$) and supersonic (red, $\Mach10\Rm6000$) turbulence. The dashed, grey line shows average for each characteristic wavenumber in the kinematic and saturated stages (see \Fig{fig:ts} for corresponding times for each stage and Mach number) and the average value with one-sigma variation is given in the legend. All length scales, except the resistive dissipation length scale for supersonic turbulence, are enhanced as the magnetic field grows and saturates. $\kpar$ and $\kbcj$ are always smaller for the subsonic turbulence and $\kbdj$ and $\krms$ are comparable in the kinematic stage but are higher for the supersonic turbulence in the saturated stage.}}
    \label{fig:charlens} 
\end{figure*}
\rev{Besides the correlation length of the magnetic field (\Fig{fig:spec}~(d)), the following characteristic magnetic length scales (or equivalently wavenumbers) can be used to further study the local magnetic field structure \citep{ZeldovichEA1984, SchekochihinEA2004},
\begin{align}
\kpar & = \left(\frac{\langle | \vec{b} \cdot \nabla \vec{b} |^{2} \rangle}{\langle b^{4} \rangle} \right)^{1/2}, \label{eq:kpar} \\
\kbdj & = \left(\frac{\langle | \vec{b} \cdot \vec{j} |^{2} \rangle}{\langle b^{4} \rangle} \right)^{1/2}, \label{eq:kbdj} \\
\kbcj & = \left(\frac{\langle | \vec{b} \times \vec{j} |^{2} \rangle}{\langle b^{4} \rangle} \right)^{1/2}, \label{eq:kbcj} \\
\krms & = \left(\frac{\langle |\nabla \vec{b}|^{2}\rangle}{\langle b^{2} \rangle}\right)^{1/2}. \label{eq:krms}
\end{align}
These wavenumbers (or length scales) can be related to the physical effects and structure of magnetic fields. $\kpar$ is a probe of variation of the magnetic field along itself and is related to the length scale associated with greatest stretching of magnetic field lines  (which maximises magnetic field amplification). $\kbcj$ is a probe of the magnetic field across itself and is related to resistive dissipation. $\kbdj$ is a probe of magnetic field variations along a direction perpendicular to both magnetic field ($\vec{b}$) and Lorentz force ($\vec{j} \times \vec{b}$) and is related to highest compressive motions (maximising the effect of compression on magnetic field). $\krms$ measures the overall variation of magnetic fields. For a subsonic fluctuation dynamo with $\Pm \gg 1$, it is shown that the magnetic field organises itself into folded structures \citep{SchekochihinEA2004, St-OngeEA2020}. Furthermore, the structures are folded sheets in the kinematic stage (identified by the condition, $\kpar \lesssim \kbdj \ll \kbcj \sim \krms$) and folded ribbons in the saturated stage (identified by the condition, $\kpar \ll \kbdj \lesssim \kbcj \sim \krms$). Since our simulations are with $\Pm \gtrsim 1$ (highest $\Pm$ is $3$, for $\Rm=6000$ and $\Re=2000$ runs), we do not necessarily expect to see such folded structures. However, we compute these characteristic scales (\Eq{eq:kpar} -- \Eq{eq:krms}) to further study the dynamics of growing and saturated magnetic fields in subsonic and supersonic turbulence.}

\rev{\Fig{fig:charlens} shows the temporal evolution of wavenumbers, $\kpar$ (a), $\kbdj$ (b), $\kbcj$ (c), and $\krms$ (d), for subsonic ($\Mach0.1\Rm6000$) and supersonic ($\Mach10\Rm6000$) turbulent flows (both at $\Pm=3$). Furthermore, the average values of these wavenumbers with one-sigma fluctuations in the kinematic and saturated stages are provided in the legend. For subsonic turbulence, in the kinematic stage, $\kpar < \kbdj \lesssim \kbcj \lesssim \krms$, and thus magnetic structures can probably be characterised as folded ribbons, but structures in the saturated stage are neither folded ribbons nor folded sheets. All four wavenumbers decrease as the magnetic field saturates for the subsonic case.}

\rev{For the supersonic turbulence, based on these wavenumbers, the structures are never folded sheets or folded ribbons. All wavenumbers except $\kbdj$ decrease as the field grows and saturates. This shows that the dissipation takes place at a smaller length scale in the saturated stage as compared to the kinematic stage for the supersonic turbulence. However, the overall magnetic wavenumber, $\krms$, still increases. The length scales associated with both the greatest stretching and compression also grows as the magnetic field saturates. On comparing all four wavenumbers between subsonic and supersonic turbulence, $\kpar$ and $\kbcj$ are always smaller for the subsonic flow, but $\kbdj$ and $\krms$ are comparable in the kinematic stage and are higher for the supersonic case in the saturated stage. This shows that the length scale associated with greatest stretching and compression grow as the field saturates for both the subsonic and supersonic cases. However, as the magnetic field grows and saturates, the resistive dissipation scale increases for the subsonic case, but decreases for the supersonic case. The overall magnetic field scale grows as the magnetic field saturates for both Mach numbers (same as the magnetic correlation length in \Fig{fig:spec}~(d)).}


\subsection{Local stretching and compression of magnetic field lines, and magnetic field diffusion}
\begin{figure*}
    \includegraphics[width=\columnwidth]{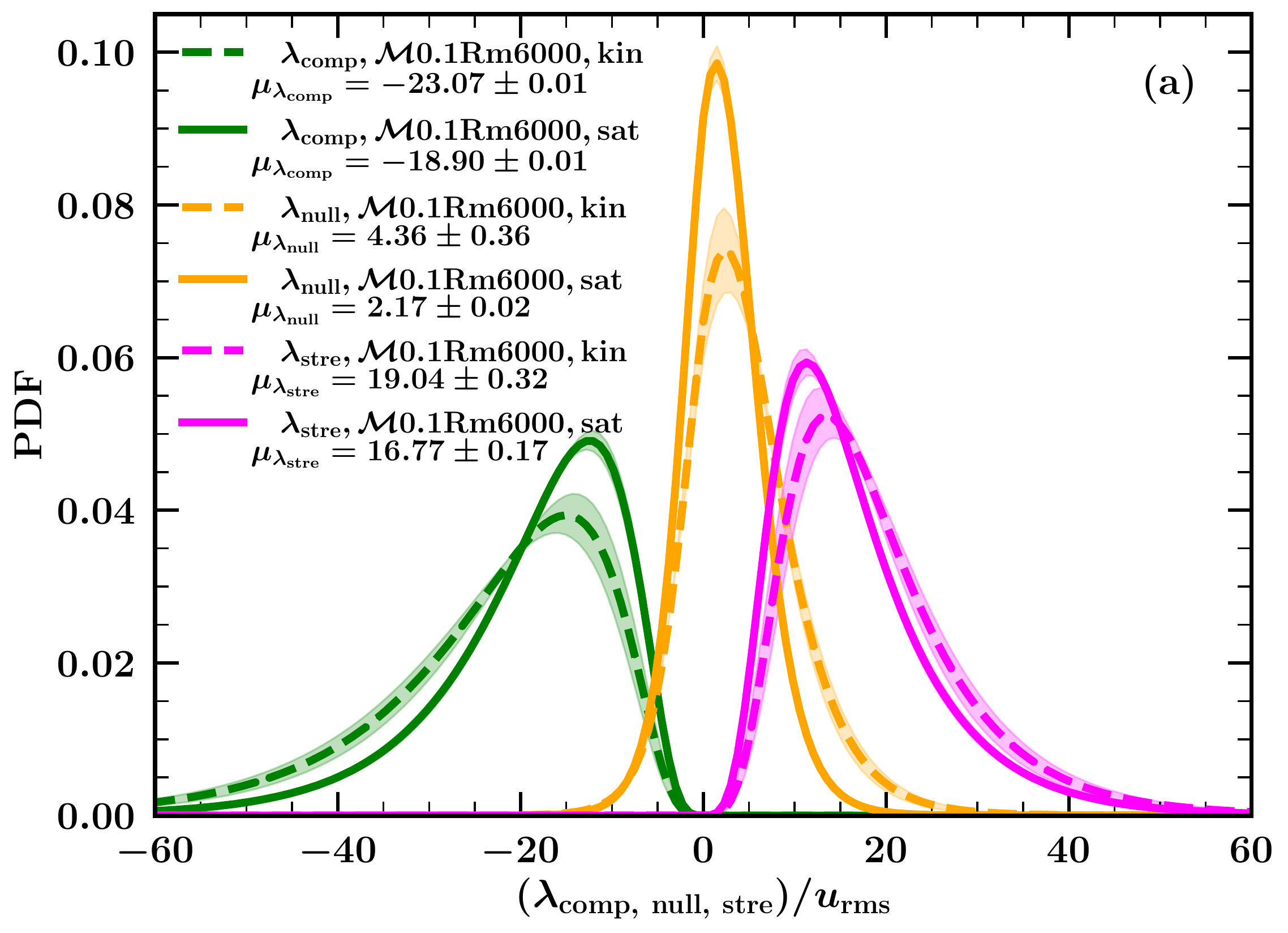}
    \includegraphics[width=\columnwidth]{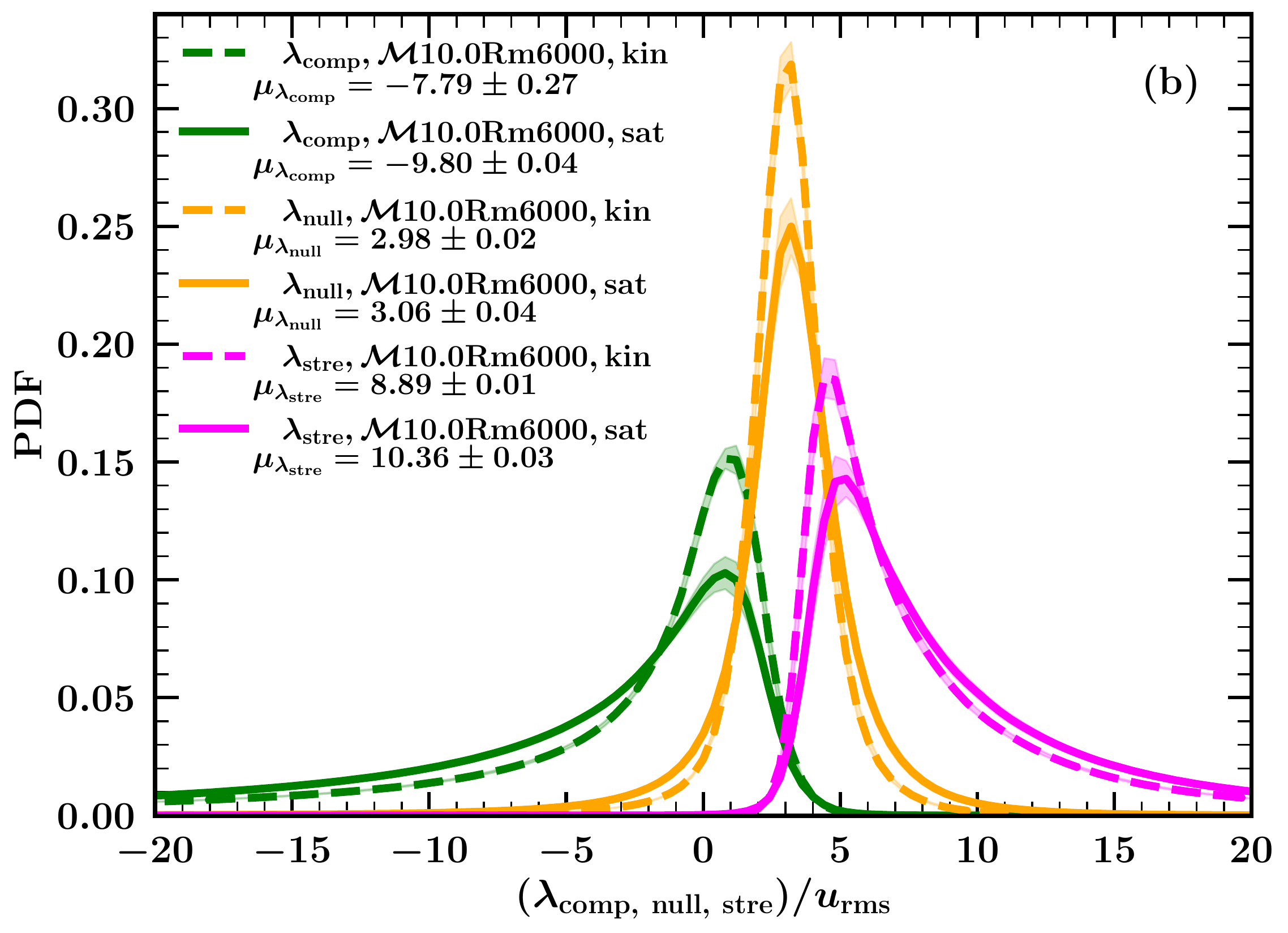}
    \caption{PDFs of the eigenvalues of the strain tensor, $\lambdacomp$ (green), $\lambdanull$ (orange), and $\lambdastre$ (magenta) normalised by $\urms$ for subsonic (a) and supersonic (b) turbulence in the kinematic (kin, dashed) and saturated (sat, solid) stages. The solid lines show the mean over ten eddy turnover times and the shaded region of the same colour shows the one-sigma variations. The legend shows the mean value (denoted by $\mu$) of the distribution for each case, where the value and the related error are obtained by time averaging. For the subsonic turbulence, $\lambdacomp$ is always less than zero (strength of local magnetic field line compression) and $\lambdastre$ is always greater than zero  (strength of local magnetic field line stretching). \rev{ This is not the case for supersonic turbulence. All three eigenvalues decreases (in a statistical sense, as also reflected by the absolute value of their mean) as the magnetic field saturates for the subsonic case but increases for the supersonic turbulence.}}
    \label{fig:evapdfs} 
\end{figure*}
We now study the local stretching and compression of magnetic field lines by the turbulent velocity using the eigenvalues and eigenvectors of the strain tensor. Such an analysis is done for isotropic subsonic turbulence \citep{ZeldovichEA1984, SetaEA2020, St-OngeEA2020}, with a different motivation for slightly supersonic turbulence \citep[$\Mach\sim2$ in][]{SurPS2014}, and for different setups such as magnetic fields in rotating convection simulations \citep{FavierB2012} and decaying magnetic field simulations \citep{BrandenburgKT2015}. Here, we aim to understand the effect of growing magnetic fields and compressibility on the local stretching and compression of magnetic field lines by analysing the strain tensor. First, we compute the strain tensor, $S_{ij} = (1/2) \, (u_{i,j} + u_{j,i})$, at each point in the domain using a sixth-order finite difference scheme and then calculate its eigenvalues ($\lambda_i$) and eigenvectors ($\vec{e}_i$). \rev{We then arrange the eigenvalues in the order \rev{$\lambda_1 > \lambda_2 > \lambda_3$} and let the corresponding eigenvectors be $\vec{e}_{1}, \vec{e}_{2},$ and $\vec{e}_{3}$. For an incompressible flow, the sum of eigenvalues, \rev{$\lambda_1 + \lambda_2 + \lambda_3 = 0$} (approximately applicable for our subsonic case) and $\lambda_1 > 0 $  and $\lambda_3 < 0$ \citep{Kerr1985,AshurstEA1987}.} This need not be the case for our supersonic runs. \rev{In general, $\vec{e}_{1}$ (positive $\lambda_1$) corresponds to the direction of local magnetic field line stretching, $\vec{e}_{3}$ (negative $\lambda_3$) corresponds to the direction of local magnetic field line compression, and $\vec{e}_{2}$ (also referred to as the `null' direction) can be either, depending on the sign of $\lambda_2$ \citep{ZeldovichEA1984, SchekochihinEA2004} (especially, see Fig.~10 in \cite{SchekochihinEA2004}). The magnitude of $\lambda_1$ (when it is positive) and $\lambda_3$ (when it is negative) can be considered as the strength of local magnetic field line stretching and compression, respectively. For clarity, throughout the rest of the paper, we refer $\lambda_1, \lambda_2,$ and $\lambda_3$ as $\lambdastre, \lambdanull$, and $\lambdacomp$ and the corresponding eigenvectors as $\estre, \enull$, and $\ecomp$.}

\Fig{fig:evapdfs} show the PDF of all three eigenvalues normalised to the rms velocity for subsonic ($\Mach0.1\Rm6000$, a) and supersonic ($\Mach10.0\Rm6000$, b) turbulence in kinematic and saturated stages. As expected, for the subsonic case, it can be seen that $\lambdacomp$ is always negative and $\lambdastre$ is always positive. This is not the case for the supersonic run, i.e., $\lambdacomp$ can be positive for $\Mach10\Rm6000$ (though the probability of this happening is low, see \Fig{fig:evapdfs}~(b)). When $\lambdanull\approx0$, the magnetic field is unaffected along the $\enull$ direction. The magnitude of $\lambdanull$ is always smaller than the other two eigenvalues. \rev{On average, $\lambdastre \sim -\lambdacomp \sim (3~\text{--}~5) \lambdanull$ and this is consistent with previous results for isotropic hydrodynamic turbulence \citep{AshurstEA1987}. The absolute value of all three eigenvalues is smaller for the supersonic case (compare $x$-axis in \Fig{fig:evapdfs}~(a) and \Fig{fig:evapdfs}~(b)). Thus the efficiency of local stretching and compression of magnetic field lines, which leads to magnetic field amplification, is lower for supersonic turbulence. This is also probably the reason for the smaller ratio of saturated magnetic to kinetic energy for compressible runs (see \Fig{fig:gammasat}~(b)). As the magnetic field saturates, the absolute value of all three eigenvalues statistically decreases for the subsonic case, but increases for the supersonic case (as shown by the mean value in the legend of \Fig{fig:evapdfs}). Thus, the local field line stretching and compression decreases for the subsonic case in the saturated stage, but increases in the supersonic case. This is probably to counter high enhancement in diffusion compared to amplification for supersonic turbulence (see \Fig{fig:apdfs} and \Tab{tab:lgd}).} 

\begin{figure*}
    \includegraphics[width=\columnwidth]{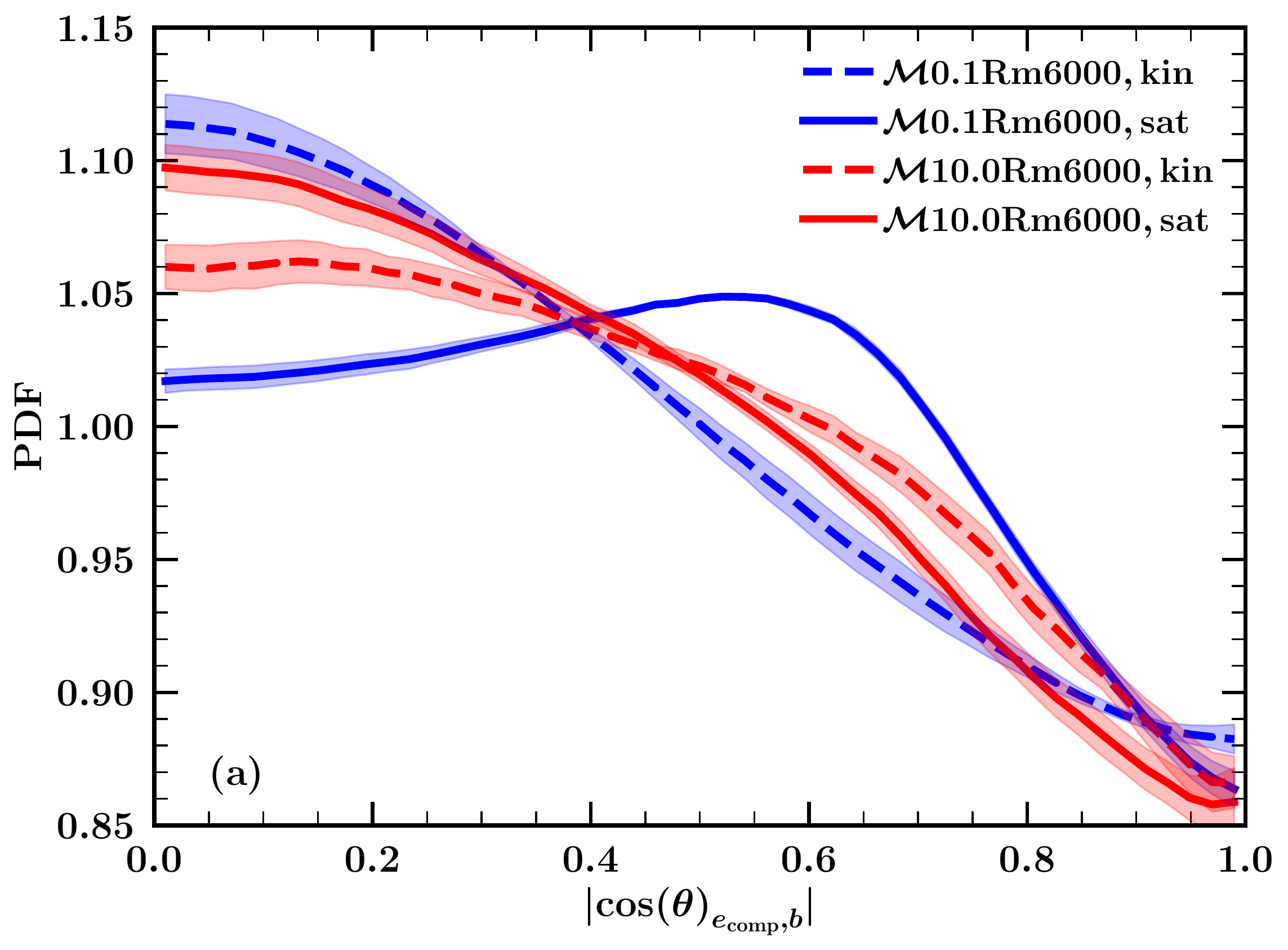}
    \includegraphics[width=\columnwidth]{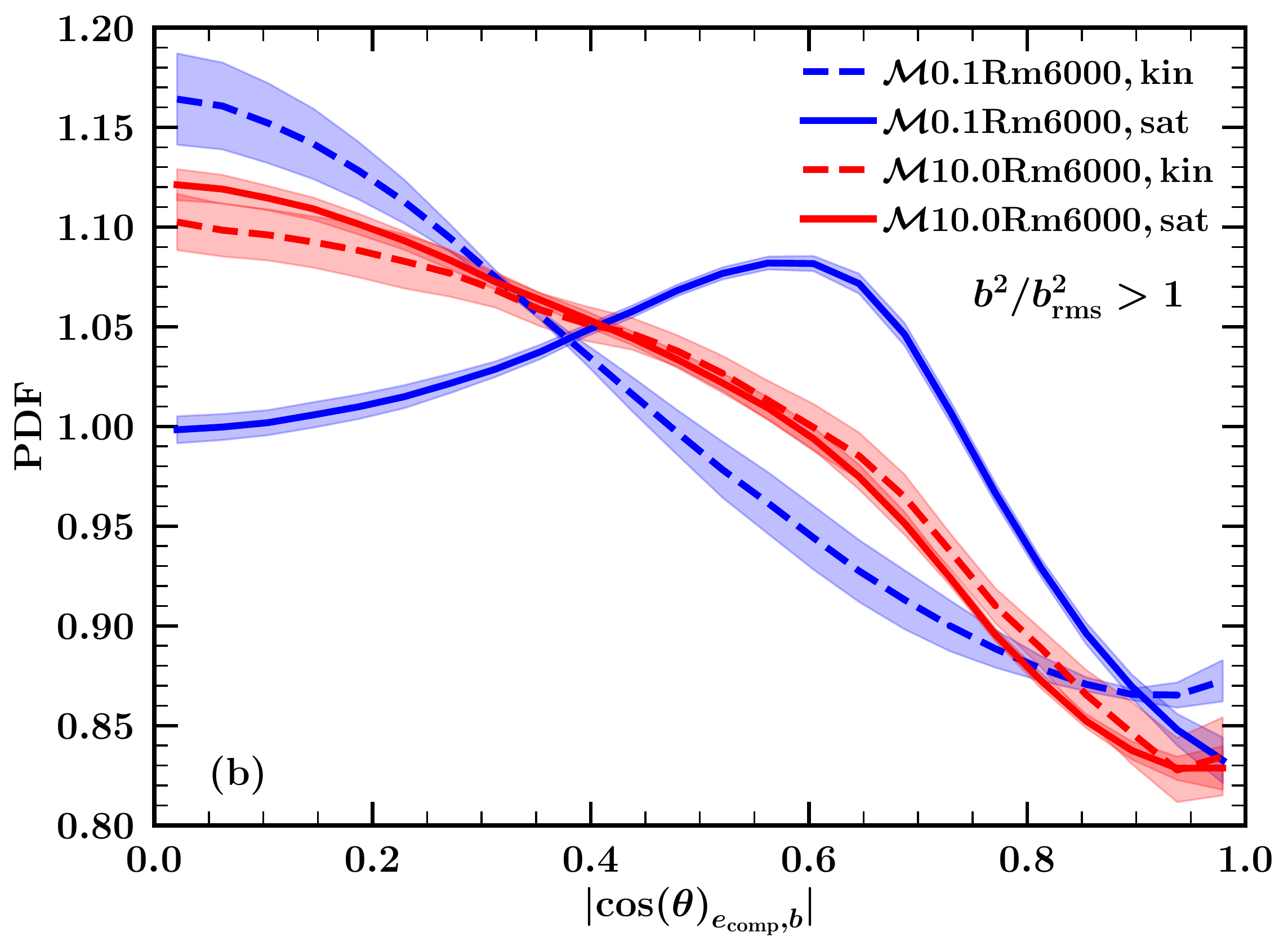}
    \includegraphics[width=\columnwidth]{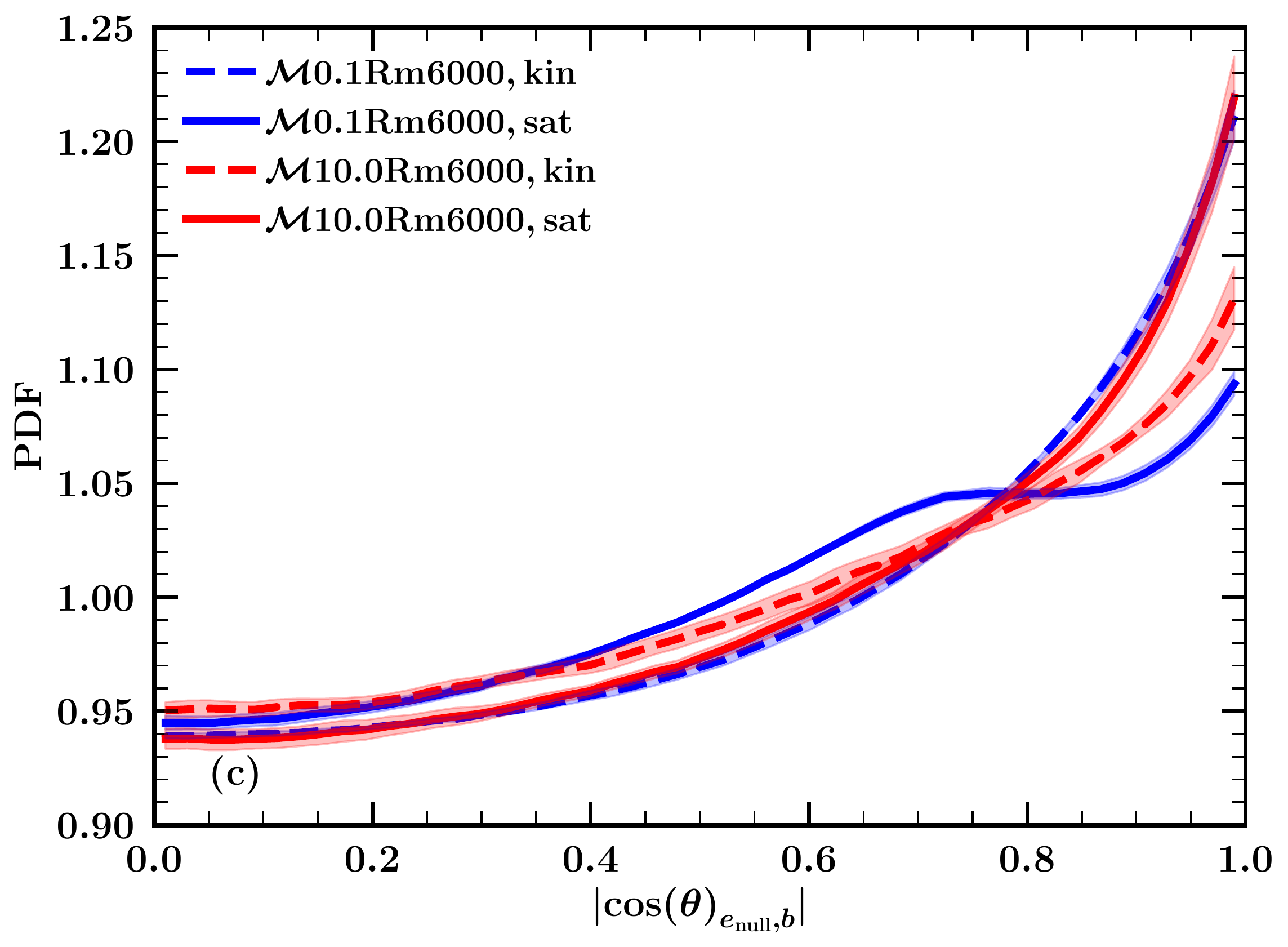}
    \includegraphics[width=\columnwidth]{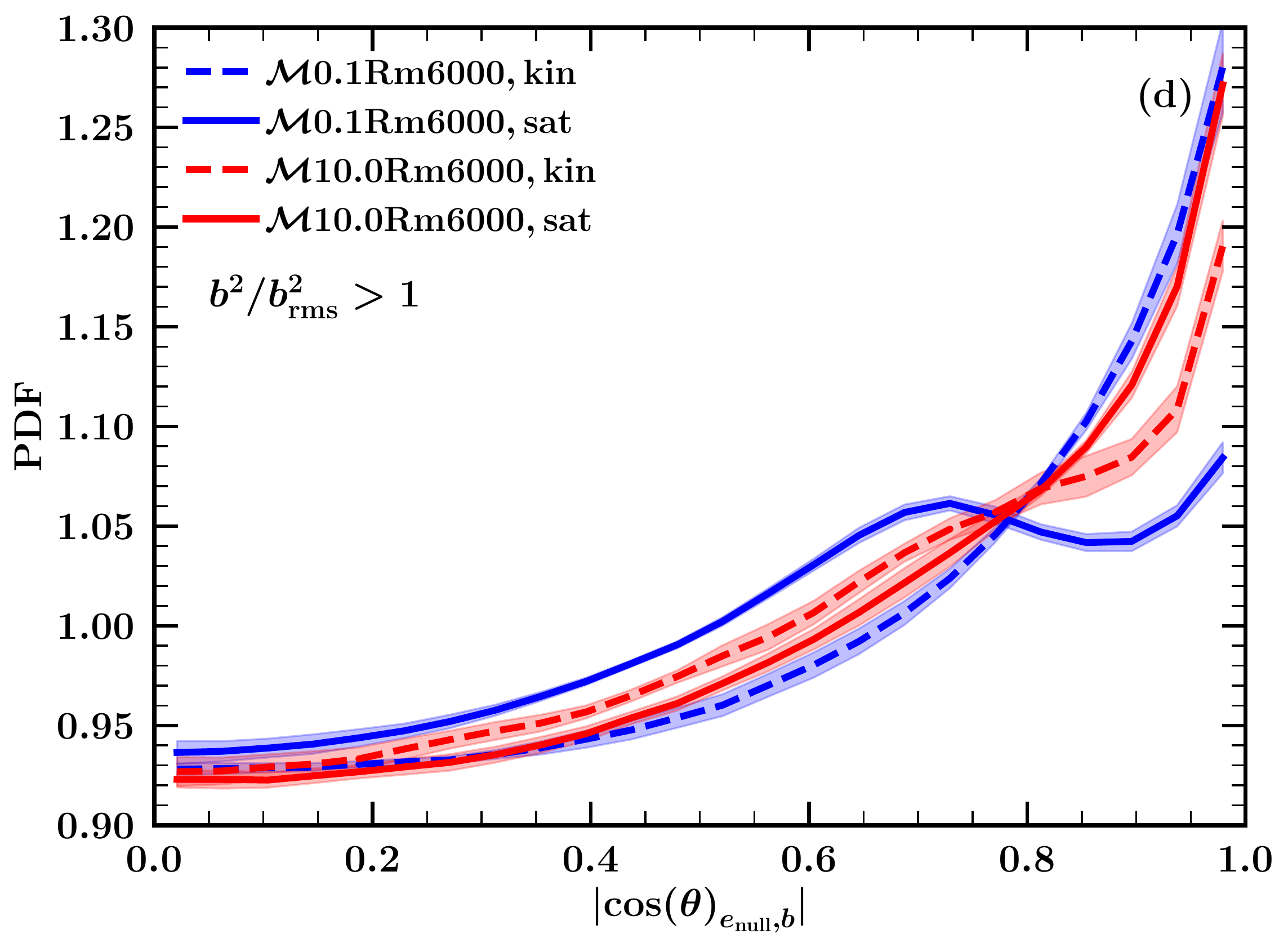}
    \includegraphics[width=\columnwidth]{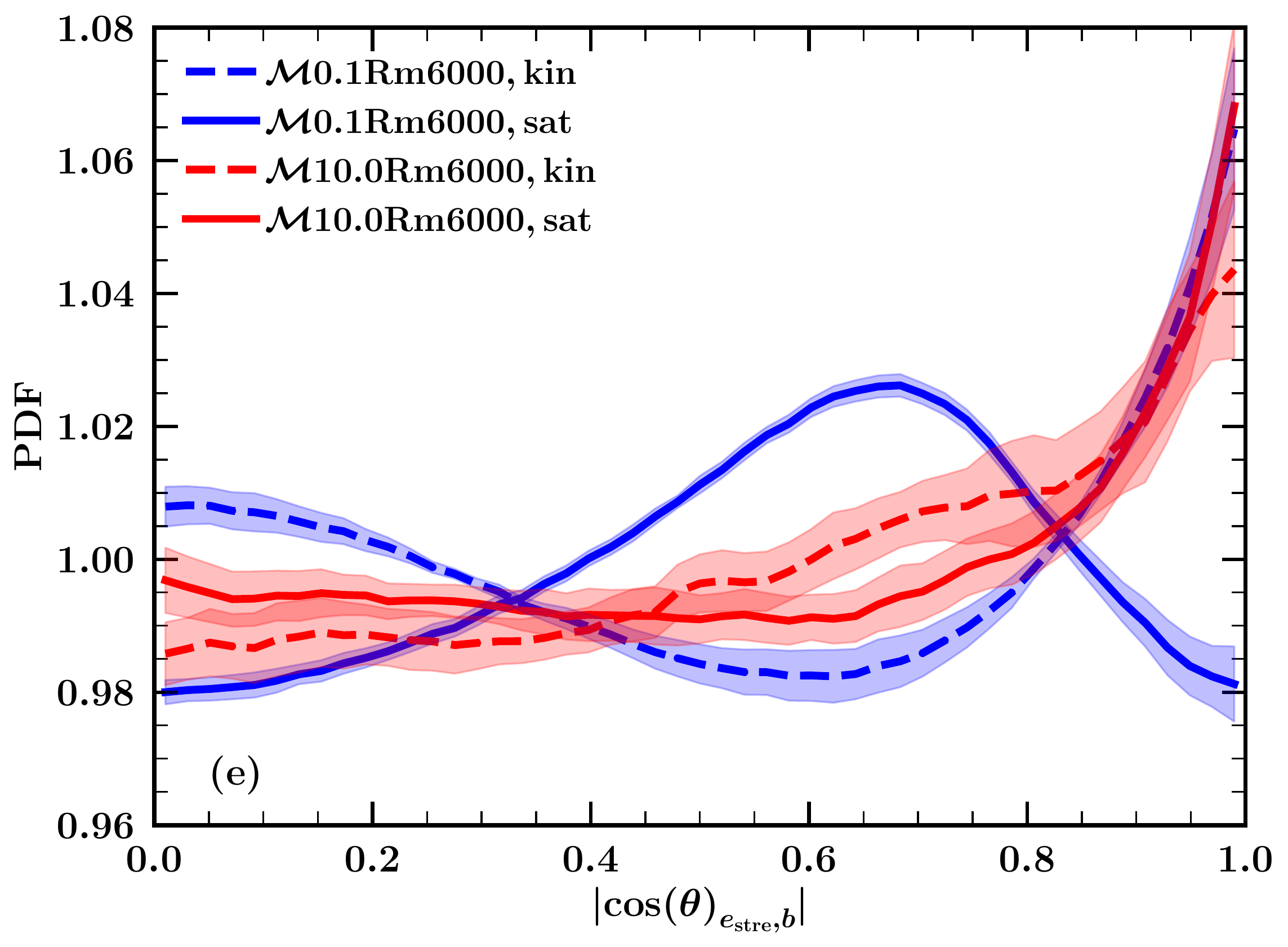}
    \includegraphics[width=\columnwidth]{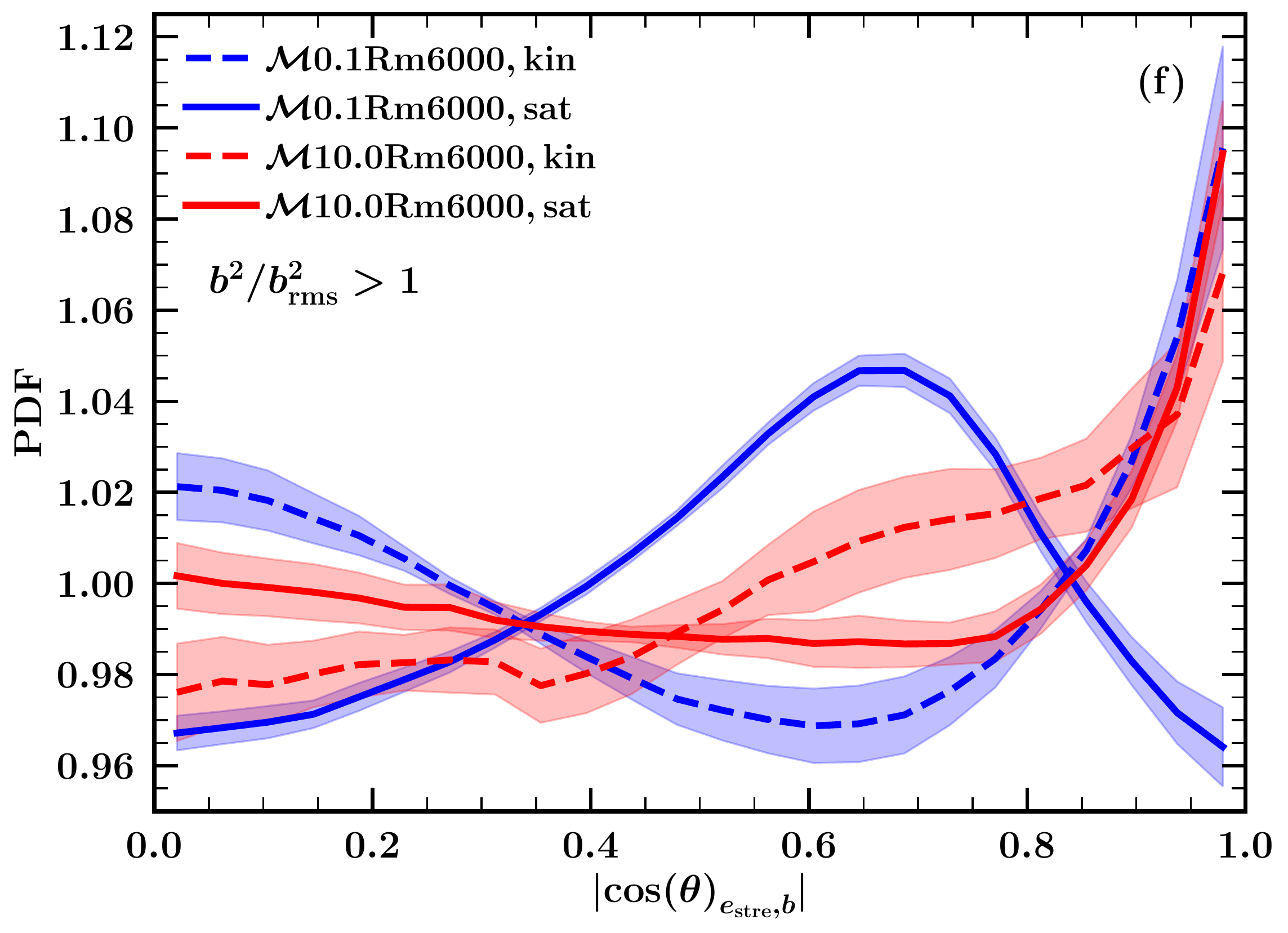}
      \caption{Total and conditional (regions with $b^2/\brms^2 > 1$) PDFs of the angle between the three eigenvectors of the rate of strain tensor ($\ecomp, \enull,$ and $\estre$) and magnetic felds ($\vec{b}$) for subsonic ($\Mach0.1\Rm6000$, blue) and supersonic ($\Mach10.0\Rm6000$, red) turbulence in kinematic (kin, dashed) and saturated (sat, solid) stages. The effect of field line compression is enhanced if the field line compression direction is perpendicular to the local magnetic field, i.e., $ |\cosecompb | \approx 0$. Magnetic field amplification is enhanced when the field line stretching direction is aligned with the local magnetic field, i.e., $|\cosestreb| \approx 1$. This is seen in the kinematic stage for the subsonic flows. As the field saturates in the subsonic case, $|\cos(\theta)|$ for both those cases moves to the range $0.6~\text{--}~0.7$, and thus the field amplification and the effect of local compression is reduced. \rev{Such a significant difference between the kinematic and saturated stages is not seen for the supersonic case. For the supersonic case, the direction of local magnetic field line compression is more aligned with the magnetic field in the saturated stage as compared to the kinematic stage (though not in the strong-field regions). Whereas, in strong-field regions only, there is a slight decrease in the level of alignment between $\estre$ and $\vec{b}$ in the saturated stage of the supersonic run.}}
       \label{fig:aevpdfs} 
 \end{figure*}

\rev{The amplification of magnetic fields due to field line stretching will be maximal when the field line stretching direction ($\estre$) is aligned with the magnetic field. Similarly, the effect of field line compression is maximal when the field line compression direction ($\ecomp$) is perpendicular to the local magnetic field.} \Fig{fig:aevpdfs} show the total (\Fig{fig:aevpdfs}~(a, c, e)) and conditional (based only on the strong-field regions, $b^2/\brms^2 > 1$, \Fig{fig:aevpdfs}~(b, d, f)) PDF of the angle between all three eigenvectors ($\ecomp, \enull,$ and $\estre$) and magnetic fields ($\vec{b}$) in the kinematic and saturated stages for subsonic and supersonic turbulent flows. For the supersonic case (see \Fig{fig:evapdfs}~(b)), we only consider those regions where $\lambdacomp < 0$ (implying local magnetic field line compression) and $\lambdastre > 0$ (implying local magnetic field line stretching).  All the possible angles between these three eigenvectors and magnetic fields are not equiprobable. 

In the kinematic stage, for the subsonic turbulence, the highest probable angle between the field line compression direction and magnetic field is $90 \degree \text{or}~270 \degree (| \cosecompb | \approx 0)$ and that between the field line stretching direction and magnetic field is $0 \degree \text{or}~180 \degree \left(| \cosestreb | \approx 1 \right)$. Thus, the magnetic field is more aligned with the field line stretching direction (and also the null direction) and perpendicular to the direction of the field line compression direction. This maximises magnetic field amplification. \rev{This also suggest that the magnetic field statistically lies more in the $\estre~\text{--}~\enull$ plane.}  As the field saturates in subsonic turbulence, the more probable angles for both $\ecomp$ and $\estre$ lie in the range of $|\cos(\theta)| = 0.6~\text{--}~0.7$ ($\enull$ is still more aligned with $\vec{b}$, but the level of alignment has decreased). Thus, the amplification due to field line stretching and the effect of local compression is reduced in the saturated stage as compared to the kinematic stage \rev{via changes in these alignments}. Moreover, these differences between the kinematic and saturated stages are enhanced in the strong-field regions (\Fig{fig:aevpdfs}~(b, d, f)). 

\rev{These trends are different for supersonic turbulence. In the kinematic stage, the magnetic field is overall more aligned with the $\estre$ and $\enull$ and also orthogonal to $\ecomp$ as in the subsonic case. Thus, here too, the magnetic field statistically lies in the $\estre~\text{--}~\enull$ plane. However, as the field saturates, the difference in the distribution of these angles is not as significant as in the subsonic case. In fact, $\ecomp$ is more aligned with $\vec{b}$ in the saturated stage (although not in the strong-field regions; see \Fig{fig:aevpdfs}~(b)), enhancing the effect of local compression regions with $b/\brms \le 1$. On the other hand, in strong-field regions (\Fig{fig:aevpdfs}~(f)), the orthogonality between $\estre$ and $\vec{b}$ is slightly enhanced in the saturated stage, but the overall distribution of $| \cosestreb |$ (\Fig{fig:aevpdfs}~(e)) is roughly the same for both the kinematic and saturated stages.}

\begin{figure*}
    \includegraphics[width=\columnwidth]{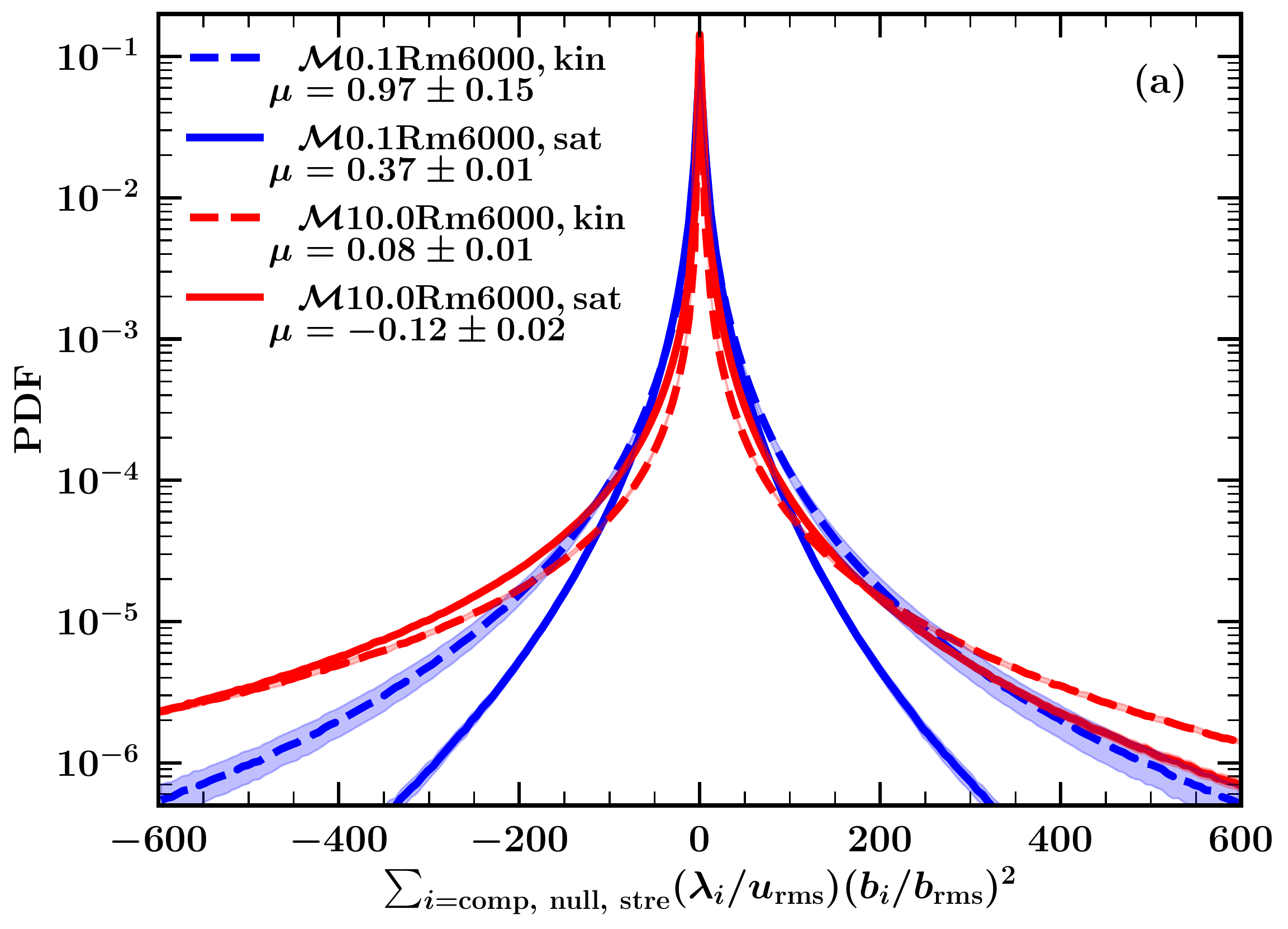}
    \includegraphics[width=\columnwidth]{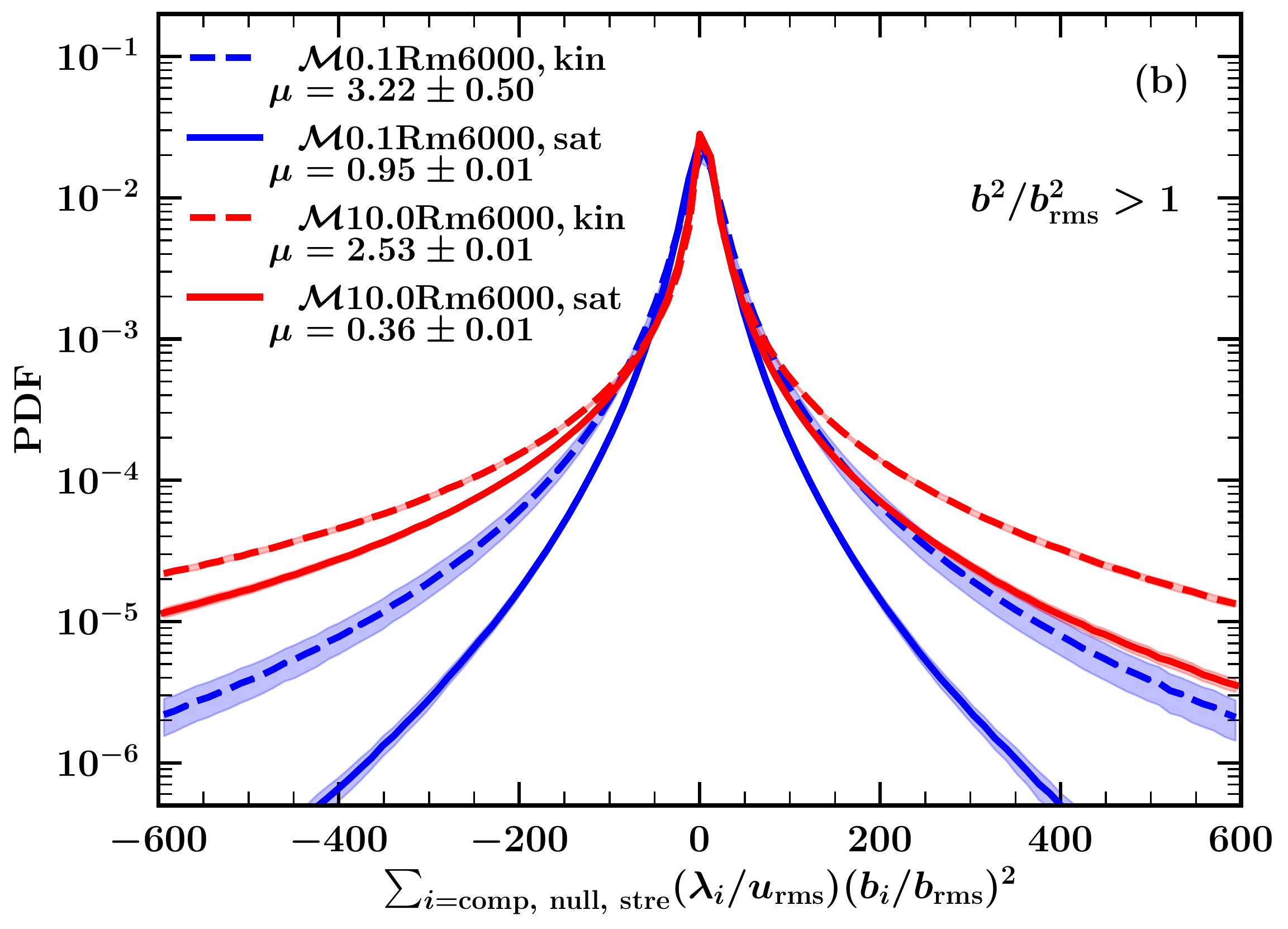}
    \includegraphics[width=\columnwidth]{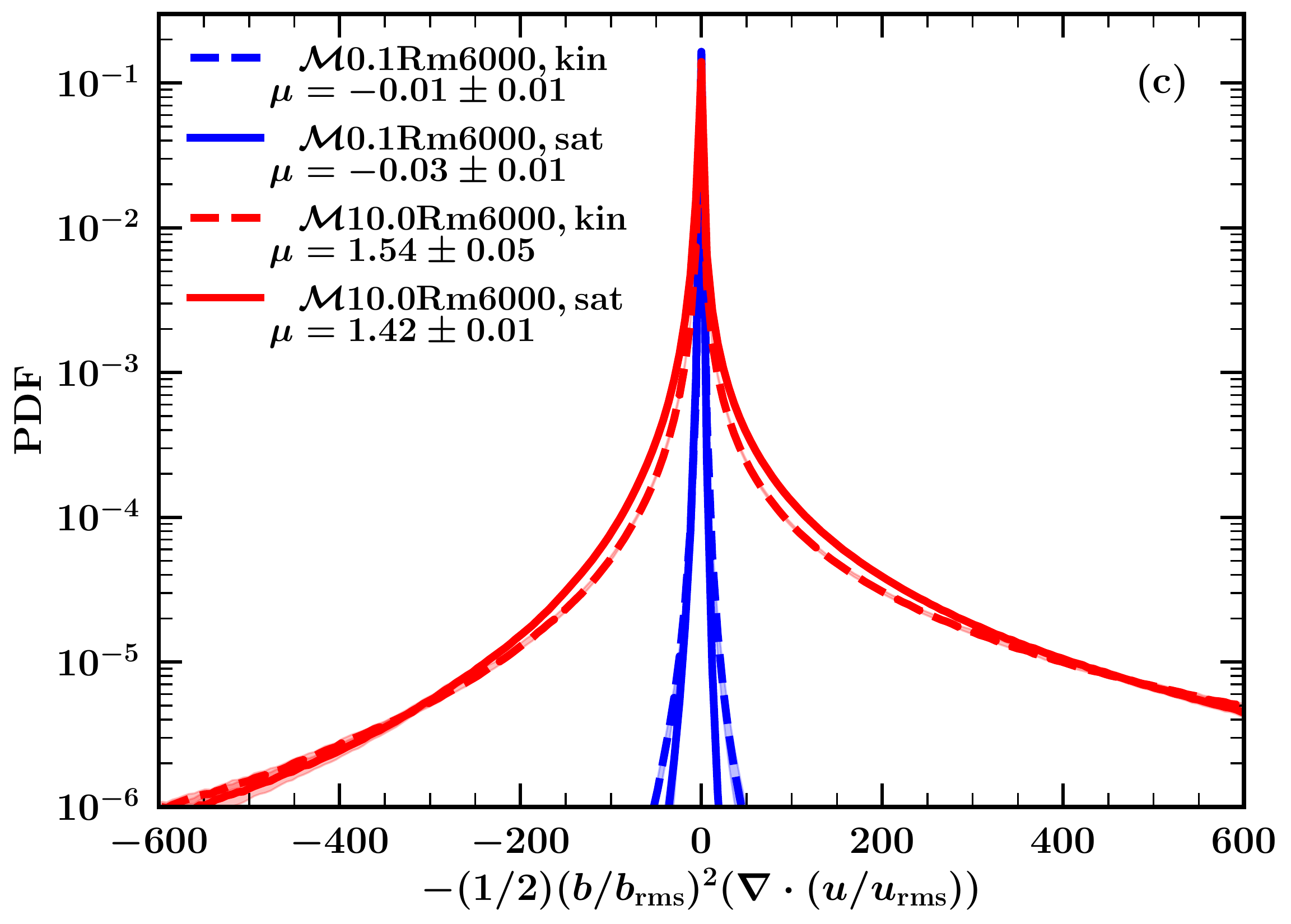}
    \includegraphics[width=\columnwidth]{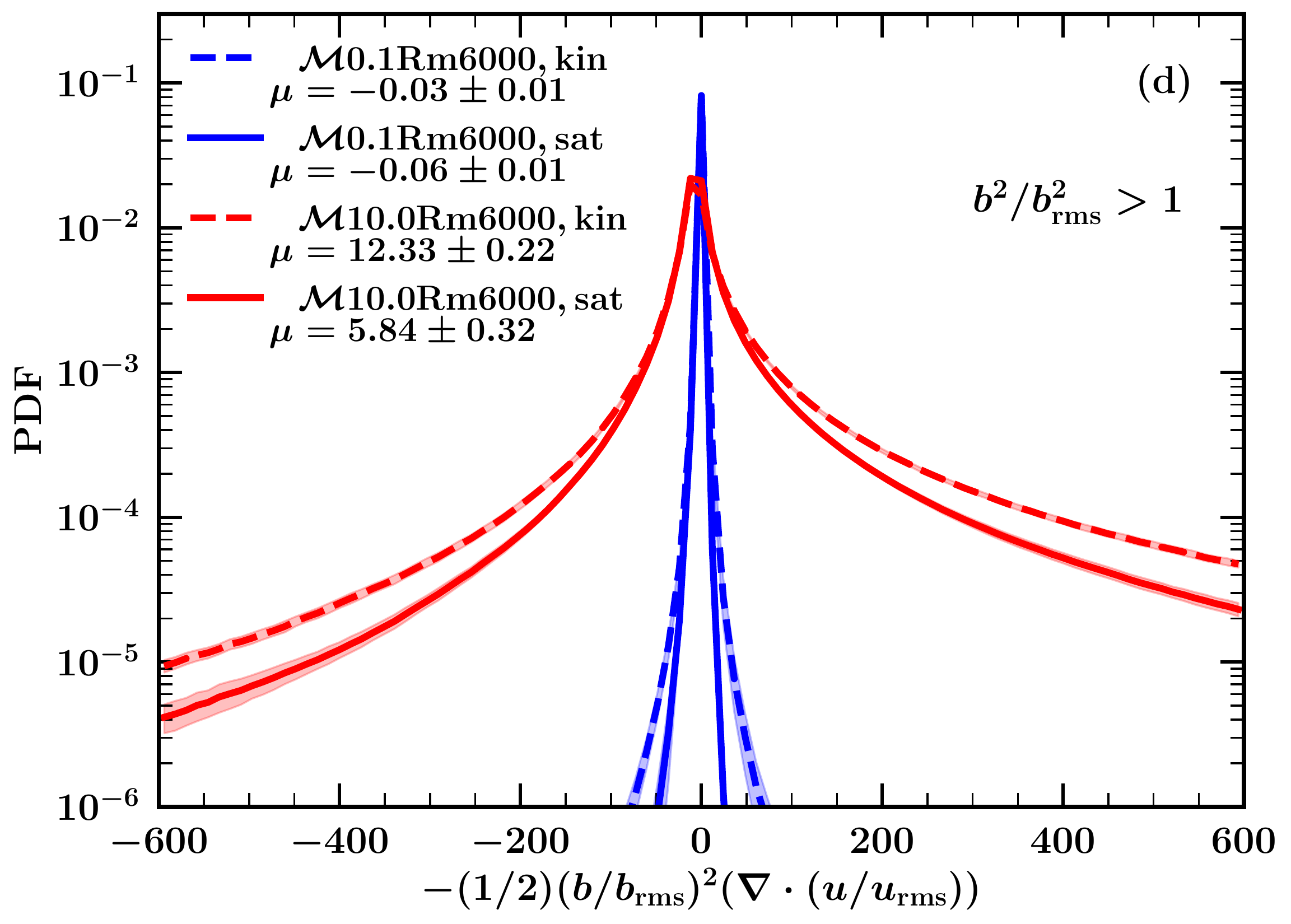}
    \includegraphics[width=\columnwidth]{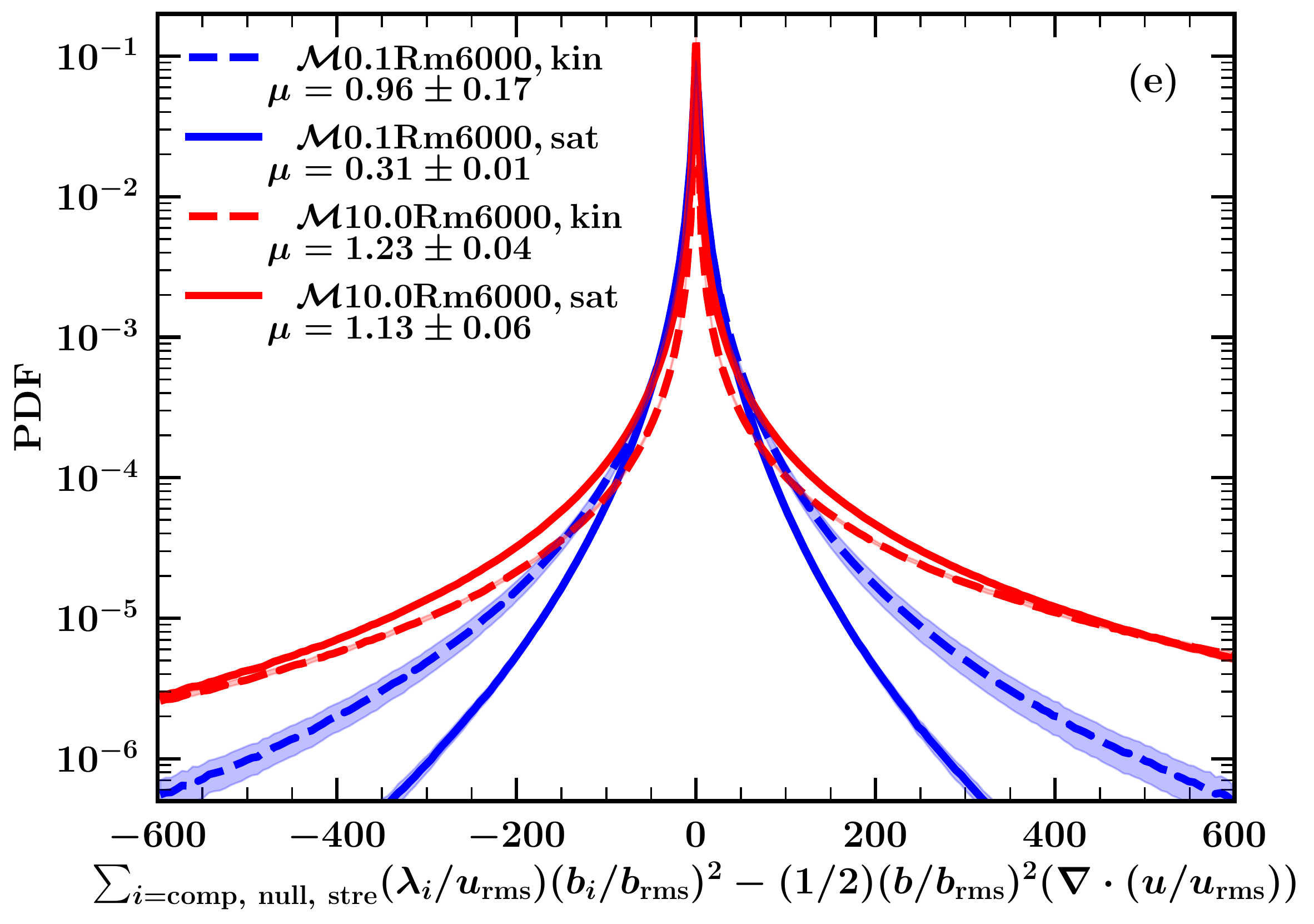}
    \includegraphics[width=\columnwidth]{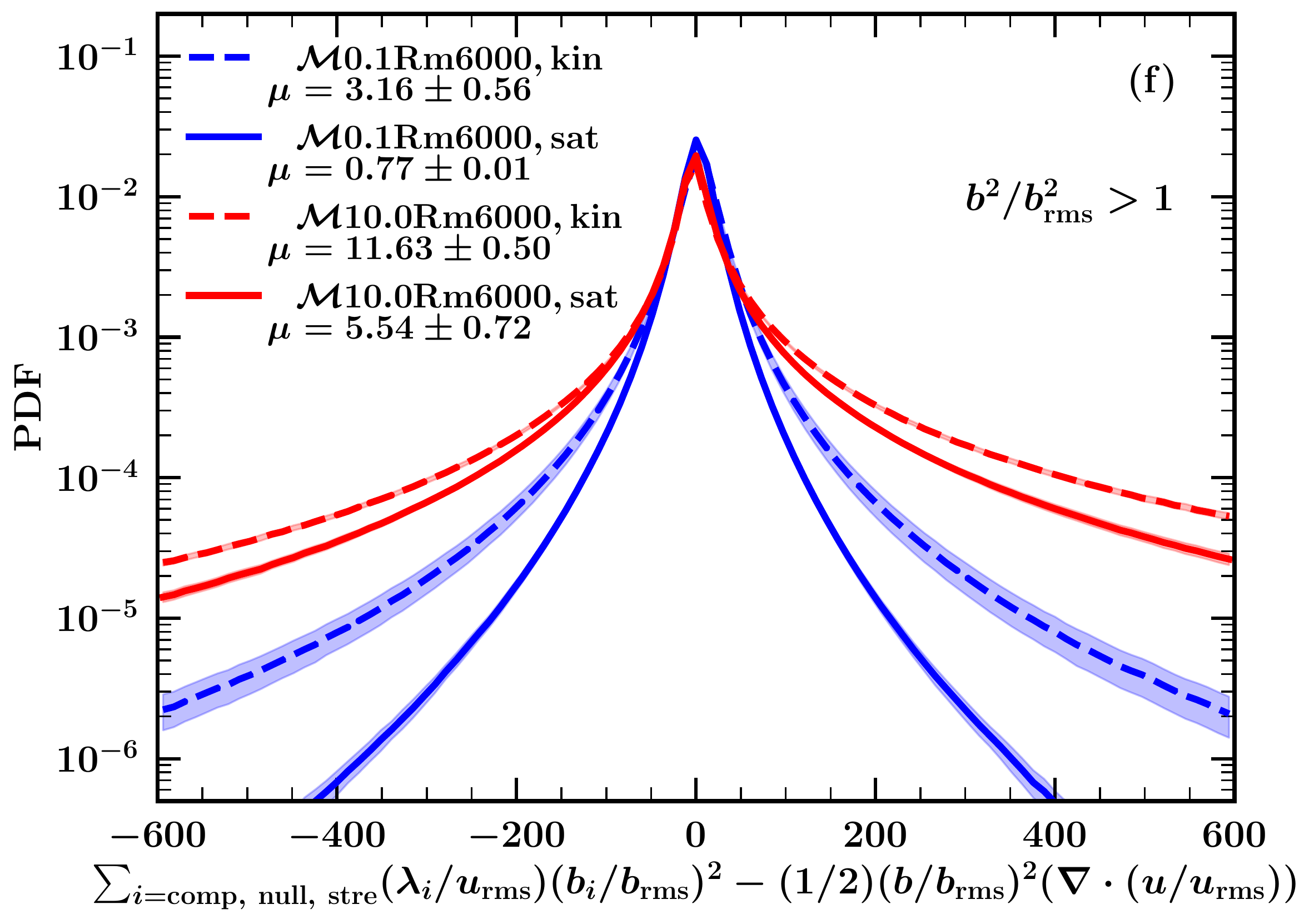}
      \caption{Total and conditional (regions with $b^2/\brms^2 > 1$) PDFs of the first $(\sum_{i=\rm comp, null, stre} (\lambda_i/\urms)(\vec{b}_i/\brms)^2)$, second $(-(1/2)(b/\brms)^2 (\nabla \cdot (\vec{u}/\urms)))$, and the combination of the first two terms in the right-hand side of the \Eq{eq:magen} for subsonic and supersonic cases ($\mu$ in the legend shows the mean of the distribution for each case). The mean of combination of both terms (e, f) is always positive. This implies that, on an average, the first two terms leads to local growth of the magnetic energy.  \rev{On saturation, the first term and first two terms combined statistically decreases for the subsonic case (as expected the second term has negligible effect). This shows reduction in amplification and this reduction is enhanced in strong-field regions. For the supersonic case, the local growth term (e) is not statistically different between the kinematic and saturated stages but is lower in the saturated stage for strong-field regions (f).}}
       \label{fig:lgpdfs} 
 \end{figure*}
\begin{figure*}
    \includegraphics[width=\columnwidth]{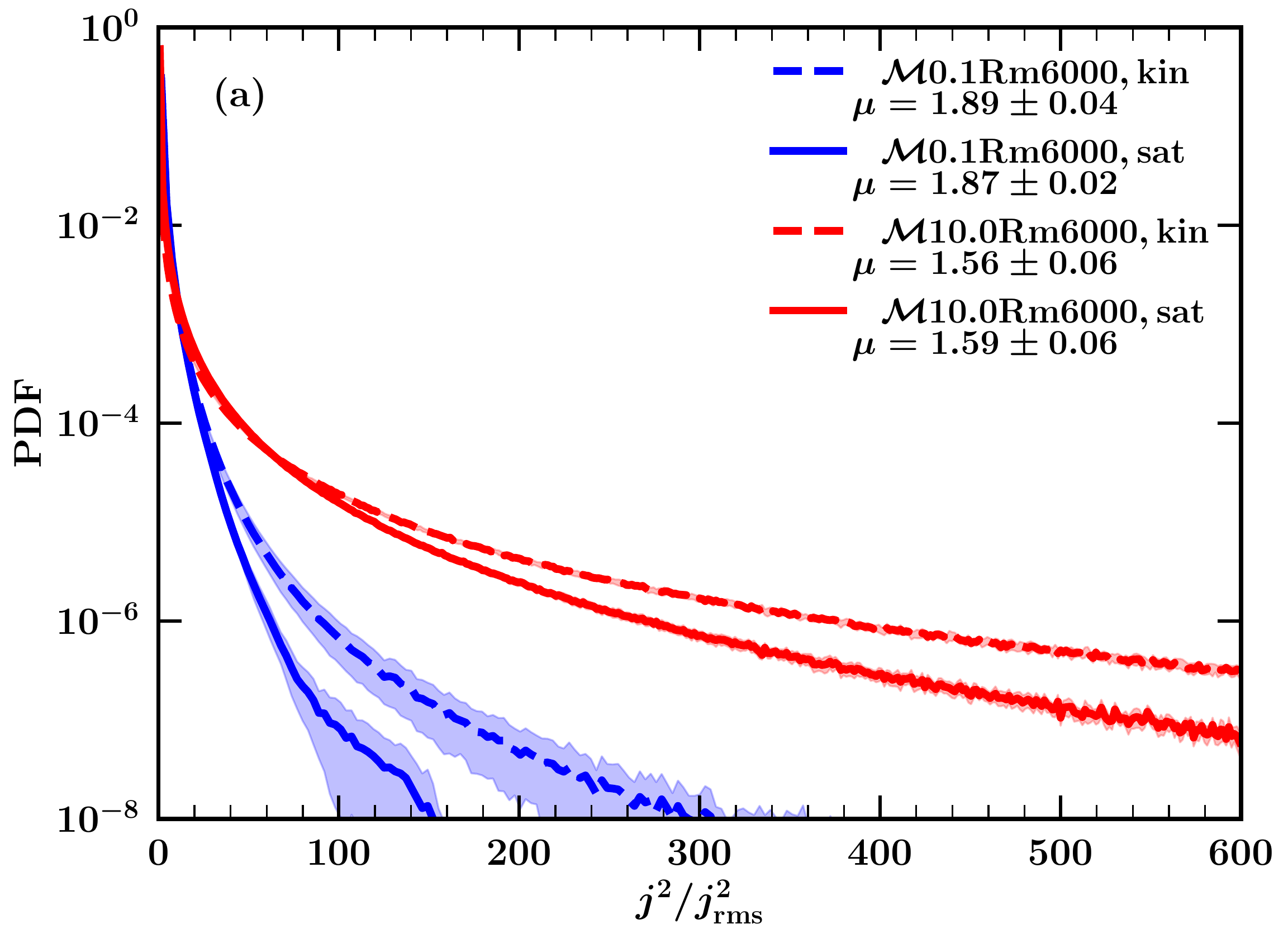}
    \includegraphics[width=\columnwidth]{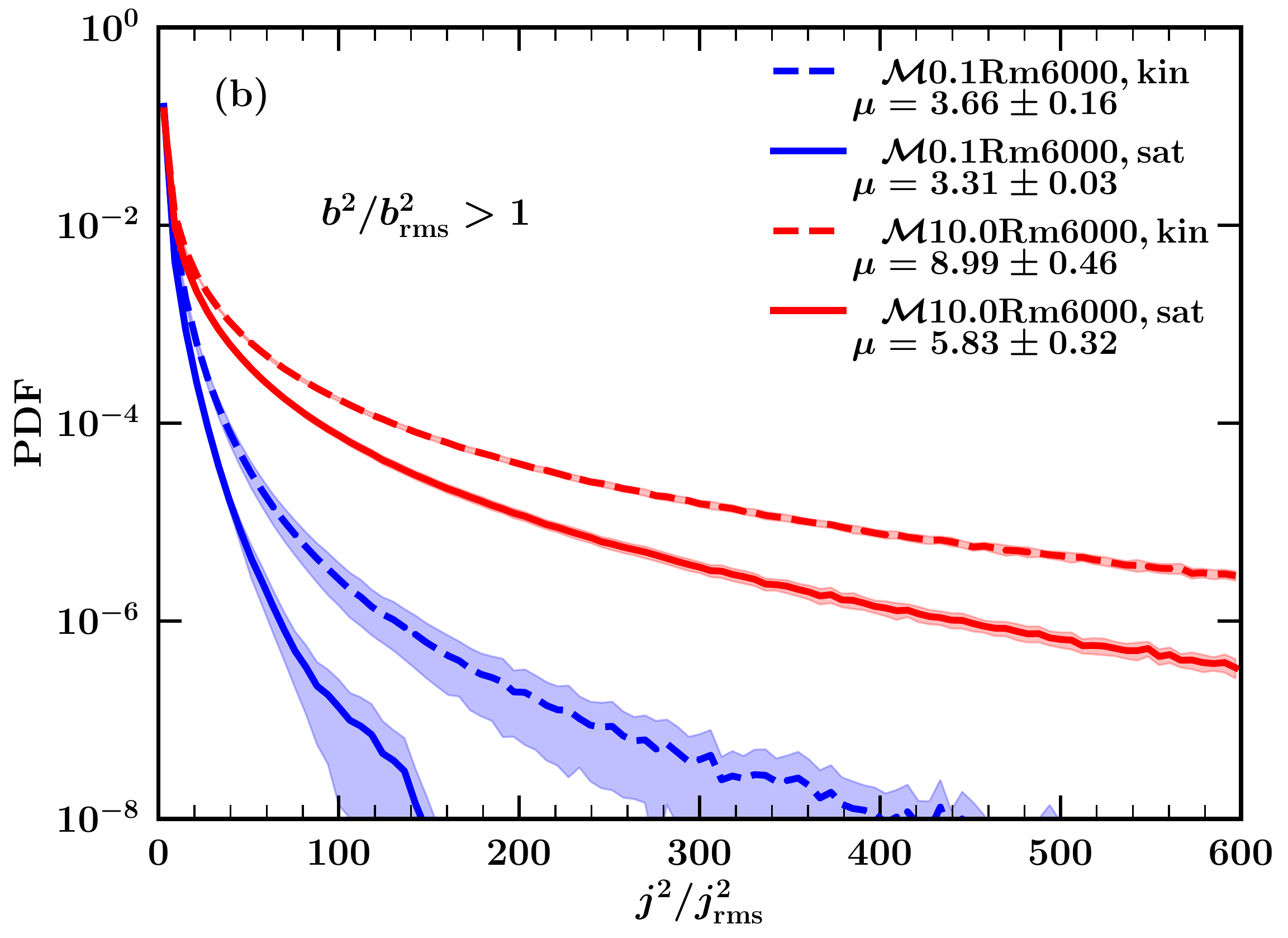}
    \caption{The total and conditional (regions with $b^2/\brms^2 > 1$) PDF of the dissipation (last) term in \Eq{eq:magen} for subsonic (blue) and supersonic (red) turbulence in the kinematic (kin, dashed) and saturated (sat, dashed) stages. The magnetic dissipation is statistically higher for the supersonic turbulence in both the kinematic and saturated stages. For both Mach numbers, the dissipation decreases as the field saturates and the decrease is higher for the supersonic case. The differences are enhanced in the strong-field regions (can also be seen via the mean of the distribution, $\mu$, in each case)}
    \label{fig:j2pdfs} 
\end{figure*}
We now explore the magnetic energy evolution equation and directly calculate the local growth and dissipation terms. The evolution of magnetic energy, for a fixed resistivity, $\eta$, can be described by the following equation (see \Sec{sec:app} for the derivation),
\begin{align}
    \int_V \frac{1}{2} \frac{\partial |\vec{b}|^2}{\partial t}~\dd V & = \int_V b_i b_j S_{ij}~\dd V 
     \nonumber \\
    & - \int_V \frac{1}{2}|\vec{b}|^2 (\nabla \cdot \vec{u}) ~\dd V  
    \nonumber \\
    & - \eta \int_V \vec{|j|}^2 ~\dd V. \label{eq:magen}
 \end{align}

The first term is due to stretching and compression of magnetic field lines by the turbulent flow and can increase (e.g., by stretching of magnetic field lines) or decrease (e.g., unstretching of magnetic field lines) the magnetic energy. Statistically, both can happen in a turbulent medium. This term is computed as follows. First, the local magnetic field is projected along the three eigenvectors of the rate of strain tensor ($\ecomp, \enull,$ and $\estre$) and let these projected vectors be $\vec{b}_{\rm comp}, \vec{b}_{\rm null}, $ and $\vec{b}_{\rm stre}$. The first term is then calculated as the sum $\sum_{i=\rm comp, null, stre} (\lambda_i/\urms)(\vec{b}_i/\brms)^2$. The second term can also reduce or enhance magnetic energy locally depending on the divergence of the velocity. \rev{As expected}, this term is negligible for the subsonic case, but does play an important role in the supersonic turbulence. The third term is the dissipative term, which always reduces the magnetic energy.

\rev{\Fig{fig:lgpdfs} shows the total and conditional (only in the strong-field regions) PDFs of the first, second, and first and second terms combined of \Eq{eq:magen} for subsonic and supersonic turbulence. For the subsonic case, the second term makes a small difference since $\nabla \cdot (\vec{u}/\urms)$ is negligible. Thus, the magnetic field growth term (first term or combination of first two terms in the magnetic energy evolution equation for the subsonic case, \Eq{eq:magen}) statistically decreases as the dynamo saturates. This implies that the growing magnetic field reduces its own amplification. The differences are further enhanced in strong-field regions (\Fig{fig:lgpdfs}~(f)). For the supersonic case, the combined term is statistically higher in comparison to the subsonic one, but there is not much difference between the corresponding kinematic and saturated stages (\Fig{fig:lgpdfs}~(e), also very similar mean value, $\mu$, for the kinematic and saturated stages). In strong-field regions, the local growth term is slightly reduced on saturation for the supersonic case. Thus, on saturation, the local growth term is reduced throughout the volume for the subsonic case, but mostly in the strong-field regions for the supersonic case. However, the mean of the local growth term ($\mu$ in \Fig{fig:lgpdfs}~(e) and  \Fig{fig:lgpdfs}~(f)) always remains positive for all cases. Thus, the magnetic field always grows. Even in the saturated state, the magnetic energy is amplified to counter diffusion.}

\Fig{fig:j2pdfs} show the PDF for the last term on the right-hand side of \Eq{eq:magen}, $j^{2}$, which probes the dissipation of magnetic energy. The dissipation is also statistically higher for the supersonic case in comparison to the subsonic one. For both cases, the magnetic dissipation statistically decreases as the dynamo saturates and the decrease is more statistically significant for the strong-field regions.

\begin{table*}
    \centering
	\caption{The ratio of the mean local magnetic growth ($\sum_{i=\rm comp, null, stre} (\lambda_i/\urms)(\vec{b}_i/\brms)^2 - (1/2)(b/\brms)^2 (\nabla \cdot (\vec{u}/\urms))$, \Fig{fig:lgpdfs} e, f) and dissipation ($j^2/\jrms^2$, \Fig{fig:j2pdfs} a, b) terms in the kinematic (kin) and saturated (sat) stages and the difference in the ratio between the two stages (kin - sat) for subsonic ($\Mach 0.1\Rm 6000$) and supersonic ($\Mach 10\Rm 6000$) turbulence. The corresponding numbers for only strong-field regions ($b^2/\brms^2 > 1$) are also provided. This shows that, even though both the growth and dissipation of the magnetic field deceases on saturation, the magnetic dissipation relative to the growth is enhanced.}
	\label{tab:lgd}
	\begin{tabular}{lcccccc} 
		\hline
		\hline
        Simulation Name & kin &  sat & (kin - sat) & kin, $b^2/\brms^2 > 1$ &  sat, $b^2/\brms^2 > 1$ & (kin - sat), $b^2/\brms^2 > 1$ 
        \\
        \hline
        $\Mach 0.1\Rm 6000$ & $0.51 \pm 0.09$ & $0.16 \pm 0.01$ & $0.35 \pm 0.09$ & $0.86 \pm 0.16$ & $0.23 \pm 0.01$ & $0.63 \pm 0.17$ \\
        $\Mach 10\Rm 6000$ & $0.78 \pm 0.04$ & $0.71 \pm 0.05$ & $0.08 \pm 0.06$ & $1.29 \pm 0.09$ & $0.95 \pm 0.13$ & $0.34 \pm 0.16$ \\
        \hline
		\hline
   \end{tabular}
\end{table*}

\Tab{tab:lgd} shows the ratio of the mean value of local growth term ($\sum_{i=\rm comp, null, stre} (\lambda_i/\urms)(\vec{b}_i/\brms)^2 - (1/2)(b/\brms)^2 (\nabla \cdot (\vec{u}/\urms))$ to the mean value of the local dissipation term ($j^2/\jrms^2$) in the kinematic and saturated stages for subsonic and supersonic turbulence (the magnetic resistivity, $\eta$, is the same for both runs). The ratio is always higher in the kinematic stage as compared to the saturated stage (see (kin-sat) in \Tab{tab:lgd}), especially in the strong-field regions. \rev{Note that, unlike the subsonic case, the difference of the ratio between the kinematic and saturated stages is small for the supersonic case and turns out to be significant only in the strong-field regions.} Thus, although both growth and dissipation of magnetic fields decrease as the field saturates, the magnetic dissipation relative to the amplification is enhanced. This leads to the saturation of the fluctuation dynamo. 


\section{Summary and Conclusions} \label{sec:con}
Using driven turbulence numerical simulations, we explore the saturation mechanism of the fluctuation dynamo. Our main aim was to study the effect of compressibility on the dynamo generated fields and the saturation mechanism. We numerically solve the equations of non-ideal compressible MHD for an isothermal gas (\Eq{eq:ce} -- \Eq{eq:div}) with very weak seed magnetic fields (random with mean zero) and primarily vary the Mach number, $\Mach$, of the turbulent driving. For all the cases, the magnetic field first amplifies exponentially (kinematic stage) and then saturates (saturated stage). We first study the global (over the entire domain) dynamo properties (growth rate, saturation level, spectra, and magnetic intermittency) as a function of $\Mach$. Then we explore the local properties and interactions of velocity and magnetic fields for subsonic ($\Mach=0.1$) and supersonic ($\Mach=10$) turbulent flows in the kinematic and saturated stages of the fluctuation dynamo. For the local study, we also isolate and study the regions with higher magnetic energy ($b^2/\brms^2>1$) as we expect that the dynamical effects of magnetic fields would be stronger in those regions. We summarize and conclude our key results below:
\begin{itemize}
    \item The growth rate of the magnetic energy decreases till $\Mach=5$ and then increases for $\Mach=10$ (\Fig{fig:gammasat}~(a)). The fraction of turbulent kinetic energy getting converted to magnetic energy, per unit time, decreases with increasing $\Mach$ (\Fig{fig:gammasat}~(b)). Thus, the overall efficiency of the dynamo decreases as the compressibility increases.
    
    \item The turbulent kinetic energy \rev{in the kinematic stage}, over a range of wavenumbers, seems to follow the Kolmogorov $k^{-5/3}$ power spectrum for subsonic turbulence and the Burgers $k^{-2}$ power spectrum for supersonic turbulence (\Fig{fig:spec}~(a)). \rev{As the field saturates, the kinetic energy power spectrum steepens for the subsonic case (also reflected in the velocity correlation length, \Fig{fig:spec}~(c)) and such a change is not that significant in the supersonic stage.} In the kinematic stage, the magnetic energy power spectrum is consistent, at larger scales, with the Kazantsev $k^{3/2}$ spectrum (the agreement is better for the subsonic turbulence). The magnetic spectra for both subsonic and supersonic flows are flatter at larger scales in the saturated stage (\Fig{fig:spec}~(b)). The computed magnetic field correlation length also increases for both Mach numbers as the dynamo saturates (\Fig{fig:spec}~(d)). 
    
    \item The velocity fields for both subsonic and supersonic turbulence roughly follows a Gaussian distribution (\Fig{fig:vxbxpdf}~(a)), the magnetic fields they amplify are non-Gaussian or spatially intermittent (\Fig{fig:vxbxpdf}~(b)). The intermittency decreases for both Mach numbers as the field saturates, but is always higher for the supersonic case (\Fig{fig:bkurt}). \rev{Furthermore, in the kinematic stage, the PDF of $b/\brms$ roughly follows a lognormal distribution and the fit is better for the supersonic case.} The effects of enhanced magnetic intermittency with compressibility must be considered while studying the effects of the fluctuation dynamo action in star-forming regions.

    \item Locally, for the subsonic turbulence, we find that the level of alignment between the velocity and vorticity, velocity and magnetic fields, and magnetic fields and current density is enhanced as the magnetic field saturates (\Fig{fig:apdfs}). \rev{However, for the supersonic case, the distribution of angles between the velocity and vorticity and vorticity and magnetic fields remains statistically the same in both stages and only the level of alignment between current density and magnetic fields is enhanced on saturation. This shows that back-reaction of the growing magnetic field on the velocity field is not that significant in supersonic turbulence in comparison to subsonic flows.}
    
    \item \rev{We also compute the evolution of following characteristic magnetic length scales: length scales associated with highest stretching of magnetic field lines (\Eq{eq:kpar}), resistive dissipation (\Eq{eq:kbdj}), highest compression of magnetic field lines (\Eq{eq:kbcj}), and overall field variation (\Eq{eq:krms}). Based on these scale, magnetic field structures in the kinematic stage of the subsonic turbulence can probably be considered as folded ribbons, but magnetic structures in the saturated stage and in both stages for the supersonic case are neither folded sheets nor ribbons. As the field saturates, all of these length scales are enhanced for the subsonic case and all but the length scale associated with the resistive dissipation is enhanced for the supersonic case (\Fig{fig:charlens}).} 
    
    \item We compute the eigenvalues and eigenvectors of the rate of strain tensor to study the local magnetic field line stretching and compression. The effect of the velocity is maximised when the direction of the magnetic field line stretching is aligned with the magnetic field and the direction of the magnetic field line compression is orthogonal to the magnetic field.  In subsonic turbulence, the level of alignment and orthogonality of direction of stretching and compression with the magnetic field, respectively, is reduced as the field saturates. However, for the supersonic case, the difference between the kinematic and saturated stages in not that significant (\Fig{fig:aevpdfs}). \rev{The magnetic field is slightly more orthogonal with the local compression direction in the saturated stage (although not in the strong-field regions), which enhances the effect of local compression.}
    
    \item Finally, we compute each term in the magnetic energy evolution equation (\Eq{eq:magen}) and show that both the local growth (\Fig{fig:lgpdfs}) and dissipation (\Fig{fig:j2pdfs}) of magnetic fields decreases as the field saturates for the subsonic case. For the supersonic case, overall, there is not much difference in the local growth between the kinematic and saturated stages, but the local growth is reduced in the strong-field regions. As in the subsonic case, the dissipation is also reduced in the saturated stage for supersonic flows (\Fig{fig:j2pdfs}). However, even though both the amplification and dissipation of magnetic fields statistically decreases as the field saturates, the dissipation relative to the amplification is enhanced (\Tab{tab:lgd}). Thus, the exponentially growing magnetic fields evolve to alter both the amplification and dissipation mechanisms of the fluctuation dynamo to achieve saturation. This also implies that a drastic change in either of them is not required. \rev{This change is significant throughout the volume for the subsonic case, but primarily occurs in strong-field regions for the supersonic turbulence.}
\end{itemize}


\begin{acknowledgments} 
 \rev{We thank both referees for their useful suggestions and comments.}   A.~S.~thanks Paul Bushby, Anvar Shukurov, and Toby Wood for useful discussions. C.~F.~acknowledges funding provided by the Australian Research Council (Discovery Project DP170100603 and Future Fellowship FT180100495), and the Australia-Germany Joint Research Cooperation Scheme (UA-DAAD). We further acknowledge high-performance computing resources provided by the Leibniz Rechenzentrum and the Gauss Centre for Supercomputing (grants~pr32lo, pr48pi, and GCS Large-scale project~10391), and the Australian National Computational Infrastructure (grant~ek9) in the framework of the National Computational Merit Allocation Scheme and the ANU Merit Allocation Scheme.
\end{acknowledgments}
\appendix 
\section{Evolution of magnetic energy in supersonic plasmas} \label{sec:app}
Here, we derive the magnetic energy evolution equation from the induction equation,
\begin{align}
     \quad \quad \frac{\partial \vec{b}}{\partial t} = \nabla \times (\vec{u} \times \vec{b}) + \eta \nabla^2{\vec{b}}.
     \label{eq:appie}
\end{align}
For a constant $\eta$, taking a dot product of \Eq{eq:appie} with $\vec{b}$ and then integrating over the volume, V, gives
\begin{align}
     \int_V \vec{b} \cdot \frac{\partial \vec{b}}{\partial t} ~\dd V = \int_V \vec{b} \cdot (\nabla \times (\vec{u} \times \vec{b}))~\dd V + \eta \int_V \vec{b} \cdot \nabla^2{\vec{b}}~\dd V, \\
     \int_V \frac{1}{2} \frac{\partial |\vec{b}|^2}{\partial t}~\dd V = \underbrace{\int_V \vec{b} \cdot (\nabla \times (\vec{u} \times \vec{b}))~\dd V}_\text{first term} + \underbrace{\eta \int_V \vec{b} \cdot \nabla^2{\vec{b}}~\dd V.}_\text{second term} 
\end{align}
The right-hand side term in the above equation represents the evolution of magnetic energy.

Simplifying the first term further,
\begin{align}
      & \int_V \vec{b} \cdot (\nabla \times (\vec{u} \times \vec{b}))~\dd V \\
      & = \int_V \vec{b} \cdot  (\vec{u}(\underbrace{\nabla \cdot \vec{b}}_\text{$=0$}) - \vec{b}(\vec{u} \cdot \nabla)\vec{b} + (\vec{b}\cdot\nabla)\vec{u} - (\vec{u}\cdot \nabla)\vec{b})~\dd V, \\
      & = \int_V   (\vec{b} \cdot(\vec{b}\cdot\nabla)\vec{u} - \vec{b}\cdot(\vec{u}\cdot\nabla)\vec{b} - \vec{b}\vec{b} (\nabla\cdot \vec{u}))~\dd V, \\
      & = \int_V (b_i (b_j \partial_j)u_i - b_i(u_j \partial_j)b_i - |\vec{b}|^2(\nabla \cdot \vec{u}))~\dd V, \\
      & = \int_V (b_i b_j (\partial_j u_i) - u_j \partial_j \frac{b_i^2}{2}-|\vec{b}|^2(\nabla \cdot \vec{u}))~\dd V, \\ 
    & \text{$S_{ij} = \partial_j u_i$ is the rate of strain tensor,} \nonumber \\
     & = \int_V (b_i b_j S_{ij} - \vec{u} \cdot \frac{1}{2} \nabla |\vec{b}|^2 - |\vec{b}|^2(\nabla \cdot \vec{u}))~\dd V, \\ 
     & = \int_V (b_i b_j S_{ij} - \left(\frac{1}{2} \nabla \cdot (\vec{u}|\vec{b}|^2) - \frac{1}{2}|\vec{b}|^2 (\nabla \cdot \vec{u})\right) - |\vec{b}|^2(\nabla \cdot \vec{u}))~\dd V, \\
     & = \int_V (b_i b_j S_{ij} - \frac{1}{2}|\vec{b}|^2 (\nabla \cdot \vec{u})) ~\dd V -\frac{1}{2} \int_V \nabla \cdot (\vec{u}|\vec{b}|^2)~\dd V, \\ 
     & = \int_V (b_i b_j S_{ij} - \frac{1}{2}|\vec{b}|^2 (\nabla \cdot \vec{u})) ~\dd V - \underbrace{\frac{1}{2} \int_S (\vec{u}|\vec{b}|^2) \cdot \hat{n}~\dd S.}_\text{$=0$ for periodic BCs}
 \end{align} 
 The last term in the above equation, being a surface integral, integrates out to zero for periodic boundary conditions. \\

Simplifying the second term further,
\begin{align}
      & \eta \int_V \vec{b} \cdot \nabla^2{\vec{b}}~\dd V  =  \eta \int_V \vec{b} \cdot (\nabla (\underbrace{\nabla \cdot \vec{b}}_\text{$=0$}) - \nabla \times (\nabla \times \vec{b}))~\dd V \\ 
      & = - \eta \int_V \vec{b} \cdot (\nabla \times (\nabla \times \vec{b}))~\dd V \\
      & = - \eta \int_V \vec{b} \cdot (\nabla \times \vec{j})~\dd V \\
      & = - \eta \int_V \vec{b} \cdot (\epsilon_{ijk} \partial_j j_k) ~\dd V \\
      & = - \eta \int_V b_i \epsilon_{ijk} \partial_j j_k ~\dd V \\
     & = - \eta \int_V j_k \underbrace{\epsilon_{ijk} \partial_j b_i}_\text{$j_k$} ~\dd V \\ 
     & = - \eta \int_V j_k^2 ~\dd V \\
     & = - \eta \int_V \vec{|j|}^2 ~\dd V  
\end{align}
Thus, the magnetic energy evolution equation is
\begin{align}
    \int_V \frac{1}{2} \frac{\partial |\vec{b}|^2}{\partial t}~\dd V & = \int_V b_i b_j S_{ij}~\dd V 
     \nonumber \\
    & - \int_V \frac{1}{2}|\vec{b}|^2 (\nabla \cdot \vec{u}) ~\dd V  
    \nonumber \\
    & - \eta \int_V \vec{|j|}^2 ~\dd V. 
 \end{align}
\bibliography{compdyn}
\end{document}